\documentclass[longauth,traditabstract]{aa}
\usepackage{etex}
\pdfoutput=1

\def\setsymbol#1#2{\expandafter\def\csname #1\endcsname{#2}}
\def\getsymbol#1{\csname #1\endcsname}

\def\Planck{\textit{Planck}}

\def\alltwentyfifteenresultspapers{\nocite{planck2014-a01, planck2014-a03, planck2014-a04, planck2014-a05, planck2014-a06, planck2014-a07, planck2014-a08, planck2014-a09, planck2014-a11, planck2014-a12, planck2014-a13, planck2014-a14, planck2014-a15, planck2014-a16, planck2014-a17, planck2014-a18, planck2014-a19, planck2014-a20, planck2014-a22, planck2014-a24, planck2014-a26, planck2014-a28, planck2014-a29, planck2014-a30, planck2014-a31, planck2014-a35, planck2014-a36, planck2014-a37, planck2014-ES}}

\newbox\tablebox    \newdimen\tablewidth
\def\leaderfil{\leaders\hbox to 5pt{\hss.\hss}\hfil}
\def\endPlancktable{\tablewidth=\columnwidth 
    $$\hss\copy\tablebox\hss$$
    \vskip-\lastskip\vskip -2pt}
\def\endPlancktablewide{\tablewidth=\textwidth 
    $$\hss\copy\tablebox\hss$$
    \vskip-\lastskip\vskip -2pt}
\def\tablenote#1 #2\par{\begingroup \parindent=0.8em
    \abovedisplayshortskip=0pt\belowdisplayshortskip=0pt
    \noindent
    $$\hss\vbox{\hsize\tablewidth \hangindent=\parindent \hangafter=1 \noindent
    \hbox to \parindent{$^#1$\hss}\strut#2\strut\par}\hss$$
    \endgroup}
\def\doubleline{\vskip 3pt\hrule \vskip 1.5pt \hrule \vskip 5pt}

\def\L2{\ifmmode L_2\else $L_2$\fi}

\def\DeltaT{\ifmmode \Delta T\else $\Delta T$\fi}
\def\deltat{\ifmmode \Delta t\else $\Delta t$\fi}
\def\fknee{\ifmmode f_{\rm knee}\else $f_{\rm knee}$\fi}
\def\Fmax{\ifmmode F_{\rm max}\else $F_{\rm max}$\fi}
\def\solar{\ifmmode{\rm M}_{\mathord\odot}\else${\rm M}_{\mathord\odot}$\fi}
\def\Msolar{\ifmmode{\rm M}_{\mathord\odot}\else${\rm M}_{\mathord\odot}$\fi}
\def\Lsolar{\ifmmode{\rm L}_{\mathord\odot}\else${\rm L}_{\mathord\odot}$\fi}
\def\inv{\ifmmode^{-1}\else$^{-1}$\fi}
\def\mo{\ifmmode^{-1}\else$^{-1}$\fi}
\def\sup#1{\ifmmode ^{\rm #1}\else $^{\rm #1}$\fi}
\def\expo#1{\ifmmode \times 10^{#1}\else $\times 10^{#1}$\fi}
\def\,{\thinspace}
\def\lsim{\mathrel{\raise .4ex\hbox{\rlap{$<$}\lower 1.2ex\hbox{$\sim$}}}}
\def\gsim{\mathrel{\raise .4ex\hbox{\rlap{$>$}\lower 1.2ex\hbox{$\sim$}}}}

\def\simprop{\mathrel{\raise .4ex\hbox{\rlap{$\propto$}\lower 1.2ex\hbox{$\sim$}}}}
\def\deg{\ifmmode^\circ\else$^\circ$\fi}
\def\pdeg{\ifmmode $\setbox0=\hbox{$^{\circ}$}\rlap{\hskip.11\wd0 .}$^{\circ}
          \else \setbox0=\hbox{$^{\circ}$}\rlap{\hskip.11\wd0 .}$^{\circ}$\fi}
\def\arcs{\ifmmode {^{\scriptstyle\prime\prime}}
          \else $^{\scriptstyle\prime\prime}$\fi}
\def\arcm{\ifmmode {^{\scriptstyle\prime}}
          \else $^{\scriptstyle\prime}$\fi}
\newdimen\sa  \newdimen\sb
\def\parcs{\sa=.07em \sb=.03em
     \ifmmode \hbox{\rlap{.}}^{\scriptstyle\prime\kern -\sb\prime}\hbox{\kern -\sa}
     \else \rlap{.}$^{\scriptstyle\prime\kern -\sb\prime}$\kern -\sa\fi}
\def\parcm{\sa=.08em \sb=.03em
     \ifmmode \hbox{\rlap{.}\kern\sa}^{\scriptstyle\prime}\hbox{\kern-\sb}
     \else \rlap{.}\kern\sa$^{\scriptstyle\prime}$\kern-\sb\fi}
\def\ra[#1 #2 #3.#4]{#1\sup{h}#2\sup{m}#3\sup{s}\llap.#4}
\def\dec[#1 #2 #3.#4]{#1\deg#2\arcm#3\arcs\llap.#4}
\def\deco[#1 #2 #3]{#1\deg#2\arcm#3\arcs}
\def\rra[#1 #2]{#1\sup{h}#2\sup{m}}

\def\dots{\relax\ifmmode \ldots\else $\ldots$\fi}
\def\WHzsr{\ifmmode $W\,Hz\mo\,sr\mo$\else W\,Hz\mo\,sr\mo\fi}
\def\mHz{\ifmmode $\,mHz$\else \,mHz\fi}
\def\GHz{\ifmmode $\,GHz$\else \,GHz\fi}
\def\mKs{\ifmmode $\,mK\,s$^{1/2}\else \,mK\,s$^{1/2}$\fi}
\def\muKs{\ifmmode \,\mu$K\,s$^{1/2}\else \,$\mu$K\,s$^{1/2}$\fi}
\def\muKRJs{\ifmmode \,\mu$K$_{\rm RJ}$\,s$^{1/2}\else \,$\mu$K$_{\rm RJ}$\,s$^{1/2}$\fi}
\def\muKHz{\ifmmode \,\mu$K\,Hz$^{-1/2}\else \,$\mu$K\,Hz$^{-1/2}$\fi}
\def\MJysr{\ifmmode \,$MJy\,sr\mo$\else \,MJy\,sr\mo\fi}
\def\MJysrmK{\ifmmode \,$MJy\,sr\mo$\,mK$_{\rm CMB}\mo\else \,MJy\,sr\mo\,mK$_{\rm CMB}\mo$\fi}
\def\microns{\ifmmode \,\mu$m$\else \,$\mu$m\fi}

\def\muK{\ifmmode \,\mu$K$\else \,$\mu$\hbox{K}\fi}
\def\microK{\ifmmode \,\mu$K$\else \,$\mu$\hbox{K}\fi}
\def\muW{\ifmmode \,\mu$W$\else \,$\mu$\hbox{W}\fi}
\def\kms{\ifmmode $\,km\,s$^{-1}\else \,km\,s$^{-1}$\fi}
\def\kmsMpc{\ifmmode $\,\kms\,Mpc\mo$\else \,\kms\,Mpc\mo\fi}
\providecommand{\sorthelp}[1]{}

\usepackage{ifthen}
\usepackage{natbib}
\usepackage{amsmath}
\usepackage{graphicx}
\usepackage[varg]{txfonts}
\usepackage{fixltx2e}
\usepackage{url}
\usepackage[breaklinks,colorlinks,citecolor=blue]{hyperref}
\usepackage{tikz}
\usetikzlibrary{decorations,arrows,shapes}
\usepackage[ugly]{units}

\bibpunct{(}{)}{;}{a}{}{,}

\newcommand{\OldPlanckLFICalPaper}{Cal13}
\newcommand{\DaCapo}{{\tt DaCapo}}
\newcommand{\Commander}{{\tt Commander}}
\newcommand{\ud}{\mathrm{d}}
\newcommand{\phid}{\phi_D}
\newcommand{\phisky}{\phi_{\mathrm{sky}}}
\newcommand{\fsl}{f_\mathrm{sl}}
\newcommand{\Tsky}{T_{\mathrm{sky}}}
\newcommand{\Tgalaxy}{T_{\mathrm{Gal}}}
\newcommand{\Tother}{T_{\mathrm{other}}}
\newcommand{\Tcmb}{T_{\mathrm{CMB}}}
\newcommand{\Bmain}{B_{\mathrm{main}}}
\newcommand{\Bsl}{B_{\mathrm{sl}}}
\newcommand{\Tskymeas}{\tilde{T}_{\mathrm{sky}}}

\def\xversor{\vec{x}}

\newenvironment{DIFnomarkup}{}{}

\graphicspath{{images/}}

\tikzstyle{every picture}+=[font=\sffamily]

\begin{document}

\title{\textit{Planck} 2015 results. V. LFI calibration}

\authorrunning{Planck Collaboration}
%This author list corresponds to \title{Author list for A06\_LFI\_calibration}
%Prepared by M. Lopez-Caniego (Marcos.Lopez.Caniego@sciops.esa.int), ESAC/ESA
%This version is from Tue Nov 17 07:55:42 2015 CET
%\subtitle{There are 208 co-authors in this list}
\author{\small
Planck Collaboration: P.~A.~R.~Ade\inst{89}
\and
N.~Aghanim\inst{61}
\and
M.~Ashdown\inst{72, 6}
\and
J.~Aumont\inst{61}
\and
C.~Baccigalupi\inst{88}
\and
A.~J.~Banday\inst{97, 9}
\and
R.~B.~Barreiro\inst{67}
\and
N.~Bartolo\inst{30, 68}
\and
P.~Battaglia\inst{32, 34}
\and
E.~Battaner\inst{98, 99}
% ...
}
\institute{\small
APC, AstroParticule et Cosmologie, Universit\'{e} Paris Diderot, CNRS/IN2P3, CEA/lrfu, Observatoire de Paris, Sorbonne Paris Cit\'{e}, 10, rue Alice Domon et L\'{e}onie Duquet, 75205 Paris Cedex 13, France\goodbreak
\and
Aalto University Mets\"{a}hovi Radio Observatory and Dept of Radio Science and Engineering, P.O. Box 13000, FI-00076 AALTO, Finland\goodbreak
\and
African Institute for Mathematical Sciences, 6-8 Melrose Road, Muizenberg, Cape Town, South Africa\goodbreak
\and
Agenzia Spaziale Italiana Science Data Center, Via del Politecnico snc, 00133, Roma, Italy\goodbreak
\and
Aix Marseille Universit\'{e}, CNRS, LAM (Laboratoire d'Astrophysique de Marseille) UMR 7326, 13388, Marseille, France\goodbreak
\and
Astrophysics Group, Cavendish Laboratory, University of Cambridge, J J Thomson Avenue, Cambridge CB3 0HE, U.K.\goodbreak
\and
CGEE, SCS Qd 9, Lote C, Torre C, 4$^{\circ}$ andar, Ed. Parque Cidade Corporate, CEP 70308-200, Bras\'{i}lia, DF,Ê Brazil\goodbreak
\and
CITA, University of Toronto, 60 St. George St., Toronto, ON M5S 3H8, Canada\goodbreak
\and
CNRS, IRAP, 9 Av. colonel Roche, BP 44346, F-31028 Toulouse cedex 4, France\goodbreak
\and
CRANN, Trinity College, Dublin, Ireland\goodbreak
\and
California Institute of Technology, Pasadena, California, U.S.A.\goodbreak
\and
Centre for Theoretical Cosmology, DAMTP, University of Cambridge, Wilberforce Road, Cambridge CB3 0WA, U.K.\goodbreak
\and
Computational Cosmology Center, Lawrence Berkeley National Laboratory, Berkeley, California, U.S.A.\goodbreak
\and
Consejo Superior de Investigaciones Cient\'{\i}ficas (CSIC), Madrid, Spain\goodbreak
\and
DSM/Irfu/SPP, CEA-Saclay, F-91191 Gif-sur-Yvette Cedex, France\goodbreak
\and
DTU Space, National Space Institute, Technical University of Denmark, Elektrovej 327, DK-2800 Kgs. Lyngby, Denmark\goodbreak
\and
D\'{e}partement de Physique Th\'{e}orique, Universit\'{e} de Gen\`{e}ve, 24, Quai E. Ansermet,1211 Gen\`{e}ve 4, Switzerland\goodbreak
\and
Departamento de Astrof\'{i}sica, Universidad de La Laguna (ULL), E-38206 La Laguna, Tenerife, Spain\goodbreak
\and
Departamento de F\'{\i}sica, Universidad de Oviedo, Avda. Calvo Sotelo s/n, Oviedo, Spain\goodbreak
\and
Department of Astronomy and Astrophysics, University of Toronto, 50 Saint George Street, Toronto, Ontario, Canada\goodbreak
\and
Department of Astrophysics/IMAPP, Radboud University Nijmegen, P.O. Box 9010, 6500 GL Nijmegen, The Netherlands\goodbreak
\and
Department of Physics \& Astronomy, University of British Columbia, 6224 Agricultural Road, Vancouver, British Columbia, Canada\goodbreak
\and
Department of Physics and Astronomy, Dana and David Dornsife College of Letter, Arts and Sciences, University of Southern California, Los Angeles, CA 90089, U.S.A.\goodbreak
\and
Department of Physics and Astronomy, University College London, London WC1E 6BT, U.K.\goodbreak
\and
Department of Physics, Florida State University, Keen Physics Building, 77 Chieftan Way, Tallahassee, Florida, U.S.A.\goodbreak
\and
Department of Physics, Gustaf H\"{a}llstr\"{o}min katu 2a, University of Helsinki, Helsinki, Finland\goodbreak
\and
Department of Physics, Princeton University, Princeton, New Jersey, U.S.A.\goodbreak
\and
Department of Physics, University of California, Santa Barbara, California, U.S.A.\goodbreak
\and
Department of Physics, University of Illinois at Urbana-Champaign, 1110 West Green Street, Urbana, Illinois, U.S.A.\goodbreak
\and
Dipartimento di Fisica e Astronomia G. Galilei, Universit\`{a} degli Studi di Padova, via Marzolo 8, 35131 Padova, Italy\goodbreak
\and
Dipartimento di Fisica e Scienze della Terra, Universit\`{a} di Ferrara, Via Saragat 1, 44122 Ferrara, Italy\goodbreak
\and
Dipartimento di Fisica, Universit\`{a} La Sapienza, P. le A. Moro 2, Roma, Italy\goodbreak
\and
Dipartimento di Fisica, Universit\`{a} degli Studi di Milano, Via Celoria, 16, Milano, Italy\goodbreak
\and
Dipartimento di Fisica, Universit\`{a} degli Studi di Trieste, via A. Valerio 2, Trieste, Italy\goodbreak
\and
Dipartimento di Matematica, Universit\`{a} di Roma Tor Vergata, Via della Ricerca Scientifica, 1, Roma, Italy\goodbreak
\and
Discovery Center, Niels Bohr Institute, Blegdamsvej 17, Copenhagen, Denmark\goodbreak
\and
Discovery Center, Niels Bohr Institute, Copenhagen University, Blegdamsvej 17, Copenhagen, Denmark\goodbreak
\and
European Space Agency, ESAC, Planck Science Office, Camino bajo del Castillo, s/n, Urbanizaci\'{o}n Villafranca del Castillo, Villanueva de la Ca\~{n}ada, Madrid, Spain\goodbreak
\and
European Space Agency, ESTEC, Keplerlaan 1, 2201 AZ Noordwijk, The Netherlands\goodbreak
\and
Facolt\`{a} di Ingegneria, Universit\`{a} degli Studi e-Campus, Via Isimbardi 10, Novedrate (CO), 22060, Italy\goodbreak
\and
Gran Sasso Science Institute, INFN, viale F. Crispi 7, 67100 L'Aquila, Italy\goodbreak
\and
HGSFP and University of Heidelberg, Theoretical Physics Department, Philosophenweg 16, 69120, Heidelberg, Germany\goodbreak
\and
Haverford College Astronomy Department, 370 Lancaster Avenue, Haverford, Pennsylvania, U.S.A.\goodbreak
\and
Helsinki Institute of Physics, Gustaf H\"{a}llstr\"{o}min katu 2, University of Helsinki, Helsinki, Finland\goodbreak
\and
INAF - Osservatorio Astrofisico di Catania, Via S. Sofia 78, Catania, Italy\goodbreak
\and
INAF - Osservatorio Astronomico di Padova, Vicolo dell'Osservatorio 5, Padova, Italy\goodbreak
\and
INAF - Osservatorio Astronomico di Roma, via di Frascati 33, Monte Porzio Catone, Italy\goodbreak
\and
INAF - Osservatorio Astronomico di Trieste, Via G.B. Tiepolo 11, Trieste, Italy\goodbreak
\and
INAF/IASF Bologna, Via Gobetti 101, Bologna, Italy\goodbreak
\and
INAF/IASF Milano, Via E. Bassini 15, Milano, Italy\goodbreak
\and
INFN, Sezione di Bologna, Via Irnerio 46, I-40126, Bologna, Italy\goodbreak
\and
INFN, Sezione di Roma 1, Universit\`{a} di Roma Sapienza, Piazzale Aldo Moro 2, 00185, Roma, Italy\goodbreak
\and
INFN, Sezione di Roma 2, Universit\`{a} di Roma Tor Vergata, Via della Ricerca Scientifica, 1, Roma, Italy\goodbreak
\and
INFN/National Institute for Nuclear Physics, Via Valerio 2, I-34127 Trieste, Italy\goodbreak
\and
ISDC, Department of Astronomy, University of Geneva, ch. d'Ecogia 16, 1290 Versoix, Switzerland\goodbreak
\and
IUCAA, Post Bag 4, Ganeshkhind, Pune University Campus, Pune 411 007, India\goodbreak
\and
Imperial College London, Astrophysics group, Blackett Laboratory, Prince Consort Road, London, SW7 2AZ, U.K.\goodbreak
\and
Infrared Processing and Analysis Center, California Institute of Technology, Pasadena, CA 91125, U.S.A.\goodbreak
\and
Institut N\'{e}el, CNRS, Universit\'{e} Joseph Fourier Grenoble I, 25 rue des Martyrs, Grenoble, France\goodbreak
\and
Institut Universitaire de France, 103, bd Saint-Michel, 75005, Paris, France\goodbreak
\and
Institut d'Astrophysique Spatiale, CNRS (UMR8617) Universit\'{e} Paris-Sud 11, B\^{a}timent 121, Orsay, France\goodbreak
\and
Institut d'Astrophysique de Paris, CNRS (UMR7095), 98 bis Boulevard Arago, F-75014, Paris, France\goodbreak
\and
Institut f\"ur Theoretische Teilchenphysik und Kosmologie, RWTH Aachen University, D-52056 Aachen, Germany\goodbreak
\and
Institute of Astronomy, University of Cambridge, Madingley Road, Cambridge CB3 0HA, U.K.\goodbreak
\and
Institute of Theoretical Astrophysics, University of Oslo, Blindern, Oslo, Norway\goodbreak
\and
Instituto de Astrof\'{\i}sica de Canarias, C/V\'{\i}a L\'{a}ctea s/n, La Laguna, Tenerife, Spain\goodbreak
\and
Instituto de F\'{\i}sica de Cantabria (CSIC-Universidad de Cantabria), Avda. de los Castros s/n, Santander, Spain\goodbreak
\and
Istituto Nazionale di Fisica Nucleare, Sezione di Padova, via Marzolo 8, I-35131 Padova, Italy\goodbreak
\and
Jet Propulsion Laboratory, California Institute of Technology, 4800 Oak Grove Drive, Pasadena, California, U.S.A.\goodbreak
\and
Jodrell Bank Centre for Astrophysics, Alan Turing Building, School of Physics and Astronomy, The University of Manchester, Oxford Road, Manchester, M13 9PL, U.K.\goodbreak
\and
Kavli Institute for Cosmological Physics, University of Chicago, Chicago, IL 60637, USA\goodbreak
\and
Kavli Institute for Cosmology Cambridge, Madingley Road, Cambridge, CB3 0HA, U.K.\goodbreak
\and
Kazan Federal University, 18 Kremlyovskaya St., Kazan, 420008, Russia\goodbreak
\and
LAL, Universit\'{e} Paris-Sud, CNRS/IN2P3, Orsay, France\goodbreak
\and
LERMA, CNRS, Observatoire de Paris, 61 Avenue de l'Observatoire, Paris, France\goodbreak
\and
Laboratoire AIM, IRFU/Service d'Astrophysique - CEA/DSM - CNRS - Universit\'{e} Paris Diderot, B\^{a}t. 709, CEA-Saclay, F-91191 Gif-sur-Yvette Cedex, France\goodbreak
\and
Laboratoire Traitement et Communication de l'Information, CNRS (UMR 5141) and T\'{e}l\'{e}com ParisTech, 46 rue Barrault F-75634 Paris Cedex 13, France\goodbreak
\and
Laboratoire de Physique Subatomique et Cosmologie, Universit\'{e} Grenoble-Alpes, CNRS/IN2P3, 53, rue des Martyrs, 38026 Grenoble Cedex, France\goodbreak
\and
Laboratoire de Physique Th\'{e}orique, Universit\'{e} Paris-Sud 11 \& CNRS, B\^{a}timent 210, 91405 Orsay, France\goodbreak
\and
Lawrence Berkeley National Laboratory, Berkeley, California, U.S.A.\goodbreak
\and
Lebedev Physical Institute of the Russian Academy of Sciences, Astro Space Centre, 84/32 Profsoyuznaya st., Moscow, GSP-7, 117997, Russia\goodbreak
\and
Max-Planck-Institut f\"{u}r Astrophysik, Karl-Schwarzschild-Str. 1, 85741 Garching, Germany\goodbreak
\and
McGill Physics, Ernest Rutherford Physics Building, McGill University, 3600 rue University, Montr\'{e}al, QC, H3A 2T8, Canada\goodbreak
\and
National University of Ireland, Department of Experimental Physics, Maynooth, Co. Kildare, Ireland\goodbreak
\and
Nicolaus Copernicus Astronomical Center, Bartycka 18, 00-716 Warsaw, Poland\goodbreak
\and
Niels Bohr Institute, Blegdamsvej 17, Copenhagen, Denmark\goodbreak
\and
Niels Bohr Institute, Copenhagen University, Blegdamsvej 17, Copenhagen, Denmark\goodbreak
\and
SISSA, Astrophysics Sector, via Bonomea 265, 34136, Trieste, Italy\goodbreak
\and
School of Physics and Astronomy, Cardiff University, Queens Buildings, The Parade, Cardiff, CF24 3AA, U.K.\goodbreak
\and
School of Physics and Astronomy, University of Nottingham, Nottingham NG7 2RD, U.K.\goodbreak
\and
Sorbonne Universit\'{e}-UPMC, UMR7095, Institut d'Astrophysique de Paris, 98 bis Boulevard Arago, F-75014, Paris, France\goodbreak
\and
Space Sciences Laboratory, University of California, Berkeley, California, U.S.A.\goodbreak
\and
Special Astrophysical Observatory, Russian Academy of Sciences, Nizhnij Arkhyz, Zelenchukskiy region, Karachai-Cherkessian Republic, 369167, Russia\goodbreak
\and
Sub-Department of Astrophysics, University of Oxford, Keble Road, Oxford OX1 3RH, U.K.\goodbreak
\and
Theory Division, PH-TH, CERN, CH-1211, Geneva 23, Switzerland\goodbreak
\and
UPMC Univ Paris 06, UMR7095, 98 bis Boulevard Arago, F-75014, Paris, France\goodbreak
\and
Universit\'{e} de Toulouse, UPS-OMP, IRAP, F-31028 Toulouse cedex 4, France\goodbreak
\and
University of Granada, Departamento de F\'{\i}sica Te\'{o}rica y del Cosmos, Facultad de Ciencias, Granada, Spain\goodbreak
\and
University of Granada, Instituto Carlos I de F\'{\i}sica Te\'{o}rica y Computacional, Granada, Spain\goodbreak
% ...
}

%\author{Planck Collaboration\thanks{Corresponding author: Maurizio~Tomasi, \url{mailto:maurizio.tomasi@unimi.it}, Skype: \texttt{zio\_{}tom78}}}

\abstract{We present a description of the pipeline used to calibrate the \Planck\ Low Frequency Instrument (LFI) timelines into thermodynamic temperatures for the \Planck\ 2015 data release, covering four years of uninterrupted operations. As in the 2013 data release, our calibrator is provided by the spin-synchronous modulation of the cosmic microwave background dipole, but we now use the orbital component, rather than adopting the Wilkinson Microwave Anisotropy Probe (WMAP) solar dipole.  This allows our 2015 LFI analysis to provide an independent Solar dipole estimate, which is in excellent agreement with that of HFI and within $1\sigma$ (0.3\,\% in amplitude) of the WMAP value. This 0.3\,\% shift in the peak-to-peak dipole temperature from WMAP and a global overhaul of the iterative calibration code increases the overall level of the LFI maps by 0.45\,\% (30\,GHz), 0.64\,\% (44\,GHz), and 0.82\,\% (70\,GHz) in temperature with respect to the 2013 \Planck\ data release, thus reducing the discrepancy with the power spectrum measured by WMAP. We estimate that the LFI calibration uncertainty is now at the level of 0.20\,\% for the 70\,GHz map, 0.26\,\% for the 44\,GHz map, and 0.35\,\% for the 30\,GHz map. We provide a detailed description of the impact of all the changes implemented in the calibration since the previous data release.}


\maketitle

\alltwentyfifteenresultspapers

%%%%%%%%%%%%%%%%%%%%%%%%%%%%%%%%%%%%%%%%%%%%%%%%%%%%%%%%%%%%%%%%%%%%%%

\section{Introduction}

This paper, one of a set associated with the 2015 release of data from the \Planck{}\footnote{\Planck\ (\url{http://www.esa.int/Planck}) is a project of the European Space Agency (ESA) with instruments provided by two scientific consortia funded by ESA member states and led by Principal Investigators from France and Italy, telescope reflectors provided through a collaboration between ESA and a scientific consortium led and funded by Denmark, and additional contributions from NASA (USA).} mission, describes the techniques we employed to calibrate the voltages measured by the Low Frequency Instrument (LFI) radiometers into a set of thermodynamic temperatures, which we refer to as ``photometric calibration.'' We expand here the work described in \citet{planck2013-p02b}, henceforth \OldPlanckLFICalPaper; we try to follow as closely as possible the structure of the earlier paper to help the reader understand what has changed between the 2013 and the 2015 \Planck\ data releases.

The calibration of both \Planck\ instruments \citep[for HFI, see][]{planck2014-a08} is now based on the small ($270\,\mu\mathrm{K}$) dipole signal induced by the annual motion of the satellite around the Sun -- the orbital dipole, which we derive from our knowledge of the orbital parameters of the spacecraft.  The calibration is thus \emph{absolute}, and not dependent on external measurements of the larger solar ($3.35\,\mathrm{mK}$) dipole, as was the case for \OldPlanckLFICalPaper.  Absolute calibration allows us both to improve the current measurement of the solar dipole (see Sect.~\ref{sec:DipoleEstimation}), and to transfer \Planck's calibration to various ground-based instruments \citep[see, e.g.,][]{perley2015} and other cosmic microwave background (CMB) experiments \citep[e.g.,][]{louis2014act}.

Accurate calibration of the LFI is crucial to ensure reliable cosmological and astrophysical results from the \Planck\ mission. Internally consistent photometric calibration of the nine \Planck\ frequency channels is essential for component separation, where we disentangle the CMB from the varoius Galactic and extragalactic foreground emission processes
\citep{planck2014-a11,planck2014-a12}. In addition, the LFI calibration directly impacts the \Planck\ polarization likelihood at low multipoles, based on the LFI\,70 GHz channel, which is extensively employed in the cosmological analysis of this 2015 release. Furthermore, a solid absolute calibration is needed to compare and combine \Planck\ data with results from other experiments, most notably with WMAP. Detailed comparisons between calibrated data from single LFI radiometers, between the three LFI frequency channels, and between LFI and HFI, allow us to test the internal consistency and accuracy of our calibration.

In this paper, we quantify both the absolute and relative accuracy in the calibration of the LFI instrument, and find an overall uncertainty of 0.35\,\% (30\,GHz map), 0.26\,\% (44\,GHz), and 0.20\,\% (70\,GHz). The level of the power spectrum near the first peak is now remarkably consistent with WMAP's. Other paper in this \Planck{} data release deal with the quality of the LFI calibration, in particular:
\begin{itemize}
\item \citet{planck2014-a12} quantifies the consistency between the calibration of the LFI/HFI/WMAP channels in the context of foreground component separation, finding that the measured discrepancies among channels are of the order of a few tenths of a percent;
\item \citet{planck2014-a13} analyses the consistency between the LFI 70\,GHz low-$\ell$ polarization map and the WMAP map in pixel space, finding no hints of inconsistencies;
\item \citet{planck2014-a15} compares the estimate for the $\tau$ and $z_\text{re}$ cosmological parameters (reionization optical depth and redshift) using either LFI 70\,GHz polarization maps or WMAP maps, finding statistically consistent values.
\end{itemize}

To achieve calibration accuracy at the few-per-thousand level requires careful attention to instrumental systematic effects and foreground contamination of the orbital dipole.  Much of this paper is devoted to a discussion of such effects and the means to mitigate them.

In this paper we do not explicitly discuss polarization-related issues. Although polarization analysis is one of the most important results of this data release, the calibration of the LFI radiometers is inherently based on temperature signals \citep{leahy2010}. Estimates of the sensitivity in polarization, as well as the impact of calibration-related systematics on it, are provided by \citet{planck2014-a04}.

\subsection{Basis of the calibration}

\begin{figure}
	\includegraphics{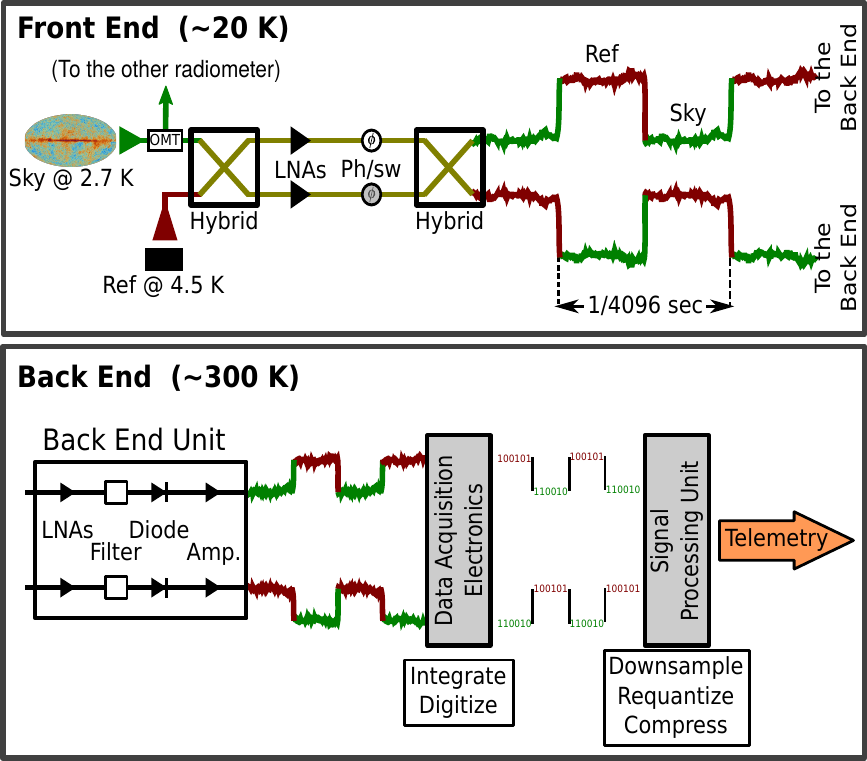}
	\caption{\label{fig:radiometerSchema} Schematic of an LFI radiometer,
taken from \protect\OldPlanckLFICalPaper. The two linearized polarization
components are separated by an orthomode transducer (OMT), and each of them
enters a twin radiometer, only one of which is shown in the figure. A first
amplification stage is provided in the cold (20\,K) focal plane, where the
signal is combined with a reference signal originating in a thermally stable
4.5-K thermal load. The radio frequency signal is then propagated through a set
of composite waveguides to the warm (300\,K) back-end, where it is further
amplified and filtered, and finally converted into a sequence of digitized
numbers by an analogue-to-digital converter. The numbers are then compressed
into packets and sent to Earth.}
\end{figure}

A schematic of the LFI pseudo-correlation receiver is shown in
Fig.~\ref{fig:radiometerSchema}. We model the output voltage $V(t)$ of each
radiometer as
\begin{equation}
\label{eq:radiometerEquation}
V(t) = G(t)\,\bigl[B * (\Tsky + D)\bigr](t) + M + N,
\end{equation}
where $G$ is the ``gain'' (measured in $\unit[]{V\,K^{-1}}$), $B$ is the beam
response, $D$ is the thermodynamic temperature of the total CMB dipole signal
(i.e., a combination of the Solar and orbital components, including the
quadrupolar relativistic corrections), which we use as a calibrator, and
$\Tsky = \Tcmb + \Tgalaxy + \Tother$ is the overall temperature of the sky (CMB
anisotropies, diffuse Galactic emission and other\footnote{Within this term we
include extragalactic foregrounds and all point sources.} foregrounds,
respectively) apart from $D$. Finally, $M$ is a constant offset and $N$ is a
noise term.  Note that in the following sections we will use
Eq.~(\ref{eq:radiometerEquation}) many times; whenever the presence of the $N$
term will not be important, it will be silently dropped. The $*$ operator
represents a convolution over the $4\pi$ sphere. We base our calibration on the
a priori knowledge of the spacecraft velocity around the Sun, producing the
orbital dipole and use the orbital dipole to accurately measure the dominant
solar dipole component. The purpose of this paper is to explain how we
implemented and validated the pipeline that estimates the ``calibration
constant'' $K \equiv G^{-1}$ (which is used to convert the voltage $V$ into a
thermodynamic temperature), to quantify the quality of our estimate for $K$,
and to quantify the impact of possible systematic calibration errors on the
\Planck/LFI data products.

\subsection{Structure of this paper}

Several improvements were introduced in the LFI pipeline for calibration
relative to \OldPlanckLFICalPaper. In Sect.~\ref{sec:beamEfficiency} we recall
some terminology and basic ideas presented in \OldPlanckLFICalPaper{} to
discuss the normalization of the calibration, i.e., what factors influence the
average value of $G$ in Eq.~\eqref{eq:radiometerEquation}.
Section~\ref{sec:Pipeline} provides an overview of the new LFI calibration
pipeline and underlines the differences with the pipeline described in
\OldPlanckLFICalPaper{}. One of the most important improvements in the 2015
calibration pipeline is the implementation of a new iterative algorithm to
calibrate the data, \DaCapo. Its principles are presented separately in a
dedicated section, Sect.~\ref{sec:DaCapo}. This code has also been used to
characterize the orbital dipole. The details of this latter analysis are
provided in Sect.~\ref{sec:DipoleEstimation}, where we present a new
characterization of the Solar dipole. These two steps are crucial for the
calibration of LFI. Section~\ref{sec:validationAndAccuracy} describes a number
of validation tests we have run on the calibration, as well as the results of a
quality assessment. This section is divided into several parts: in
Sect.~\ref{sec:absoluteCalibration} we compare the overall level of the
calibration in the 2015 LFI maps with those in the previous data release; in
Sect.~\ref{sec:fitErrors} we provide a brief account of the simulations
described in \citet{planck2014-a04}, which assess the calibration error due to
the white noise and approximations in the calibration algorithm itself; in
Sect.~\ref{sec:beamUncertainties} we describe how uncertainties in the shape of
the beams might affect the calibration;
Sects.~\ref{sec:interChannelConsistency} and
\ref{sec:interFrequencyConsistency} measure the agreement between radiometers
and groups of radiometers in the estimation of the $TT$ power spectrum;
and Sect.~\ref{sec:nullTests} provides a reference to the discussion of
null tests provided in \citet{planck2014-a04}. Finally, in
Sect.~\ref{sec:planets}, we derive an independent estimate of the LFI
calibration from our measurements of Jupiter and discuss its consistency with
our nominal dipole calibration.

%%%%%%%%%%%%%%%%%%%%%%%%%%%%%%%%%%%%%%%%%%%%%%%%%%%%%%%%%%%%%%%%%%%%%%

\section{Handling beam efficiency}
\label{sec:beamEfficiency}
In this section we develop a mathematical model to relate the absolute level of the calibration (i.e., the average level of the raw power spectrum $\tilde C_\ell$ for an LFI map) to a number of instrumental parameters related to the beams and the scanning strategy.

The beam response $B(\theta, \varphi)$ is a dimensionless function defined over the $4\pi$ sphere. In Eq.~\eqref{eq:radiometerEquation}, $B$ appears in the convolution
\begin{equation}
\label{eq:beamConvolutionWithT}
B * (\Tsky + D) = \frac{\int_{4\pi} B(\theta, \varphi)\,(\Tsky + D)
 (\theta,\varphi)\,\ud\Omega}{\int_{4\pi} B(\theta,\varphi)\,\ud\Omega},
\end{equation}
whose value changes with time because of the change of orientation of the spacecraft. Since no time-dependent optical effects are evident from the data taken from October 2009 to February 2013 \citep{planck2014-a05}, we assume there is no intrinsic change in the shape of $B$ during the surveys.

In the previous data release, we approximated $B$ as a Dirac delta function (a ``pencil beam'') when modeling the dipole signal seen by the LFI radiometers. The same assumption has been used for all the WMAP data releases \citep[see, e.g.,][]{hinshaw2009}, as well as in the HFI pipeline \citep{planck2014-a09}. However, the real shape of $B$ deviates from the ideal case of a pencil beam because of two factors: (1) the main beam is more like a Gaussian with an elliptical section, whose FWHM (full width half maximum) ranges between $13\arcm$ and $33\arcm$ in the case of the LFI radiometers; and (2) farther than $5^\circ$ from the beam axis, the presence of far sidelobes further dilutes the signal measured through the main beam and induces an axial asymmetry on $B$. Previous studies\footnote{Apart from the use of appropriate window functions \protect\citep[e.g.,][]{page2003a}, the WMAP team implemented a number of other corrections to further reduce systematic errors due to the non-ideality of their beams. In their first data release, the WMAP team estimated the contribution of the Galaxy signal picked up through the sidelobes at the map level \protect\citep{barnes2003} and then subtracted them from the maps. Starting from the third year release, they estimated a multiplicative correction, called the ``recalibration factor,'' assumed constant throughout the survey, by means of simulations. This constant accounts for the sidelobe pickup and has been applied to the TODs \protect\citep{jarosik2007}. The deviation from unity of this factor ranges from 0.1\,\% to 1.5\,\%.} tackled the first point by applying a window function to the power spectrum computed from the maps in order to correct for the finite size of the main beam. However, the presence of far sidelobes might cause the presence of stripes in maps. For this 2015 data release, we use the full shape of $B$ in computing the dipole signal adopted for the calibration. No significant variation in the level of the CMB power spectra with respect to the previous data release is expected, since we are basically subtracting power during the calibration process instead of reducing the level of the power spectrum by means of the window function. However, this new approach improves the internal consistency of the data, since the beam shape is taken into account from the very first stages of data processing (i.e., the signal measured by each radiometer is fitted with its own calibration signal $B_{\mathrm{rad}} * D$); see Sect.~\ref{sec:nullTests}. The definition of the beam window function has been changed accordingly; see \citet{planck2014-a05}.

In \OldPlanckLFICalPaper{} we introduced the two quantities $\phid$ and $\phisky$ as a way to quantify the impact of a beam window function on the calibration\footnote{The definition of $\protect\phid$ provided in \protect\OldPlanckLFICalPaper{} was not LFI-specific: it can be applied to any experiment that uses the dipole signal for the calibration.} and on the mapmaking process, respectively. Here we briefly summarize the theory, and we introduce new equations that are relevant for understanding the normalization of the new \Planck-LFI results in this data release.

\begin{figure*}
	\centering
	\includegraphics{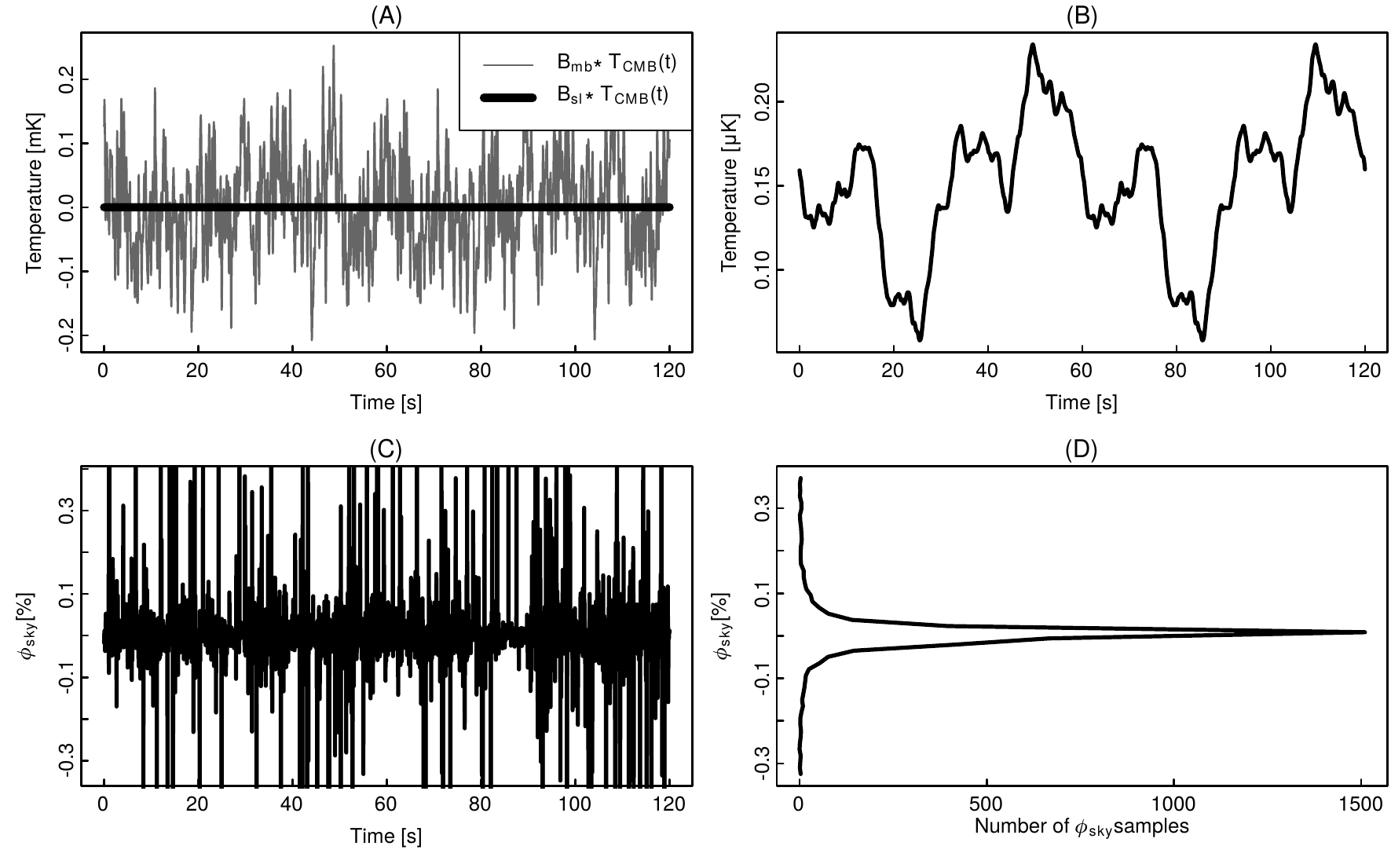}
	\caption{\label{fig:phiskyExample} Quantities used in the determination of the value of $\phisky$ (Eq.~\protect\ref{eq:phisky}) for radiometer LFI-27M (30\,GHz) during a short time span (2\,min). \textit{Panel A}: the quantities $B_{\mathrm{mb}} * T_{\mathrm{CMB}}$ and $B_{\mathrm{sl}} * T_{\mathrm{CMB}}$ are compared. Fluctuations in the latter term are much smaller than those in the former. \textit{Panel B}: the quantity $B_{\mathrm{sl}} * T_{\mathrm{CMB}}$ shown in the previous panel is replotted here to highlight the features in its tiny fluctuations. The fact that the pattern of fluctuations repeats twice depends on the scanning strategy of \Planck, which observes the sky along the same circle many times. \textit{Panel C}: Value of $\phisky$ calculated using Eq.~\protect\eqref{eq:phisky}. There are several values that diverge to infinity, which is due to the denominator in the equation going to zero. \textit{Panel D}: distribution of the values of $\phisky$ plotted in panel C. The majority of the values fall around the number $+0.02\,\%$.}
\end{figure*}

The Solar dipole $D$, due to the motion of the Solar System in the CMB rest frame, is given by
\begin{equation}
\label{eq:dipole}
D(\xversor, t) = T_\mathrm{CMB}\left(\frac1{\gamma(t)\,
\bigl(1 - \vec{\beta}(t)\cdot \xversor\bigr)} - 1\right),
\end{equation}
where $T_\mathrm{CMB}$ is the CMB monopole, $\vec{\beta} = \vec{v}/c$ is the velocity of the spacecraft, and $\gamma = (1 - \beta^2)^{-1/2}$.  Each radiometer measures the signal $D$ convolved with the beam response $B$, according to Eq.~\eqref{eq:beamConvolutionWithT}; therefore, in principle, each radiometer has a different calibration signal. Under the assumption of a Dirac delta shape for $B$, \OldPlanckLFICalPaper{} shows that the estimate of the gain constant $\tilde G$ is related to the true gain $G$ by the formula
\begin{equation}
\label{eq:gainConstantAndPhidForDiracDelta}
\tilde G^\mathrm{pen} = G \bigl(1 - \fsl\bigr) \bigl(1 + \phid\bigr),
\end{equation}
where
\begin{equation}
\fsl = \frac{\int_{\theta > 5^\circ} B\,\ud\Omega}{\int_{4\pi} B\,\ud\Omega}
\end{equation}
is the fraction of power entering the sidelobes (i.e., along directions farther than $5^\circ$ from the beam axis), and
\begin{equation}
\label{eq:phid}
\phid = \frac{\partial_t \Bsl * D}{\partial_t \Bmain * D}
\end{equation}
is a time-dependent quantity that depends on the shape of $B = \Bmain + \Bsl$ and its decomposition into a ``main'' ($\theta < 5^\circ$) and ``sidelobe'' part, on the signal $D$, and on the scanning strategy because of the time dependence of the stray light; the notation $\partial_t$ indicates a time derivative. Once the timelines are calibrated, traditional mapmaking algorithms approximate\footnote{\protect\citet{keihanen2012} provide a deconvolution code that can be used to produce maps potentially free of this effect.} $B$ as a Dirac delta \citep[e.g.,][]{hinshaw2003a,jarosik2007,keihanen2010}, thus introducing a new systematic error. In this case, the mean temperature $\Tskymeas$ of a pixel in the map would be related to the true temperature $\Tsky$ by the formula
\begin{equation}
\label{eq:TskyandPhiskyForDiracDelta}
\Tsky = \Tskymeas^\mathrm{pen} (1 - \phisky + \phid),
\end{equation}
which applies to timelines and should be considered valid only when considering details on angular scales larger than the width of the main beam. \OldPlanckLFICalPaper{} defines the quantity $\phisky$ using the following equation:
\begin{equation}
\label{eq:phisky}
\phisky = \frac{\Bsl * \Tsky}{\Bmain * \Tsky}
 \left(\frac{\Tsky}{\Tskymeas}\right) = \frac{\Bsl * \Tsky}\Tskymeas.
\end{equation}
See Fig.~\ref{fig:phiskyExample} for an example showing how $\phisky$ is computed.

In this 2015 \Planck\ data release, we take advantage of our knowledge of the shape of $B$ to compute the value of Eq.~\eqref{eq:beamConvolutionWithT} and use this as our calibrator. Since the term $B * \Tsky = (\Bmain + \Bsl) * \Tsky$ is unknown, we apply the following simplifications:
\begin{enumerate}
\item we apply the point source and 80\,\% Galactic masks \citep{planck2014-ES}, in order not to consider the $\Bmain * \Tsky$ term in the computation of the convolution;
\item we assume that $\Bsl * \Tsky \approx \Bsl * \Tgalaxy$ and subtract it from the calibrated timelines, using an estimate for $\Tgalaxy$ computed by means of models of the Galactic emission \citep{planck2014-a11,planck2014-a12}.
\end{enumerate}
The result of such transformations is a new timeline $V_\mathrm{out}'$. Under the hypothesis of perfect knowledge of the beam $B$ and of the dipole signal $D$, these steps are enough to estimate the true calibration constant without bias\footnote{It is easy to show this analytically. Alternatively, it is enough to note that considering the full $4\pi$ beam makes $\fsl = 0$, and $\phi_D$ is identically zero because there are no ``sidelobes'' falling outside the beam. With these substitutions, Eq.~\protect\eqref{eq:gainConstantAndPhidForDiracDelta} becomes Eq.~\protect\eqref{eq:gainConstantAndPhid}.} (unlike Eq.~\ref{eq:gainConstantAndPhidForDiracDelta}):
\begin{equation}
\label{eq:gainConstantAndPhid}
\tilde G^{4\pi} = G,
\end{equation}
which should be expected; since no systematic effects caused by the shape of $B$ affect the estimate of the gain $G$. In order to see how Eq.~\eqref{eq:TskyandPhiskyForDiracDelta} changes in this case, we write the measured temperature $\Tskymeas$ as
\begin{equation}
\Tskymeas = B * \Tsky + M = \Bmain * \Tsky + \Bsl * \Tsky + M.
\end{equation}
Since in this 2015 data release we remove $\Bsl * \Tgalaxy$, the contribution of the pickup of Galactic signal through the sidelobes \citep{planck2014-a03}, the equation can be rewritten as
\begin{equation}
\label{eq:GalacticPickupRemoval}
\Tskymeas^{4\pi} = \Bmain * \Tsky + \Bsl * (\Tcmb + \Tother) + M.
\end{equation}
If we neglect details at angular scales smaller than the main beam size, then
\begin{equation}
\Bmain * \Tsky \approx (1 - \fsl)\,\Tsky,
\end{equation}
so that
\begin{equation}
\label{eq:tskymeas}
\Tskymeas^{4\pi} = (1 - \fsl)\,\Tsky + \Bsl * (\Tcmb + \Tother) + M.
\end{equation}
We modify Eq.~\eqref{eq:phisky} in order to introduce a new term $\phisky'$:
\begin{equation}
\phisky' = \frac{\Bsl * (\Tcmb + \Tother)}\Tskymeas;
\end{equation}
solving for $\Tsky$, Eq.~\eqref{eq:tskymeas} can be rewritten as
\begin{equation}
\label{eq:TskyandPhiskyForFourPi}
\Tsky = \Tskymeas^{4\pi}\,\frac{1 - \phisky'}{1 - \fsl} + T_0,
\end{equation}
where $T_0 = M / \bigl(1 - \fsl\bigr)$ is a constant offset that is of little relevance for pseudo-differential instruments like LFI. Eq.~\eqref{eq:TskyandPhiskyForFourPi} is the equivalent of Eq.~\eqref{eq:TskyandPhiskyForDiracDelta} in the case of a calibration pipeline that takes into account the $4\pi$ shape of $B$, as is the case for the \Planck-LFI pipeline used for the 2015 data release.

Since one of the purposes of this paper is to provide a quantitative comparison of the calibration of this \Planck\ data release with the previous one, we provide now a few formulae that quantify the change in the average level of the temperature fluctuations and of the power spectrum between the 2013 and 2015 releases. The variation in temperature can be derived from Eq.~\eqref{eq:TskyandPhiskyForDiracDelta} and Eq.~\eqref{eq:TskyandPhiskyForFourPi}:
\begin{equation}
\label{eq:fourPiPencilTRatio}
\frac{\Tskymeas^{2015}}{\Tskymeas^\mathrm{2013}} =
 \frac{\bigl(1 - \phisky + \phid\bigr)\bigl(1 - \fsl\bigr)}{1 - \phisky'}
 \approx \frac{1 - \fsl - \phisky + \phid}{1 - \phisky'}.
\end{equation}
If we consider the ratio between the power spectra $\tilde C_\ell^{2015}$ and $\tilde C_\ell^\mathrm{2013}$, the quantity becomes
\begin{equation}
\label{eq:fourPiPencilClRatio}
\frac{\tilde C_\ell^{2015}}{\tilde C_\ell^\mathrm{2013}}
 \approx \left(\frac{1 - \fsl - \phisky + \phid}{1 - \phisky'}\right)^2.
\end{equation}
In Sect.~\ref{sec:absoluteCalibration} we will provide quantitative estimates of $\fsl$, $\phid$, $\phisky$, and $\phisky'$, as well as the ratios in Eq.~\eqref{eq:fourPiPencilTRatio} and in Eq.~\eqref{eq:fourPiPencilClRatio}.

%%%%%%%%%%%%%%%%%%%%%%%%%%%%%%%%%%%%%%%%%%%%%%%%%%%%%%%%%%%%%%%%%%%%%%

\section{The calibration pipeline}
\label{sec:Pipeline}

\begin{figure*}
	\centering
	\begin{tikzpicture}[>=latex,line join=bevel,]
\definecolor{fillcolor}{rgb}{0.85,0.85,0.85};
  \node (sat_pos) at (267bp,300bp) [draw,fill=fillcolor,ellipse] {\parbox{8em}{\centering Satellite
                      position and velocity}};
  \definecolor{fillcolor}{rgb}{0.85,0.85,0.85};
  \node (solar_dipole) at (147bp,206bp) [draw,fill=fillcolor,ellipse] {Solar dipole};
  \definecolor{fillcolor}{rgb}{0.85,0.85,0.85};
  \node (beams) at (147bp,300bp) [draw,fill=fillcolor,ellipse] {Beams ($B$)};
  \node (smoother) at (107bp,112bp) [draw,rectangle] {Smoothing filter};
  \node (da_capo_unc) at (147bp,253bp) [draw,rectangle] {\parbox{8em}{\centering
                          DaCapo (unconstrained)}};
  \definecolor{fillcolor}{rgb}{0.85,0.85,0.85};
  \node (toi) at (221bp,18bp) [draw,fill=fillcolor,ellipse] {\parbox{8em}{\centering Calibrated timelines}};
  \node (da_capo_cons) at (69bp,159bp) [draw,rectangle] {\parbox{6em}{\centering DaCapo (constrained)}};
  \node (calibrate) at (221bp,65bp) [draw,rectangle] {\parbox{12em}{\centering Calibrate and remove
                        dipole and Galactic pickup}};
  \definecolor{fillcolor}{rgb}{0.85,0.85,0.85};
  \node (hk_info) at (159bp,159bp) [draw,fill=fillcolor,ellipse] {\parbox{6em}{\centering Housekeeping
                      information}};
  \definecolor{fillcolor}{rgb}{0.85,0.85,0.85};
  \node (data) at (27bp,300bp) [draw,fill=fillcolor,ellipse] {\parbox{12em}{\centering Radiometric
                   data and\\pointing information}};
  \draw [->] (solar_dipole) ..controls (118.24bp,188.41bp) and (112.92bp,185.34bp)  .. (da_capo_cons);
  \draw [->] (hk_info) ..controls (140.25bp,141.77bp) and (137.51bp,139.41bp)  .. (smoother);
  \draw [->] (sat_pos) ..controls (252.88bp,257.13bp) and (232.6bp,210.54bp)  .. (198bp,188bp) .. controls (168.67bp,168.89bp) and (154.08bp,185.01bp)  .. (120bp,177bp) .. controls (119.67bp,176.92bp) and (119.33bp,176.84bp)  .. (da_capo_cons);
  \draw [->] (data) ..controls (7.4594bp,251.14bp) and (-13.206bp,183.53bp)  .. (18bp,141bp) .. controls (26.72bp,129.12bp) and (40.486bp,122.14bp)  .. (smoother);
  \draw [->] (da_capo_unc) ..controls (147bp,234.76bp) and (147bp,234.54bp)  .. (solar_dipole);
  \draw [->] (beams) ..controls (114.87bp,285.55bp) and (105.33bp,279.27bp)  .. (99bp,271bp) .. controls (80.375bp,246.67bp) and (73.3bp,211.61bp)  .. (da_capo_cons);
  \draw [->] (beams) ..controls (147bp,281.76bp) and (147bp,281.54bp)  .. (da_capo_unc);
  \draw [->] (smoother) ..controls (161.21bp,89.602bp) and (173.46bp,84.766bp)  .. (calibrate);
  \draw [->] (data) ..controls (35.197bp,259.72bp) and (44.343bp,220.43bp)  .. (56bp,188bp) .. controls (56.159bp,187.56bp) and (56.322bp,187.11bp)  .. (da_capo_cons);
  \draw [->] (calibrate) ..controls (221bp,46.763bp) and (221bp,46.539bp)  .. (toi);
  \draw [->] (beams) ..controls (178.56bp,285.09bp) and (188.17bp,278.87bp)  .. (195bp,271bp) .. controls (215.11bp,247.8bp) and (221bp,237.7bp)  .. (221bp,207bp) .. controls (221bp,207bp) and (221bp,207bp)  .. (221bp,158bp) .. controls (221bp,136.34bp) and (221bp,111.81bp)  .. (calibrate);
  \draw [->] (da_capo_cons) ..controls (84.217bp,139.98bp) and (85.082bp,138.96bp)  .. (smoother);
  \draw [->] (data) ..controls (64.389bp,284.98bp) and (83.17bp,277.94bp)  .. (da_capo_unc);
  \draw [->] (sat_pos) ..controls (228.13bp,284.43bp) and (210.19bp,277.7bp)  .. (da_capo_unc);
  \draw [->] (sat_pos) ..controls (293.89bp,266.17bp) and (313bp,236.46bp)  .. (313bp,207bp) .. controls (313bp,207bp) and (313bp,207bp)  .. (313bp,158bp) .. controls (313bp,124.86bp) and (282.79bp,99.492bp)  .. (calibrate);
\end{tikzpicture}
	\caption{\label{fig:calPipelineDiagram} Diagram of the pipeline used to
produce the LFI frequency maps in the 2015 \Planck\ data release. The grey
ovals represent input/output data for the modules of the calibration pipeline,
which are represented as white boxes. The product of the pipeline is a set of
calibrated timelines that are passed as input to the mapmaker.}
\end{figure*}
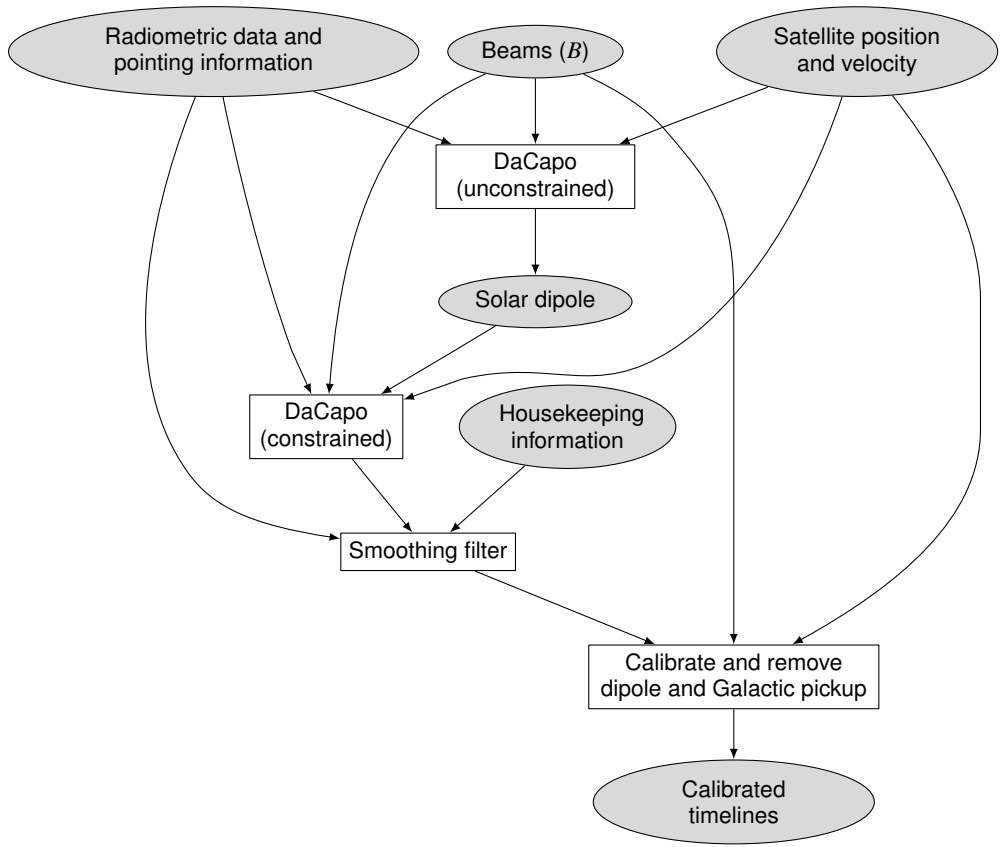

\begin{figure*}
	\centering
	\includegraphics{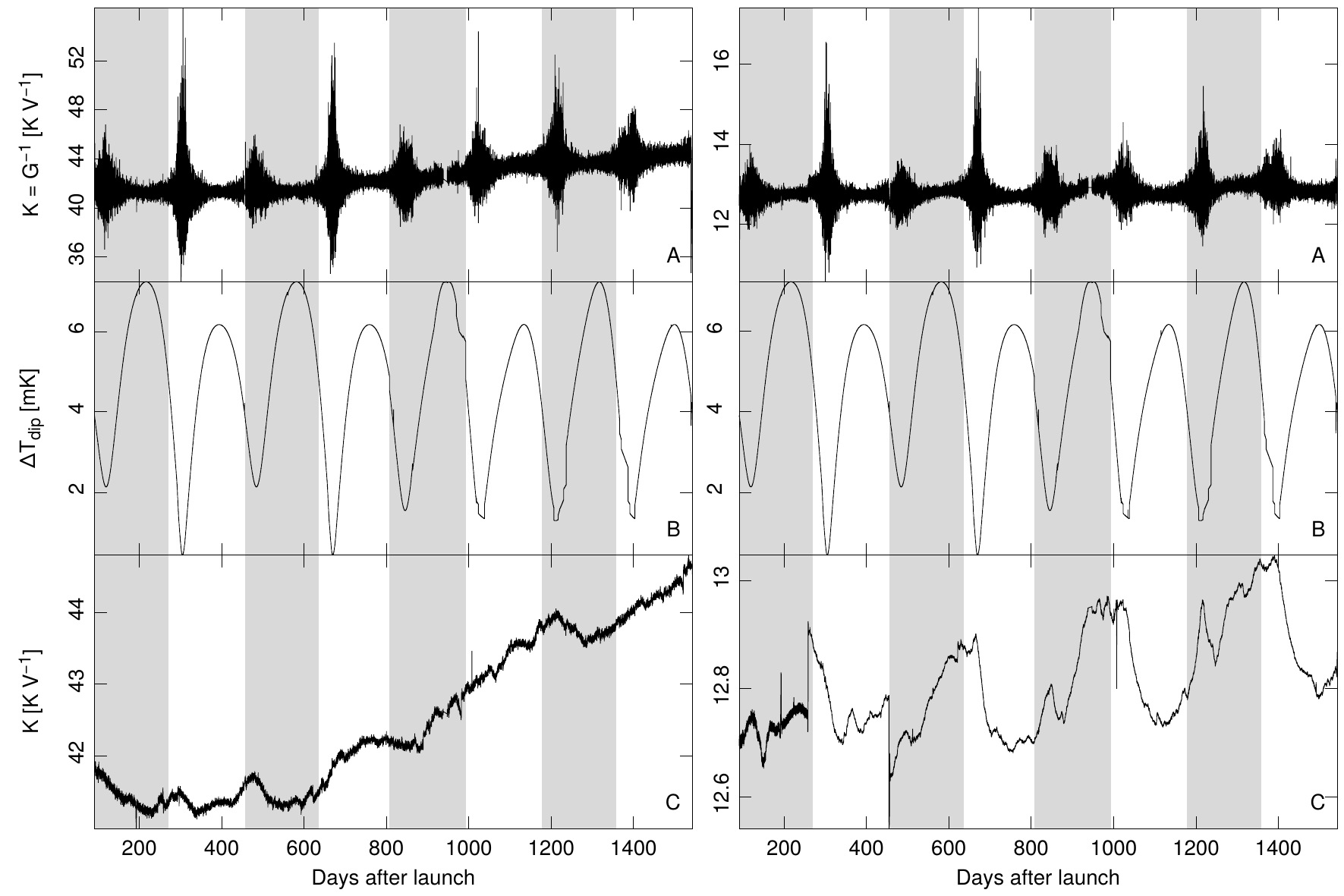}
	\caption{\label{fig:gainCurve} Variation in time of a few quantities relevant for calibration, for radiometer LFI-21M (70\,GHz, left) and LFI-27M (30\,GHz, right). Grey/white bands mark complete sky surveys. All temperatures are thermodynamic. \textit{Panel A}: calibration constant $K$ estimated using the expected amplitude of the CMB dipole. Note that the uncertainty associated with the estimate changes with time, according to the amplitude of the dipole as seen in each ring. \textit{Panel B}: expected peak-to-peak difference of the dipole signal (solar + orbital). The shape of the curve depends on the scanning strategy of \Planck, and it is strongly correlated with the uncertainty in the gain constant (see panel A). Note that the deepest minima happen during Surveys~2 and 4; because of the higher uncertainties in the calibration (and the consequent bias in the maps), these surveys have been neglected in some of the analyses in this \Planck\ data release \protect\citep[see e.g.,][]{planck2014-a15}. \textit{Panel C:} the calibration constants $K$ used to actually calibrate the data for this \Planck\ data release are derived by applying a smoothing filter to the raw gains in panel A. Details regarding the smoothing filter are presented in Appendix~\ref{sec:smoothing}.}
\end{figure*}

In this section we briefly describe the implementation of the calibration
pipeline. Readers interested in more detail should refer to
\citet{planck2014-a03}.

Evaluating the calibration constant $K$ (see Eq.~\ref{eq:radiometerEquation}) requires us to fit the timelines of each radiometer with the expected signal $D$ induced by the dipole as \Planck\ scans the sky. This process provides the conversion between the voltages and the measured thermodynamic temperature.

As discussed in Sect.~\ref{sec:beamEfficiency}, we have improved the model used for $D$, since we are now computing the convolution of $D$ with each beam $B$ over the full $4\pi$ sphere. Moreover, we are considering the $\Bsl * \Tgalaxy$ term in the fit, in order to reduce the bias due to the pickup of Galactic signal by the beam far sidelobes. The model of the dipole $D$ now includes the correct\footnote{Because of a bug in the implementation of the pipeline, the previous data release had a spurious factor that led to a residual quadrupolar signal of $\sim 1.9\,\mu\mathrm{K}$, as described in \OldPlanckLFICalPaper.} quadrupolar corrections required by special relativity. The quality of the beam estimate $B$ has been improved as well: we are now using all the seven Jupiter transits observed in the full 4-year mission, and we account for the optical effects due to the variation of the beam shape across the band of the radiometers. It is important to underline that these new beams do not follow the same normalization convention as in the first data release (now $\int_{4\pi} B(\theta, \varphi)\,\ud\Omega \not= 1$), as numerical inaccuracies in the simulation of the $4\pi$ beams cause a loss of roughly 1\,\% of the signal entering the sidelobes\footnote{This loss was present in the beams used for the 2013 release too, but in that case we applied a normalization factor to $B$. The reason why we removed this normalization is that it had the disadvantage of uniformly spreading the 1\,\% sidelobe loss over the whole $4\pi$ sphere.}: see \citet{planck2014-a05} for a discussion of this point.

As was the case in the 2013 data release, the calibration constant $K$ is estimated once per each pointing period, i.e., the period during which the spinning axis of the spacecraft holds still and the spacecraft rotates at a constant spinning rate of $1/60\,\mathrm{Hz}$. The code used to estimate $K$,  named \DaCapo, has been completely rewritten; it is able to run in two modes, one of which (the so-called \emph{unconstrained mode}) is able to produce an estimate of the solar dipole signal, and the other one (the \emph{constrained mode}) which requires the solar dipole parameters as input. We have used the unconstrained mode to assess the characteristics of the solar dipole, which have then been used as an input to the constrained mode of \DaCapo{} for producing the actual calibration constants.

We smooth the calibration constants produced by \DaCapo{} by means of a running mean, where the window size has a variable length. That length is chosen so that every time there is a sudden change in the state of the instrument (e.g., because of a change in the thermal environment of the front-end amplifiers) that discontinuity is not averaged out. However, this kind of filter removes any variation in the calibration constants, whose timescale is smaller than a few weeks. One example of this latter kind of fluctuation is the daily variation measured in the radiometer back-end gains during the first survey, which was caused by the continuous turning on-and-off of the transponder\footnote{This operating mode was subsequently changed and the transponder has been kept on for the remainder of the mission starting from 272 days after launch, thus removing the origin of this kind of gain fluctuations.} while sending the scientific data to Earth once per day. In order to keep track of such fluctuations, we have estimated the calibration constants $K$ using the signal of the 4.5\,K reference load in a manner similar to that described in \OldPlanckLFICalPaper{} under the name of ``4\,K calibration'', and we have added this estimate to the \DaCapo{} gains after having applied a high-pass filter to them, as shown in Fig.~\ref{fig:osgtvSketch}. Details about the implementation of the smoothing filter are provided in Appendix~\ref{sec:smoothing}.

\begin{figure*}
    \centering
    \includegraphics[width=170mm]{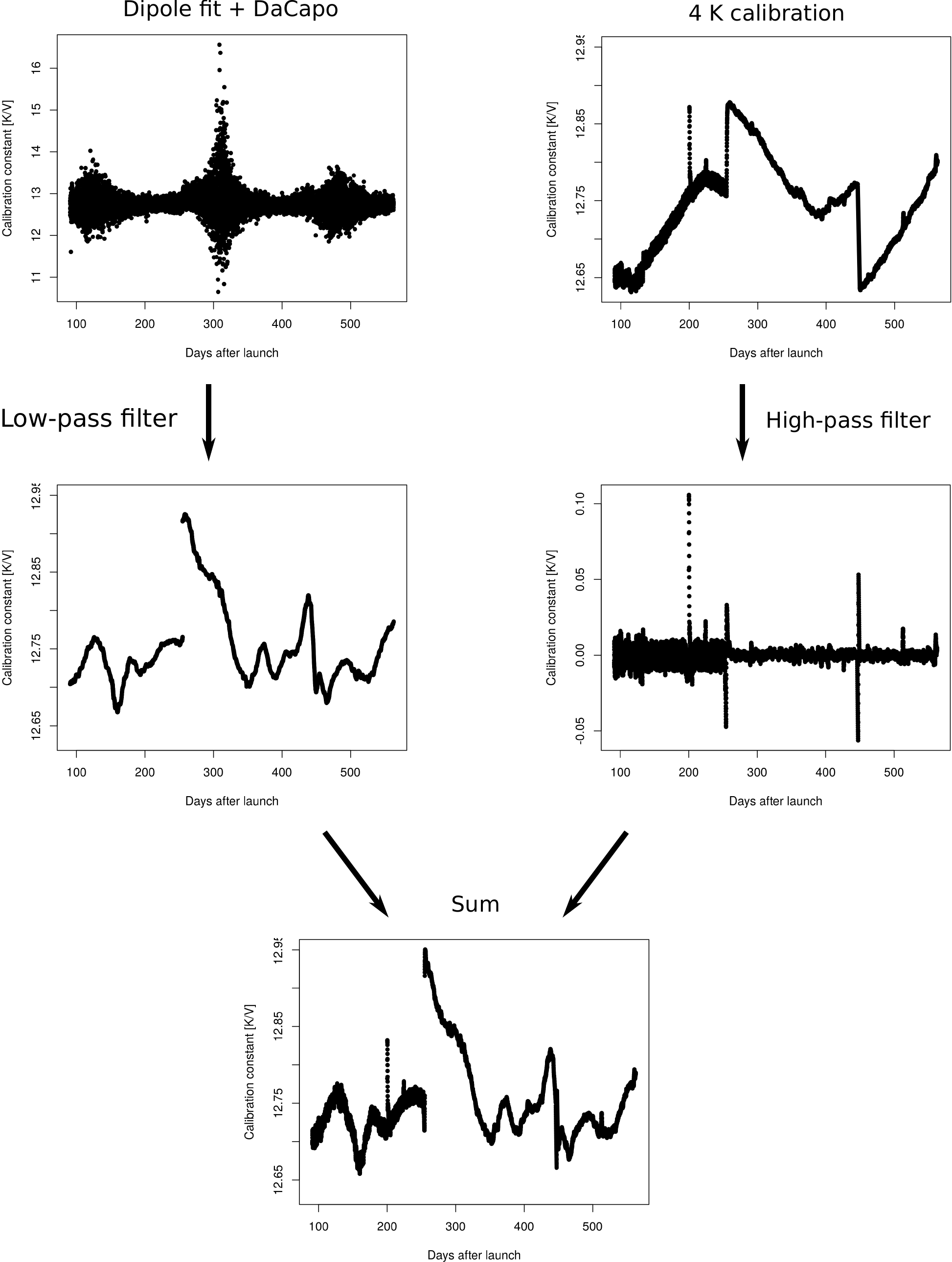}
    \caption{\label{fig:osgtvSketch} Visual representation of the algorithms
used to filter the calibration constants produced by \DaCapo{} (top-left plot;
see Sect.~\protect\ref{sec:DaCapo}). The example in the figure refers to
radiometer LFI-27M (30\,GHz) and only shows the first part of the data (roughly
three surveys).}
\end{figure*}

Once the smoothing filter has been applied to the calibration constants $K$, we multiply the voltages by $K$ in order to convert them into thermodynamic temperatures, and we remove the term $B * D + \Bsl * \Tsky$ from the result, thus removing the dipole and the Galactic signal captured by the far sidelobes from the data. The value for $\Tsky$ has been taken from a sum of the foreground signals considered in the simulations described in \citet{planck2014-a11}; refer to \citet{planck2014-a03} for further details.

\section{The calibration algorithm}
\label{sec:DaCapo}

\newcommand{\ve}[1]{\mathbf{#1}}

\DaCapo{} is an implementation of the calibration algorithm we have used in this data release to produce an estimate of the calibration constant $K$ in Eq.~\eqref{eq:radiometerEquation}. In this section we describe the model on which \DaCapo{} is based, as well as a few details of its implementation.

\subsection{Unconstrained algorithm}

Let $V_{i}$ be the $i$th sample of an uncalibrated data stream, and $k(i)$ the pointing period to which the sample belongs. Following Eq.~\eqref{eq:radiometerEquation} and assuming the usual mapmaking convention of scanning the sky, $T_\mathrm{sky}$, using a pencil beam, we model the uncalibrated time stream as
\begin{equation}
  V_{i} = G_{k(i)}(T_{i} + B*D_{i}) + b_{k(i)} + N_{i},
\end{equation}
where we write $B * D_i \equiv (B*D)_i$ and use the shorthand notation $T_i = \bigl(\Tsky\bigr)_i$. The quantity $G_{k}$ is the unknown gain factor for $k$th pointing period, $n_{i}$ represents white noise, and $b_{k}$ is an offset\footnote{The offset absorbs noise at frequencies lower than the inverse of the pointing period length (typically 40 min).  The process of coadding scanning rings effciently reduces noise at higher frequencies. We treat the remaining noise as white.} which captures the correlated noise component. We denote by $T_{i}$ the sky signal, which includes foregrounds and the CMB sky apart from the dipole, and by $B * D_{i}$ the dipole signal as seen by the beam $B$. The dipole includes both the Solar and orbital components, and it is convolved with the full $4\pi$ beam. The beam convolution is carried out by an external code, and the result is provided as input to \DaCapo.

The signal term is written with help of a pointing matrix $P$ as
\begin{equation}
  T_{i} = \sum_{p}P_{ip} m_{p}.
\end{equation}
Here $P$ is a pointing matrix which picks the time-ordered signal from the unknown sky map $m$. The current implementation takes into consideration only the temperature component. In radiometer-based calibration, however, the polarisation signal is partly accounted for, since the algorithm interprets as temperature signal whatever combination of the Stokes parameters $(I,Q,U)$ the radiometer records.  In regions that are scanned in one polarisation direction only, this gives a consistent solution that does not induce any error on the gain.  A small error can be expected to arise in those regions where the same sky pixel is scanned in vastly different directions of polarisation sensitivity.  The error is proportional to the ratio of the polarisation signal and the total sky signal, including the dipole.

We determine the gains by minimising the quantity
\begin{equation}
  \chi^{2} = \sum_{i}\frac{1}{\sigma_{i}^{2}} \bigl( V_{i} - V_{i}^{\mathrm{mod}}\bigr)^{2},  \label{eq:chi2}
\end{equation}
where
\begin{equation}
  V_{i}^{\mathrm{mod}} = G_{k(i)}\left(\sum_{p}P_{ip} m_{p} + B*D_{i}\right) +b_{k(i)}. \label{eq:ymodel}
\end{equation}
and $\sigma_{i}^{2}$ is the white noise variance. The unknowns of the model are $m$, $G$, $b$, and $n$ (while ee assume that the beam $B$ is perfectly known). The dipole signal $D$ and pointing matrix $P$ are assumed to be known.

To reduce the uncertainty that arises from beam effects and subpixel variations in signal, we apply a galactic mask and include in the sum in Eq.~\eqref{eq:chi2} only those samples that fall outside the mask.

Since Eq.~\eqref{eq:ymodel} is quadratic in the unknowns, the minimisation of $\chi^2$ requires iteration. To linearise the model we first rearrange it as
\begin{equation}
\begin{split}
  V_{i}^{\mathrm{mod}} &= G_{k(i)}(B * D_{i}+\sum_{p}P_{ip}m_{p}^{0}) \\
     & + G^{0}_{k(i)}\sum_{p}P_{ip} (m_{p}-m_{p}^{0})      \\
     & +\left[ (G_{k(i)}-G^0_{k(i)})(m_{p}-m_{p}^{0}) \right] +b_{k(i)}.
\end{split}
\end{equation}
Here $G^{0}$ and $m^{0}$ are the gains and the sky map from the previous iteration step. We drop the quadratic term in brackets and obtain that
\begin{equation}
\begin{split}
  V_{i}^{\mathrm{mod}} &= G_{k(i)}\left(B * D_{i}+\sum_{p}P_{ip}m_{p}^{0}\right) \\
     & + G^{0}_{k(i)}\sum_{p}P_{ip} \tilde m_p    \label{eq:ymodlin}  +b_{k(i)}.
\end{split}
\end{equation}
Here
\begin{equation}
  \tilde m_p = (m_{p}-m_{p}^{0})
\end{equation}
is a correction to the map estimate from the previous iteration step. Eq.~\eqref{eq:ymodlin} is linear in the unknowns $\hat m$, $G$ and $b$. We run an iterative procedure, where at every step we minimize $\chi^2$ with the linearized model in Eq.~\eqref{eq:ymodlin}, update the map and the gains as $m^{0}\rightarrow m^0+\tilde m$ and $G^{0}\rightarrow g$, and repeat until convergence. The iteration is started from $G^0=m^0=0$. Thus at the first step we are fitting just the dipole model and a baseline $G_{k(i)}\,B * D_{i}+b_k$, and we obtain the first estimate for the gains. The first map estimate is obtained in the second iteration step.

\DaCapo{} solves the gains for two radiometers of a horn at the same time. Two map options are available. Either the radiometers have their own sky maps, or both see the same sky. In the former case the calibrations become independent.

\subsubsection{Solution of the linear system}

Minimisation of $\chi^2$ yields a large linear system. The number of unknowns is dominated by the number of pixels in map $m$. It is possible, however, to reformulate the problem as a much smaller system as follows.

We first rewrite the model using matrix notation. We combine the first and last terms of Eq.~\eqref{eq:ymodlin} formally into
\begin{equation}
  G_{k(i)}\left(B * D_{i} +\sum_{p}P_{ip}m_{p}^{0}\right)  +b_{k(i)} = \sum_{j} F_{ij} a_{j}.
\end{equation}
The vector $a_{j}$ contains the unknowns $b$ and $G$, and the matrix $F$ spreads them into a time-ordered data stream. The dipole signal $B * D$ seen by the beam $B$, and a signal picked from map $m^{0}$, are included in $F$.

Eq.~\eqref{eq:ymodel} can now be written in matrix notation as
\begin{equation}
  \ve{V}^{\mathrm{model}} = \tilde P\tilde{\ve m} + F\ve a.
\end{equation}
Gains $G^{0}$ have been transferred inside matrix $\tilde P$,
\begin{equation}
  \tilde P_{ip} = G^{0}_{k(i)}P_{ip}. \label{eq:Ptilde}
\end{equation}
Using this notation, Eq.~\eqref{eq:chi2} becomes
\begin{equation}
\label{eq:chi2mat}
  \chi^{2} = ( \ve V -\tilde P\tilde{\ve m} -F\ve a)^{T} C_{n}^{-1}
             ( \ve V -\tilde P\tilde{\ve m} -F\ve a),
\end{equation}
where $C_{n}$ is the white noise covariance.

Eq.~\eqref{eq:chi2mat} is equivalent to the usual destriping problem of map-making \citep{planck2014-a07}, only the interpretation of the terms is slightly different. In place of the pointing matrix $P$ we have $\tilde P$, which contains the gains from previous iteration step, and $\ve a$ contains the unknown gains beside the usual baseline offsets.

We minimize Eq.~\eqref{eq:chi2mat} with respect to $\tilde{\vec m}$, insert the result back into Eq.~\eqref{eq:chi2mat}, and minimize with respect to $\vec a$. The solution is identical to the destriping solution
\begin{equation}
\label{eq:destr}
 \hat{\vec a} = ( \tens{F}^{\rm T}C_{\rm n}^{-1}\tens{Z}\tens{F})^{-1}
 \tens{F}^{\rm T}C_{\rm n}^{-1}\tens{Z}\vec V,  \end{equation}
where
\begin{equation}
 \tens{Z} = \tens{I} - \tens{\tilde P}(\tens{\tens P}^{\rm T}C_{\rm n}^{-1}
 \tens{\tilde P})^{-1} \tens{\tilde P}^{\rm T} C_{\rm n}^{-1}.
 \label{eq:Zmatrix}
\end{equation}
We use a hat to indicate that $\hat{\vec a}$ is an estimate of the true $\vec a$. We are here making use of the sparse structure of the pointing matrix, which allows us to invert matrix $\tens{\tilde P}^{\rm T}C_{\rm n}^{-1} \tens{\tilde P}$ through non-iterative methods. For a detailed solution of an equivalent problem in mapmaking, see \cite{keihanen2010} and references therein. The linear system in Eq.~\eqref{eq:destr} is much smaller than the original one. The rank of the system is of the order of the number of pointing periods, which is 44\,070 for the full four-year mission.

Eq.~\eqref{eq:destr} can be solved by conjugate gradient iteration. The map correction  is obtained as
\begin{equation}
  \hat{\ve m} = (\tilde P^{T}C_{n}^{-1} \tilde P)^{-1} \tilde P^{T}C_{n}^{-1}(\ve V - F\hat{\ve a}).
\end{equation}
Matrix $ \tilde P^{T}C_{n}^{-1} \tilde P$ is diagonal, and inverting it is a trivial task.

A lower limit for the gain uncertainty, based on radiometer white noise only, is given by the covariance matrix
\begin{equation}
  C_{\hat a} =   ( F^{T}C_{n}^{-1}F)^{-1}.
\end{equation}

\subsection{Constrained algorithm}
\label{sec:DaCapoConstrained}

\subsubsection{Role of the Solar dipole}

The dipole signal is a sum of the Solar and orbital contributions. The Solar dipole can be thought of as being picked from an approximately\footnote{It is not exactly constant, as the dipole signal is $B*D$. Since the orientation of $B$ changes with time, any deviation from axial symmetry in $B$ (ellipticity, far sidelobes\ldots) falsifies this assumption. However, when convolving a large-scale signal such as the CMB dipole with the LFI beams, such asymmetries are a second-order effect.} constant dipole map, while the orbital component depends on beam orientation and satellite velocity. The latter can be used as an independent and absolute calibration. As we will discuss in Sect.~\ref{sec:DipoleEstimation}, this has allowed us to determine the amplitude and direction of the Solar dipole and decouple the \Planck{} absolute calibration from that of WMAP.

The Solar dipole can be interpreted either as part of the dipole signal $B * D$ or part of the sky map $m$. This has important consequences. The advantage is that we can calibrate using only the orbital dipole, which is better known than the solar component and can be measured absolutely (it only depends on the temperature of the CMB monopole and the velocity of the \Planck{} spacecraft). When the unconstrained \DaCapo{} algorithm is run with erroneous dipole parameters, the difference between the input dipole and the true dipole simply leaks into the sky map $m$. The map can then be analysed to yield an estimate for the Solar dipole parameters.

The drawback from the degeneracy is that the overall gain level is weakly constrained, since it is determined from the orbital dipole alone. In the absence of the orbital component, a constant scaling factor applied to the gains would be fully compensated by an inverse scaling applied to the signal. It would then be impossible to determine the overall scaling of the gain. The orbital dipole breaks the degeneracy, but leaves the overall gain level weakly constrained compared with the relative gain fluctuations.

The degeneracy is not perfect, since the signal seen by a radiometer is modified by the beam response $B$. In particular, a beam sidelobe produces a strongly orientation-dependent signal. This is however, a small correction to the full dipole signal.

\subsubsection{Dipole constraint}

Because of the degeneracy between the overall gain level and the map dipole, it makes sense to constrain the map dipole to zero. For this to work, two conditions must be fulfilled: 1) the Solar dipole must be known, and 2) the contribution  of foregrounds (outside the mask) to the dipole of the sky must either be negligible, or it must be known and included in the dipole model.

In the following we assume that both the orbital and the Solar dipole are known. We aim at deriving a modified version of the \DaCapo{} algorithm, where we impose the additional constraint $\ve m_{D}^{T}\ve m=0$. Here $\ve m_{D}$ is a a map representing the Solar dipole component. We are thus requiring that the dipole {\it in the direction of the Solar dipole}  is completely included in the dipole model $D$, and nothing is left for the sky map. Note that $\ve m_D$ only includes the pixels outside the mask.

It turns out that condition $\ve m_D^{T}\ve m=0$ alone is not sufficient, since there is another degeneracy in the model that must be taken into account. The monopole of the sky map is not constrained by data, since it cannot be distinguished from a global noise offset $b_k$=constant. It is therefore possible to satisfy the condition $\ve m_D^{T}\ve m=0$ by adjusting simultaneously the baselines and the monopole of the map, with no cost in $\chi^2$. To avoid this pitfall, we constrain simultaneously the dipole and the monopole of the map. We require $\ve m_D^{T}\ve m=0$ and $\ve 1^{T}\ve m=0$, and combine them into one constraint
\begin{equation}
\ve m_{c}^{T}\ve m=0,  \label{eq:constraint}
\end{equation}
where $\ve m_c$ now is a two-column object.

We add to Eq.~\eqref{eq:chi2mat} an additional prior term
\begin{equation}
\begin{split}
  \chi^{2} &=  ( \ve V -\tilde P\tilde{\ve m} -F\ve a)^{T}C_{n}^{-1} ( \ve V -\tilde P\tilde{\ve m} -F\ve a) \\
    & + \tilde{\ve m}^{T}\ve m_cC_{d}^{-1}\ve m_c^T\tilde{\ve m} \label{eq:chi2c}
\end{split}
\end{equation}
and aim at taking $C_{D}^{-1}$ to infinity. This will drive $\ve m_{c}^{T}\ve m$ to zero.

Minimisation of Eq.~\eqref{eq:chi2c} yields the solution
\begin{equation}
  \hat{\ve a} =   ( F^{T}C_{n}^{-1}ZF)^{-1} F^{T}C_{n}^{-1}Z\ve V, \label{eq:solu1}
\end{equation}
with
\begin{equation}
  Z = I- \tilde P(M +\ve m_cC_{D}^{-1}\ve m_c^T)^{-1} \tilde P^{T} C_{n}^{-1}, \label{eq:solu2}
\end{equation}
where we have written for brevity
\begin{equation}
  M = \tilde P^{T}C_{n}^{-1} \tilde P.  \label{eq:solu3}
\end{equation}
This differs from the original solution (Equations \ref{eq:destr}-\ref{eq:Zmatrix}) by the term $\ve m_cC_{D}^{-1}\ve m_c^T$ in the definition of $Z$.

Eq.~\eqref{eq:solu2} is unpractical due to the large size of the matrix to be inverted.
To proceed, we apply the Sherman-Morrison formula and let $C_{D}\rightarrow0$, yielding
\begin{equation}
  \label{eq:solu4}
  (M +\ve m_cC_{D}^{-1}\ve m_c^T)^{-1} =
   M^{-1}   -M^{-1} \ve m_{c} (\ve m_{c}^{T} M^{-1} \ve m_{c})^{-1}
               \ve m_{c}^{T}M^{-1}.
\end{equation}
The middle matrix $\ve m_{c}^{T} M^{-1} \ve m_{c}$ is a 2x2 block diagonal matrix, and is easy to invert.

Equations~\eqref{eq:solu1}--\eqref{eq:solu4} are the basis of the constrained \DaCapo{} algorithm. The system is solved using a conjugate-gradient method, similarly to the unconstrained algorithm.

The map correction becomes
\begin{equation}
  \hat{\ve m} = (M +\ve m_cC_{D}^{-1}\ve m_c^T)^{-1} \tilde P^{T}C_{n}^{-1}(\ve V - F\hat{\ve a}).
\end{equation}
One readily sees that $\hat{\ve m}$ fulfills the condition expressed by Eq.~\eqref{eq:constraint}, and thus so does the full map $\ve m$.

The constraint breaks the degeneracy between the gain and the signal, but also makes the gains again dependent on the Solar dipole, which must be known beforehand.

\subsection{Use of unconstrained and constrained algorithms}

We have used the unconstrained and constrained versions of the algorithm together to obtain a self-consistent calibration, and to obtain an independent estimate for the Solar dipole.

We have first run the unconstrained algorithm, using the known orbital dipole and an initial guess for the Solar dipole. The results depend on the Solar dipole only through the beam correction. The difference between the input dipole and the true dipole are absorbed in the sky map.

We have estimated the Solar dipole from these maps (Sect.~\ref{sec:DipoleEstimation}); since the Solar dipole is same for all radiometers, we have combined data from all the 70\,GHz radiometers to reduce the error bars. (Simply running the unconstrained algorithm, fixing the dipole, and running the constrained version with same combination of radiometers would just have yielded the same solution.)

Once we have produced an estimate of the Solar dipole, we reran \DaCapo{} in constrained mode to determine the calibration coefficients $K$ more accurately.

%%%%%%%%%%%%%%%%%%%%%%%%%%%%%%%%%%%%%%%%%%%%%%%%%%%%%%%%%%%%%%%%%%%%%%

\section{Characterization of the Orbital and Solar dipoles}
\label{sec:DipoleEstimation}

In this section, we explain in detail how the Solar dipole was obtained for use
in the final \DaCapo\ run mentioned above.  We also compare the LFI
measurements with the \Planck\ nominal dipole parameters, and with the WMAP
values given by \citet{hinshaw2009}.

\subsection{Analysis}

When running \DaCapo\ in ``constrained'' mode to compute the calibration constants $K$ (Eq.~\ref{eq:radiometerEquation}), the code needs an estimate of the Solar dipole in order to calibrate the data measured by the radiometers (see Sect.~\ref{sec:Pipeline} and especially Fig.~\ref{fig:calPipelineDiagram}), since the signal produced by the orbital dipole is ten times weaker. We have used \DaCapo\ to produce this estimate from the signal produced by the orbital dipole. We limited our analysis to the 70\,GHz radiometric data, since this is the cleanest frequency in terms of foregrounds.  The pipeline was provided by a self-contained version of the \DaCapo\ program, run in unconstrained mode (see Sect.~\ref{sec:DaCapoConstrained}), in order to make the orbital dipole the only source of calibration, while the Solar dipole is left in the residual sky map.

We bin the uncalibrated differenced time-ordered data into separated rings, with one ring per pointing period.  These data are then binned according to the direction and orientation of the beam, using a {\tt Healpix}\footnote{\url{http://healpix.sourceforge.net}} \citep{gorski2005} map of resolution $N_{\rm side}=1024$ and 256 discrete bins for the orientation angle $\psi$. The far sidelobes would prevent a clean dipole from being reconstructed in the sky model map, since the signal in the timeline is convolved with the beam $B$ over the full sphere. To avoid this, an estimate for the pure dipole is obtained by subtracting the contribution due to far sidelobes using an initial estimate of the Solar dipole, which in this case was the WMAP dipole \citep{hinshaw2009}. We also subtract the orbital dipole at this stage. The bias introduced by using a different dipole is of second order and is discussed in the section on error estimates.

\DaCapo\ builds a model sky brightness distribution that is used to clean out the polarized component of the CMB and foreground signals to leave just noise, the orbital dipole, and far-sidelobe pickup. This sky map is assumed to be unpolarized, but since radiometers respond to a single linear polarization, the data will contain a polarized component, which is not compatible with the sky model and thus leads to a bias in the calibration. An estimate of the polarized signal, mostly CMB $E$-modes and some synchrotron, needs to be subtracted from the timelines. To bootstrap the process we need an intial gain estimate, which is provided by \DaCapo\ constrained to use the WMAP dipole. We then used the inverse of these gains to convert calibrated polarization maps from the previous LFI data release (which also used WMAP dipole calibration) into voltages and unwrap the map data into the timelines using the pointing information for position and boresight rotation. This polarized component due to E modes, which is $\sim 3.5\,\muK$ RMS at small angular scales plus an additional large-scale contribution of the North Galactic Spur of amplitude $\sim 3\,\muK$, was then subtracted from the time-ordered data. Further iterations using the cleaned timelines were found to make a negligible difference.

\subsection{Results}

To make maps of the dipole, a second \DaCapo\ run was made in the unconstrained mode for each LFI 70\,GHz detector using the polarization-cleaned timelines and the 30\,GHz {\tt Madam} mask, which allows 78\% of the sky to be used. The extraction of the dipole parameters (Galactic latitude, longitude, and temperture amplitude) in the presence of foregrounds was achieved with a simple Markov chain Monte Carlo (MCMC) template-fitting scheme. Single detector hit maps, together with the white noise in the LFI reduced instrument model (RIMO), were used to create the variance maps needed to construct the likelihood estimator for the MCMC samples. \Commander\ maps \citep{planck2014-a12} were used for synchrotron, free-free, and thermal dust for the template maps with the MCMC fitting for the best amplitude scaling factor to clean the dipole maps. The marginalized distribution of the sample chains between the 16th and 84th percentiles were used to estimate the statistical errors, which were 0\pdeg004, 0\pdeg009, and 0.16$\,\mu$K for latitude, longitude, and amplitude respectively. The 50\,\% point was taken as the best parameter value, as shown in Table~\ref{tab:dipoleParameters}. To estimate the systematic errors on the amplitude in calibration process due to white noise, $1/f$ noise, gain fluctuations, and ADC corrections, simulated time-ordered data were generated with these systematic effects included. These simulated timelines were then calibrated by \DaCapo\ in the same way as the data. The standard deviation of the input to output gains were taken as the error in absolute calibration with an average value of 0.11\,\%. 

\begin{figure}
	\includegraphics[width=88mm]{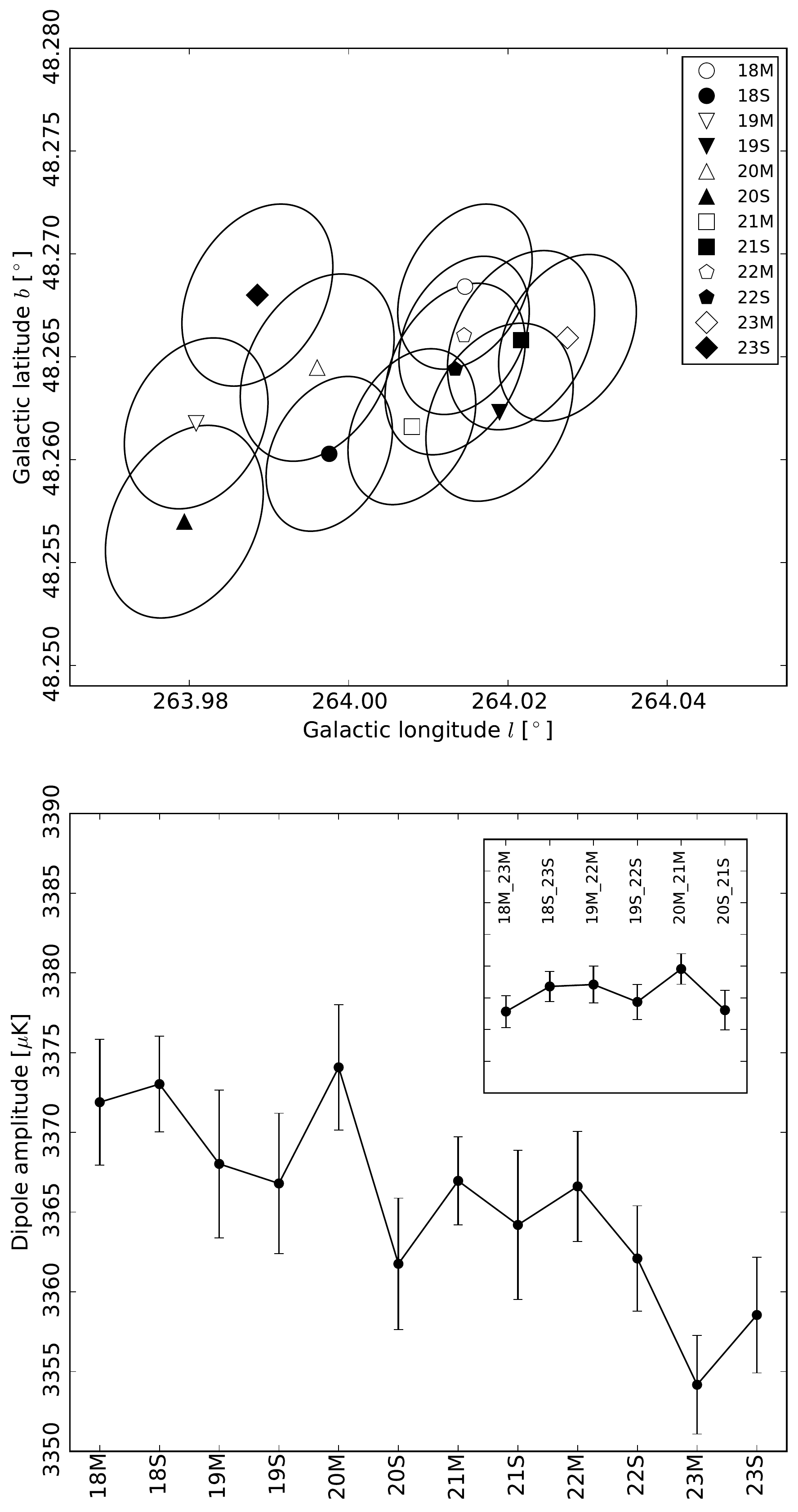}
        \caption{\label{fig:dipoleParametersPlot} Dipole amplitudes and
directions for different radiometers.  \emph{Top}: errors in the
estimation of the Solar dipole direction are represented as ellipses.
\emph{Bottom}: estimates of the amplitude of the Solar dipole signal; the
errors here are dominated by gain uncertainties. \emph{Inset}: the linear
trend (recalling that the numbering of horns is approximately from left to
right in the focal plane with respect to the scan direction),
most likely caused by a slight symmetric sidelobe residual, is removed
when we pair the 70\,GHz horns.}
\end{figure}

Plots of the dipole amplitudes with these errors are shown in Fig.~\ref{fig:dipoleParametersPlot}, together with the error ellipses for the dipole direction. As can be seen, the scatter is greater than the statistical error. Therefore, we take a conservative limit by marginalizing over all the MCMC samples for all the detectors, which results in an error ellipse ($\pm$0\pdeg02, $\pm$0\pdeg05) centred on Galactic latitude and longitude (48\pdeg26, 264\pdeg01). The dipole amplitudes exhibit a trend in focal plane position, which is likely due to residual, unaccounted-for power in far sidelobes, which would be symmetrical between horn pairs. These residuals, interacting with the Solar dipole, would behave like an orbital dipole, but in opposite ways on either side of the focal plane. This residual therefore cancels out to first order in each pair of symmetric horns in the focal plane, i.e., horns 18 with 23, 19 with 22, and 20 with 21 (see inset in Fig.~\ref{fig:dipoleParametersPlot}). As the overall dipole at 70\,GHz is calculated combining all the horns, the residual effect of far sidelobes is reduced.

\begin{DIFnomarkup}
\begin{table}
\begingroup
\newdimen\tblskip \tblskip=5pt
\caption{\label{tab:dipoleParameters} Dipole characterization from 70\,GHz
radiometers.}
\nointerlineskip
\vskip -3mm
\footnotesize
\setbox\tablebox=\vbox{
   \newdimen\digitwidth
   \setbox0=\hbox{\rm 0}
   \digitwidth=\wd0
   \catcode`*=\active
   \def*{\kern\digitwidth}
   \newdimen\signwidth
   \setbox0=\hbox{+}
   \signwidth=\wd0
   \catcode`!=\active
   \def!{\kern\signwidth}
\halign{\tabskip 0pt\hbox to 2.0cm{#\leaderfil}\tabskip 0.75em&
    \hfil$#$\hfil \tabskip 0.75em&
    \hfil$#$\hfil \tabskip 0.75em&
    \hfil$#$\hfil \tabskip 0pt\cr
\noalign{\doubleline}
\omit&&\multispan2\hfil\sc Galactic Coordinates [deg]\hfil\cr
\noalign{\vskip -3pt}
\omit&\omit\hfil Amplitude\hfil&\multispan2\hrulefill\cr
\noalign{\vskip 3pt} 
\omit Radiometer\hfil& [\unit{\mu K}_{\mathrm{CMB}}]& l& b\cr
\noalign{\vskip 3pt\hrule\vskip 5pt}
\noalign{\vskip 2pt}
18M& 3371.89\pm0.15& 264\pdeg014\pm0\pdeg008& 48\pdeg268\pm0\pdeg004\cr
18S& 3373.03\pm0.15& 263\pdeg998\pm0\pdeg008& 48\pdeg260\pm0\pdeg004\cr
19M& 3368.02\pm0.17& 263\pdeg981\pm0\pdeg009& 48\pdeg262\pm0\pdeg004\cr
19S& 3366.80\pm0.16& 264\pdeg019\pm0\pdeg009& 48\pdeg262\pm0\pdeg004\cr
20M& 3374.08\pm0.17& 264\pdeg000\pm0\pdeg010& 48\pdeg264\pm0\pdeg005\cr
20S& 3361.75\pm0.17& 263\pdeg979\pm0\pdeg010& 48\pdeg257\pm0\pdeg005\cr
21M& 3366.96\pm0.16& 264\pdeg008\pm0\pdeg008& 48\pdeg262\pm0\pdeg004\cr
21S& 3364.19\pm0.16& 264\pdeg022\pm0\pdeg009& 48\pdeg266\pm0\pdeg004\cr
22M& 3366.61\pm0.14& 264\pdeg014\pm0\pdeg008& 48\pdeg266\pm0\pdeg004\cr
22S& 3362.09\pm0.16& 264\pdeg013\pm0\pdeg009& 48\pdeg264\pm0\pdeg004\cr
23M& 3354.17\pm0.16& 264\pdeg027\pm0\pdeg009& 48\pdeg266\pm0\pdeg004\cr
23S& 3358.55\pm0.18& 263\pdeg989\pm0\pdeg009& 48\pdeg268\pm0\pdeg004\cr
\noalign{\vskip 0.5em}
Statistical& 3365.87\pm0.05& 264\pdeg006\pm0\pdeg003& 48\pdeg264\pm0\pdeg001\cr
Systematic& 3365.5\pm3.0& 264\pdeg01\pm0\pdeg05& 48\pdeg26\pm0\pdeg02\cr
\noalign{\vskip 0.5em}
Nominal$^{\rm a}$& 3364.5\pm2.0& 264\pdeg00\pm0\pdeg03& 48\pdeg24\pm0\pdeg02\cr
\noalign{\vskip 5pt\hrule\vskip 10pt}}}
\endPlancktable                    % ends one-column \halign
\tablenote {\textrm{a}} This estimate was produced combining the LFI and HFI
dipoles, and it is the one used to calibrate the LFI data delivered in the 2015
data release.\par
\endgroup
\end{table}
\end{DIFnomarkup}

%%%%%%%%%%%%%%%%%%%%%%%%%%%%%%%%%%%%%%%%%%%%%%%%%%%%%%%%%%%%%%%%%%%%%%

\section{Validation of the calibration and accuracy assessment}
\label{sec:validationAndAccuracy}

\begin{table*}[tb]
\begingroup
\newdimen\tblskip \tblskip=5pt
\caption{\label{tab:accuracyBudgetResult} Accuracy in the calibration of LFI data.}
\nointerlineskip
\vskip -3mm
\footnotesize
\setbox\tablebox=\vbox{
   \newdimen\digitwidth 
   \setbox0=\hbox{\rm 0} 
   \digitwidth=\wd0 
   \catcode`*=\active 
   \def*{\kern\digitwidth}
   \newdimen\signwidth 
   \setbox0=\hbox{+} 
   \signwidth=\wd0 
   \catcode`!=\active 
   \def!{\kern\signwidth}
\halign{
% Template
\hbox to 7.5cm{#\leaderfil}\tabskip 2.0em &
\hfil #\hfil &
\hfil #\hfil &
\hfil #\hfil\tabskip=0pt\cr
\noalign{\doubleline}
\omit\hfil Type of uncertainty\hfil &
\omit\hfil 30\,GHz\hfil &
\omit\hfil 44\,GHz\hfil &
\omit\hfil 70\,GHz\hfil\cr
\noalign{\vskip 3pt\hrule\vskip 5pt}
Solar dipole& 0.10\,\%& 0.10\,\%& 0.10\,\% \cr
Spread among independent radiometers$^a$& 0.25\,\%& 0.16\,\%& 0.10\,\%\cr
Overall error& 0.35\,\%& 0.26\,\%& 0.20\,\%\cr 
\noalign{\vskip 5pt\hrule\vskip 3pt}}}
\endPlancktablewide
\tablenote a This is the discrepancy in the measurement of the height of the first peak in the TT spectrum ($100 \leq \ell \leq 250$), as described in Sect.~\ref{sec:interChannelConsistency}.\par
\endgroup
\end{table*}

In this section we present the results of a set of checks we have run on the data that comprise this new \Planck{} release. Table~\ref{tab:accuracyBudgetResult} quantifies the uncertainties that affect the calibration of the LFI radiometers.

\subsection{Absolute calibration}
\label{sec:absoluteCalibration}

\begin{figure}
	\centering
	\includegraphics{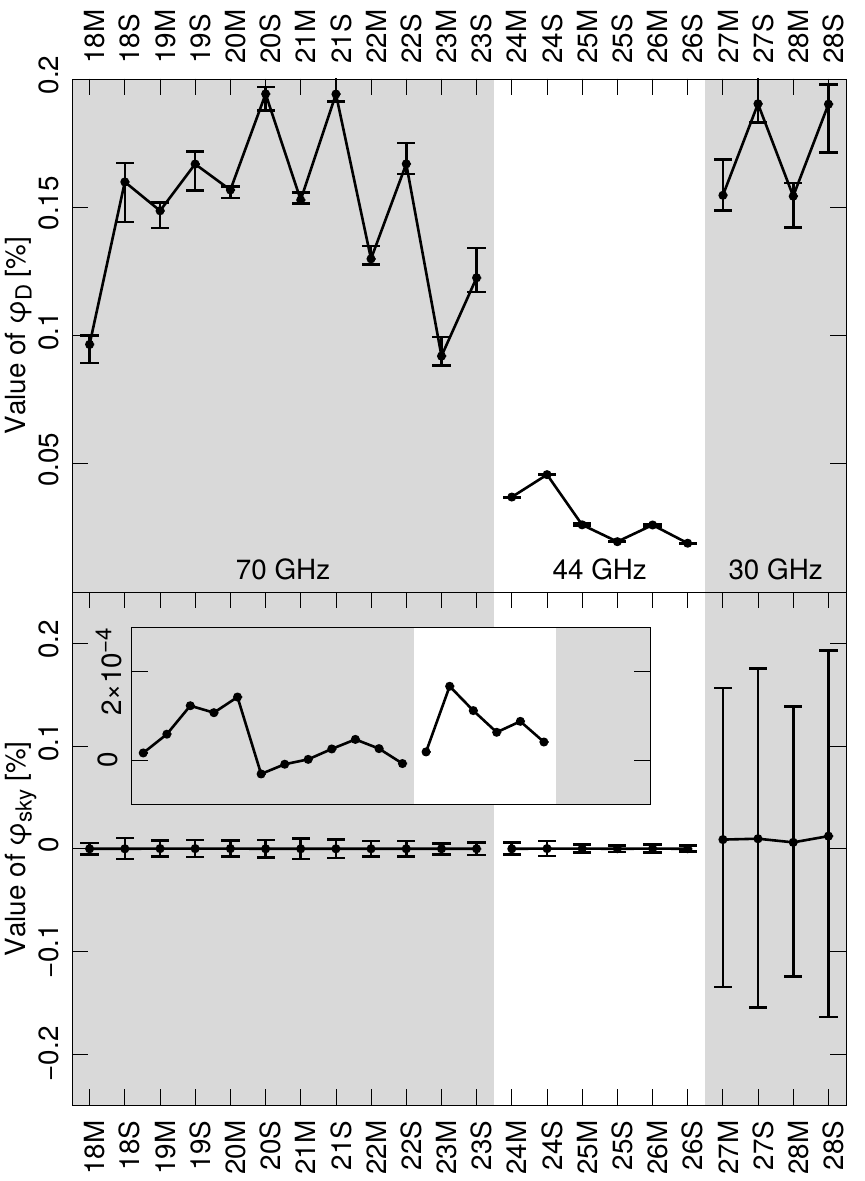}
	\caption{\label{fig:phiDPlot} \emph{Top}: estimate of the value of
$\phid$ (Eq.~\protect\ref{eq:phid}) for each LFI radiometer during the whole
mission. The plot shows the median value of $\phid$ over all the samples and
the 25th and 75th percentiles (upper and lower bar). Such bars provide an idea
of the range of variability of the quantity during the mission; they are not an
error estimate of the quantity itself. \emph{Bottom}: estimate of the value of
$\phisky$ (Eq.~\protect\ref{eq:phisky}). The points and bars have the same
meaning as in the plot above. Because of the smallness of the value for the 44
and 70\,GHz channels, the inset shows a zoom of their median values. The large
bars for the 30\,GHz channels are motivated by the coupling between the
stronger foregrounds and the relatively large power falling in the sidelobes.}
\end{figure}

\begin{DIFnomarkup}
\begin{table}
\begingroup
\newdimen\tblskip \tblskip=5pt
\caption{\label{tbl:FslAndPhiD}Optical parameters$^{\rm a}$ of the 22 LFI
 beams.}
\nointerlineskip
\vskip -3mm
\footnotesize
\setbox\tablebox=\vbox{
\newdimen\digitwidth
\setbox0=\hbox{\rm 0}
\digitwidth=\wd0
\catcode`*=\active
\def*{\kern\digitwidth}
\newdimen\signwidth
\setbox0=\hbox{+}
\signwidth=\wd0
\catcode`!=\active
\def!{\kern\signwidth}
\halign{
% Template
\tabskip 0pt\hbox to 2.0cm{#\leaderfil}\tabskip 1.0em&
\hfil#\hfil\tabskip 1.0em&
\hfil#\hfil\tabskip 1.0em&
\hfil#\hfil\tabskip 0pt\cr
\noalign{\doubleline}
% Heading
\omit Radiometer\hfil& $\fsl$~[\%]& $\phid$~[\%]& $\phisky$~[\%]\cr
\noalign{\vskip 3pt\hrule\vskip 5pt}
% Table (generated by scripts/plot_quantiles.r
\noalign{\vskip 3pt}
\multispan4 \hfil{\bf 70\,GHz}\hfil\cr
\noalign{\vskip 5pt}
18M&$0.38$& $0.097*^{+0.003*}_{-0.007*}$& $!0.0000*^{+0.0057*}_{-0.0058*}$\cr
\noalign{\vskip 3pt}                                                      
18S&$0.62$& $0.160*^{+0.007*}_{-0.016*}$& $!0.0001*^{+0.0102*}_{-0.0099*}$\cr
\noalign{\vskip 3pt}                                                      
19M&$0.60$& $0.149*^{+0.003*}_{-0.007*}$& $!0.0001*^{+0.0078*}_{-0.0077*}$\cr
\noalign{\vskip 3pt}                                                      
19S&$0.58$& $0.167*^{+0.005*}_{-0.010*}$& $!0.0001*^{+0.0083*}_{-0.0080*}$\cr
\noalign{\vskip 3pt}                                                      
20M&$0.63$& $0.157*^{+0.001*}_{-0.003*}$& $!0.0001*^{+0.0082*}_{-0.0079*}$\cr
\noalign{\vskip 3pt}                                                      
20S&$0.70$& $0.194*^{+0.003*}_{-0.006*}$& $-0.0000*^{+0.0084*}_{-0.0085*}$\cr
\noalign{\vskip 3pt}                                                      
21M&$0.59$& $0.153*^{+0.003*}_{-0.001*}$& $-0.0000*^{+0.0098*}_{-0.0100*}$\cr
\noalign{\vskip 3pt}                                                      
21S&$0.70$& $0.194*^{+0.007*}_{-0.003*}$& $!0.0000*^{+0.0093*}_{-0.0090*}$\cr
\noalign{\vskip 3pt}                                                      
22M&$0.44$& $0.130*^{+0.005*}_{-0.002*}$& $!0.0000*^{+0.0074*}_{-0.0073*}$\cr
\noalign{\vskip 3pt}                                                      
22S&$0.50$& $0.167*^{+0.008*}_{-0.004*}$& $!0.0000*^{+0.0076*}_{-0.0078*}$\cr
\noalign{\vskip 3pt}                                                      
23M&$0.35$& $0.092*^{+0.008*}_{-0.004*}$& $!0.0000*^{+0.0054*}_{-0.0056*}$\cr
\noalign{\vskip 3pt}                                                      
23S&$0.43$& $0.122*^{+0.012*}_{-0.006*}$& $-0.0000*^{+0.0062*}_{-0.0061*}$\cr
\noalign{\vskip 7pt}                                                      
\multispan4 \hfil{\bf 44\,GHz}\hfil\cr                                  
\noalign{\vskip 5pt}                                                      
24M&$0.15$& $0.0370^{+0.0001}_{-0.0001}$& $!0.0000*^{+0.0059*}_{-0.0058*}$\cr
\noalign{\vskip 3pt}                                                      
24S&$0.15$& $0.0457^{+0.0001}_{-0.0000}$& $!0.0002*^{+0.0073*}_{-0.0073*}$\cr
\noalign{\vskip 3pt}                                                      
25M&$0.08$& $0.0262^{+0.0006}_{-0.0004}$& $!0.0001*^{+0.0040*}_{-0.0040*}$\cr
\noalign{\vskip 3pt}                                                      
25S&$0.06$& $0.0196^{+0.0001}_{-0.0001}$& $!0.0001*^{+0.0032*}_{-0.0032*}$\cr
\noalign{\vskip 3pt}                                                      
26M&$0.08$& $0.0261^{+0.0004}_{-0.0006}$& $!0.0001*^{+0.0038*}_{-0.0037*}$\cr
\noalign{\vskip 3pt}                                                      
26S&$0.05$& $0.0190^{+0.0001}_{-0.0001}$& $!0.0000*^{+0.0030*}_{-0.0030*}$\cr
\noalign{\vskip 7pt}                                                      
\multispan4 \hfil{\bf 30\,GHz}\hfil\cr                                  
\noalign{\vskip 5pt}                                                      
27M&$0.64$& $0.155*^{+0.014*}_{-0.006*}$& $!0.0090*^{+0.1475*}_{-0.1439*}$\cr
\noalign{\vskip 3pt}                                                      
27S&$0.76$& $0.190*^{+0.017*}_{-0.007*}$& $!0.0098*^{+0.1656*}_{-0.1644*}$\cr
\noalign{\vskip 3pt}                                                      
28M&$0.62$& $0.154*^{+0.005*}_{-0.012*}$& $!0.0063*^{+0.1325*}_{-0.1310*}$\cr
\noalign{\vskip 3pt}                                                      
28S&$0.83$& $0.190*^{+0.008*}_{-0.019*}$& $!0.0123*^{+0.1810*}_{-0.1762*}$\cr
\noalign{\vskip 5pt\hrule\vskip 3pt}
		}
	}
	\endPlancktable
	\tablenote {{\rm a}} The values for $\phid$ and $\phisky$ are the
medians computed over the whole mission. Upper and lower bounds provide the
distance from the 25th and 75th percentiles and are meant to estimate the range
of variability of the quantity over the whole mission; they are \emph{not\/} to
be interpreted as error bars. We do not provide estimates for $\phisky'$, as
they can all be considered equal to zero.\par
	\endgroup
\end{table}
\end{DIFnomarkup}

\begin{table*}
\begingroup
\newdimen\tblskip \tblskip=5pt
\caption{\label{tbl:absoluteCalibrationChange}Changes in the
calibration level between this (2015) \Planck-LFI data release and the previous
(2013) one.}
\nointerlineskip
\vskip -3mm
\footnotesize
\setbox\tablebox=\vbox{
\newdimen\digitwidth
\setbox0=\hbox{\rm 0}
\digitwidth=\wd0
\catcode`*=\active
\def*{\kern\digitwidth}
\newdimen\signwidth
\setbox0=\hbox{+}
\signwidth=\wd0
\catcode`!=\active
\def!{\kern\signwidth}
\halign{
% Template
\tabskip 0pt\hbox to 2.0cm{#\leaderfil}\tabskip 1.0em&
\hfil#\hfil\tabskip 1.0em&
\hfil#\hfil\tabskip 1.0em&
\hfil#\hfil\tabskip 1.0em&
\hfil#\hfil\tabskip 1.0em&
\hfil#\hfil\tabskip 1.0em&
\hfil#\hfil\tabskip 0pt\cr
\noalign{\doubleline}
% Heading
\omit& Beam& Pipeline& Galactic& Reference& Estimated& Measured\cr
\omit Radiometer\hfil& convolution~[\%]& upgrades~[\%]& sidelobe removal~[\%]&
 dipole~[\%]& change~[\%]&change~[\%]\cr
\noalign{\vskip 3pt\hrule\vskip 5pt}
% Table (generated by scripts/plot_quantiles.r
\noalign{\vskip 6pt}
\multispan7 \hfil{\bf 70\,GHz}\hfil\cr
\noalign{\vskip 5pt}
18M& $-0.277$& $0.16$& $!0.000$& $0.271$& $!0.15$& $!0.19$\cr
18S& $-0.438$& $0.21$& $!0.000$& $0.271$& $!0.04$& $!0.28$\cr
19M& $-0.443$& $0.16$& $!0.000$& $0.271$& $-0.01$& $!0.09$\cr
19S& $-0.398$& $0.21$& $-0.002$& $0.271$& $!0.08$& $!0.20$\cr
20M& $-0.469$& $0.18$& $!0.000$& $0.271$& $-0.02$& $!0.18$\cr
20S& $-0.506$& $0.19$& $!0.000$& $0.271$& $-0.05$& $!0.09$\cr
21M& $-0.439$& $0.14$& $!0.000$& $0.271$& $-0.03$& $!0.01$\cr
21S& $-0.510$& $0.21$& $!0.000$& $0.271$& $-0.03$& $-0.30$\cr
22M& $-0.328$& $0.20$& $!0.000$& $0.271$& $!0.14$& $!0.08$\cr
22S& $-0.352$& $0.20$& $!0.000$& $0.271$& $!0.12$& $!0.21$\cr
23M& $-0.269$& $0.32$& $!0.000$& $0.271$& $!0.32$& $-0.03$\cr
23S& $-0.320$& $0.35$& $!0.000$& $0.271$& $!0.30$& $!0.44$\cr
\noalign{\vskip 7pt}
\multispan7 \hfil{\bf 44\,GHz}\hfil\cr
\noalign{\vskip 5pt}
24M& $-0.110$& $0.28$& $!0.000$& $0.271$& $!0.44$& $!0.71$\cr
24S& $-0.101$& $0.26$& $!0.000$& $0.271$& $!0.43$& $!0.34$\cr
25M& $-0.046$& $0.30$& $!0.000$& $0.271$& $!0.72$& $!0.27$\cr
25S& $-0.055$& $0.31$& $!0.000$& $0.271$& $!0.74$& $!0.62$\cr
26M& $-0.054$& $0.19$& $!0.000$& $0.271$& $!0.63$& $!0.52$\cr
26S& $-0.033$& $0.17$& $!0.000$& $0.271$& $!0.56$& $!0.44$\cr
\noalign{\vskip 7pt}
\multispan7 \hfil{\bf 30\,GHz}\hfil\cr
\noalign{\vskip 5pt}
27M& $-0.498$& $0.24$& $-0.015$& $0.271$& $-0.01$& $-0.15$\cr
27S& $-0.583$& $0.14$& $-0.019$& $0.271$& $-0.19$& $-0.56$\cr
28M& $-0.440$& $0.32$& $-0.003$& $0.271$& $!0.15$& $!0.35$\cr
28S& $-0.601$& $0.32$& $-0.004$& $0.271$& $-0.02$& $!0.22$\cr
\noalign{\vskip 5pt\hrule\vskip 3pt}
}}
\endPlancktablewide
\endgroup
\end{table*}

In this section we provide an assessment of the change in the absolute level of the calibration since the first \Planck\ data release, in terms of its impact on the maps and power spectra. Generally speaking, a change in the average value of $G$ in Eq.~\eqref{eq:radiometerEquation} of the form
\begin{equation}
\left<G\right> \rightarrow \left<G\right> (1 + \delta_G),\
 \text{with $\delta_G \ll 1$}
\end{equation}
leads to a change $\left<T\right> \rightarrow \left<T\right>(1 + \delta_G)$ in the average value of the pixel temperature $T$, and to a change $C_\ell \rightarrow C_\ell (1 + 2\delta_G)$ in the average level of the measured power
spectrum $C_\ell$, before the application of any window function. Our aim is to quantify the value of the variation $\delta_G$ from the previous \Planck-LFI data release to the current one. We have done this by comparing the level of power spectra in the $\ell=100$--250 multipole range consistently with \OldPlanckLFICalPaper.

There have been several improvements in the calibration pipeline that have led to a change in the value of $\left<G\right>$:
\begin{enumerate}
\item the peak-to-peak temperature difference of the reference dipole $D$ used in Eq.~\eqref{eq:radiometerEquation} has changed by $+0.27\,\%$ (see Sect.~\ref{sec:DipoleEstimation}), because we now use the Solar dipole parameters calculated from our own \Planck\ measurements \citep{planck2014-a01};
\item in the same equation, we no longer convolve the dipole $D$ with a beam $B$ that is a delta function, but instead use the full profile of the beam over the sphere (see Sect.~\ref{sec:beamEfficiency});
\item the beam normalization has changed, since in this data release $B$ is such that \citep{planck2014-a05}
\begin{equation}
\int_{4\pi} B(\theta, \varphi)\,\ud\Omega \lneqq 1;
\end{equation}
\item the old dipole fitting code has been replaced with a more robust algorithm, \DaCapo\ (see Sect.~\ref{sec:DaCapo}).
\end{enumerate}

\begin{figure}[tb]
	\includegraphics{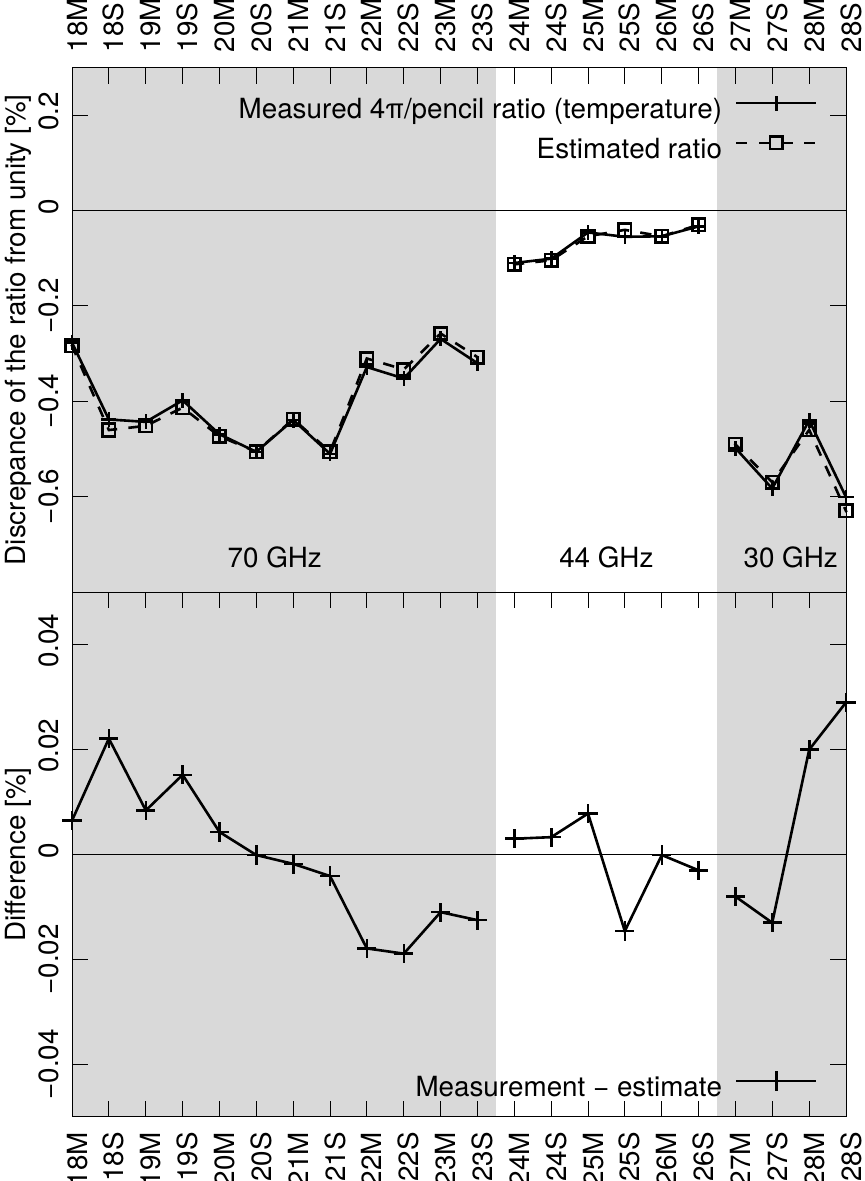}
	\caption{\label{fig:fourPiPencilRatio} \emph{Top}: comparison between
the measured and estimated ratios of the $a_{\ell m}$ harmonic coefficients for
the nominal maps (produced using the full knowledge of the beam $B$ over the
$4\pi$ sphere) and the maps produced under the assumption of a pencil beam. The
estimate has been computed using Eq.~\protect\eqref{eq:GalacticPickupRemoval}.
\emph{Bottom}: difference between the measured ratio and the estimate. The
agreement is better than 0.03\,\% for all 22 LFI radiometers.}
\end{figure}

Table~\ref{tbl:absoluteCalibrationChange} lists the impact of such effects on
the amplitude of fluctuations in the temperature $\left<T\right>$ of the 22 LFI
radiometer maps. The numbers in this table have been computed by re-running the
calibration pipeline on all the 22 LFI radiometer data with the following
setup:
\begin{enumerate}
\item a pencil-beam approximation for $B$ in Eq.~\eqref{eq:radiometerEquation}
has been used, instead of the full $4\pi$ convolution (``Beam convolution''
column), with the impact of this change quantified by
Eq.~\eqref{eq:fourPiPencilTRatio}, and the comparison between the values
predicted by this equation with the measured change in the $a_{\ell m}$
harmonic coefficients shown in Fig.~\ref{fig:fourPiPencilRatio} (the agreement
is excellent, better than 0.03\,\%);
\item the old calibration code has been used instead of the \DaCapo\ algorithm
described in Sect.~\ref{sec:DaCapo} (``Pipeline upgrades'' column);
\item the $B * \Tgalaxy$ term has \emph{not\/} been removed, as in the
discussion surrounding Eq.~\eqref{eq:GalacticPickupRemoval} (``Galactic
sidelobe removal'' column);
\item the signal $D$ used in Eq.~\eqref{eq:radiometerEquation} has been
modelled using the dipole parameters published in \citet{hinshaw2009}, as was
done in \OldPlanckLFICalPaper\ (``Reference dipole'' column).
\end{enumerate}
We have measured the actual change in the absolute calibration level by
considering the radiometric maps (i.e., maps produced using data from one
radiometer) of this data release (indicated with a prime) and of the previous
data release and averaging the ratio:
\begin{equation}
\label{eq:measuredChangeInT}
\Delta^{x,x'}_\ell = \left<\frac{C_{x' \times y}}{C_{x \times y}}\right>_y - 1,
\end{equation}
where $y$ indicates a radiometer at the same frequency as $x$ and $x'$, such
that $y \not= x$. The average is meant to be taken over all the possible
choices for $y$ (thus 11 choices for 70\,GHz radiometers, 5 for 44\,GHz, and 3
for 30\,GHz) in the multipole range $100 \leq \ell \leq 250$. The way that
cross-spectra are used in Eq.~\eqref{eq:measuredChangeInT} ensures that the
result does not depend on the white noise level. Of course, this result
quantifies the ratio between the temperature fluctuations (more correctly,
between the $a_{\ell m}$ coefficients of the expansion of the temperature map
in spherical harmonics) in the two data releases, and not between the power
spectra.

\begin{figure}[tb]
	\includegraphics{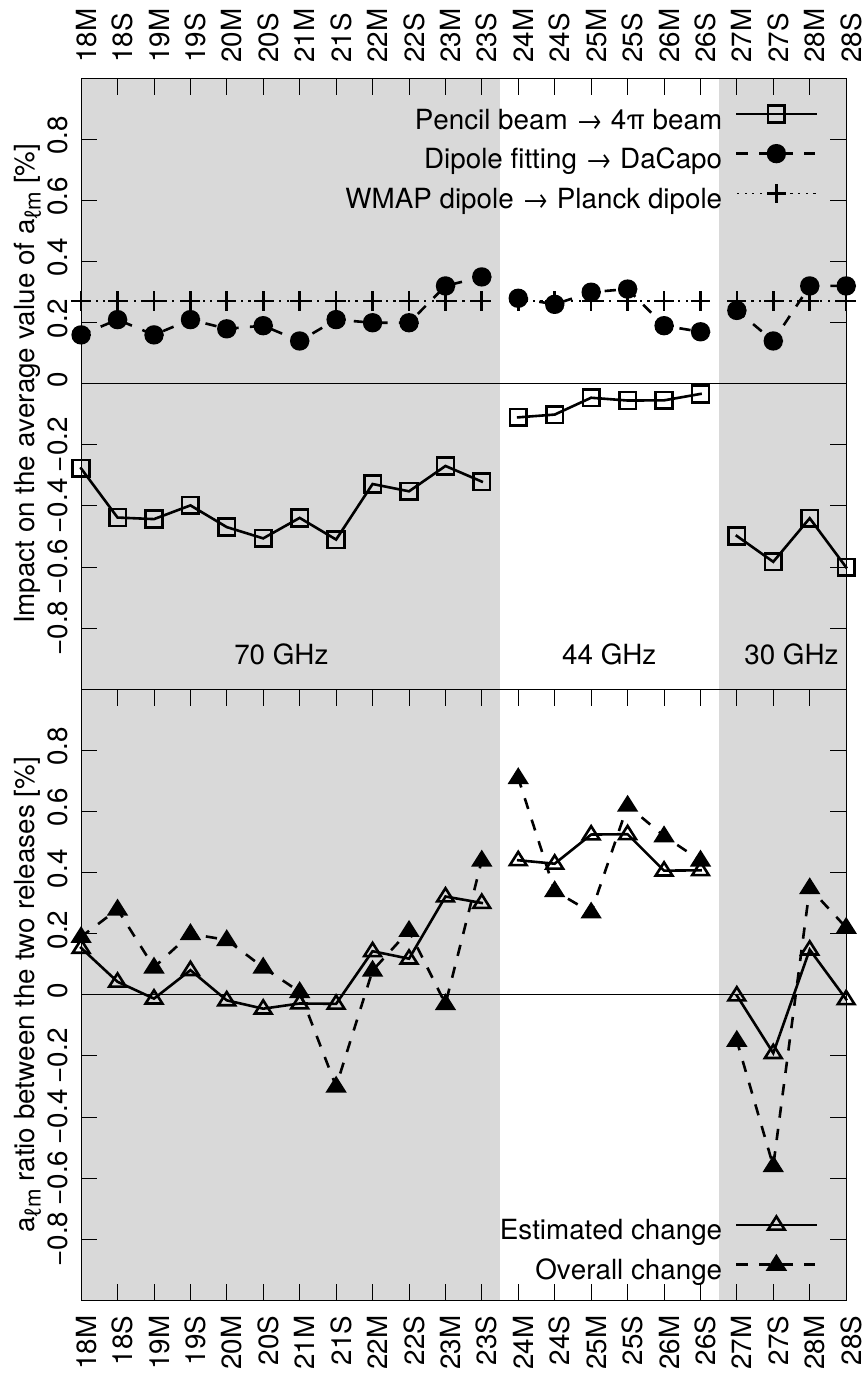}
	\caption{\label{fig:absCalContributions} \emph{Top}: impact on the
average value of the $a_{\ell m}$ spherical harmonic coefficients (computed
using Eq.~\ref{eq:measuredChangeInT}, with $100 \leq \ell \leq 250$) due to a
number of improvements in the LFI calibration pipeline, from the first to the
second data release. \emph{Bottom}: measured change in the $a_{\ell m}$
harmonic coefficients between the first and the second data release. No beam
window function has been applied. These values are compared with the estimates
produced using Eq.~\eqref{eq:estimatedChangeInT}, which assumes perfect
independence among the effects.}
\end{figure}

The column labeled ``Estimated change'' in
Table~\ref{tbl:absoluteCalibrationChange} contains a simple combination of all
the numbers in the table:
\begin{equation}
\label{eq:estimatedChangeInT}
\begin{split}
\text{Estimated change} =\ &(1 + \epsilon_\mathrm{beam})\,
 (1 + \epsilon_\mathrm{pipeline})\,\times \\
& (1 + \epsilon_\mathrm{Gal})\,(1 + \epsilon_D) - 1,
\end{split}
\end{equation}
where the $\epsilon$ factors are the numbers shown in the same table and
discussed above. This formula assumes that all the effects are mutually
independent. This is of course an approximation; however, the comparison
between this estimate and the measured value (obtained by applying
Eq.~\ref{eq:measuredChangeInT} to the 2013 and 2015 release maps) can give an
idea of the amount of interplay of such effects in producing the observed shift
in temperature. Fig.~\ref{fig:absCalContributions} shows a plot of the
contributions discussed above, as well as a visual comparison between the
measured change in the temperature and the estimate from
Eq.~\eqref{eq:estimatedChangeInT}.

\subsection{Noise in dipole fitting}
\label{sec:fitErrors}

We have performed a number of simulations that quantify the impact of white noise in the data on the estimation of the calibration constant, as well as the ability of our calibration code to retrieve the true value of the calibration constants. Details of this analysis are described in \citet{planck2014-a04}. We do not include such errors as an additional element in Table~\ref{tab:accuracyBudgetResult}, as the statistical error is already included in the row ``Spread among independent radiometers.''

\subsection{Beam uncertainties}
\label{sec:beamUncertainties}

As discussed in \citet{planck2014-a05}, the beams $B$ used in the LFI pipeline are very similar to those presented in \cite{planck2013-p02d}; they are computed with \texttt{GRASP}, properly smeared to take into account the satellite motion. Simulations have been performed using the optical model described in \cite{planck2013-p02d}, which was derived from the \Planck\ radio frequency Flight Model \citep{tauber2010b} by varying some optical parameters within the nominal tolerances expected from the thermoelastic model, in order to reproduce the measurements of the LFI main beams from seven Jupiter transits. This is the same procedure adopted in the 2013 release \citep{planck2013-p02d}; however, unlike the case presented in \citet{planck2013-p02d}, a different beam normalization is introduced here to properly take into account the actual power entering the main beam (typically about 99\,\% of the total power). 
This is discussed in more detail in \citet{planck2014-a05}.

Given the broad usage of beam shapes $B$ in the current LFI calibration pipeline, it is extremely important to assess their accuracy and how errors in $B$ propagate down to the estimate of the calibration constants $K$ in Eq.~\eqref{eq:radiometerEquation}.

In the previous data release we did not use our knowledge of the bandpasses of each radiometer to produce an in-band model of the beam shape, but instead estimated $B$ by means of a monochromatic approximation (see \OldPlanckLFICalPaper{}). In that case, we estimated the error induced in the calibration as the variation of the dipole signal when using either a monochromatic or a band-integrated beam, as we believe the latter to be a more realistic model.

In this data release, we have switched to the full bandpass-integrated beams produced using {\tt GRASP}, which represents our best knowledge of the beam \citep{planck2014-a05}. We have tested the ability of \DaCapo\ to retrieve the correct calibration constants $K$ for LFI19M (a 70\,GHz radiometer) when the large-scale component ($\ell = 1$) of the beam's sidelobes is: (1) rotated arbitrarily by an angle $-160^\circ \leq \theta \leq 160^\circ$; or (2) scaled by $\pm 20\,\%$.  We have found that such variations alter the calibration constants by approximately $0.1\,\%$. However, we do not list such a number as an additional source of uncertainty in Table~\ref{tab:accuracyBudgetResult}, since we believe that this is already captured by the scatter in the points shown in Fig.~\ref{fig:spectrumLevelRad}, which were used to produce the numbers in the row ``Inconsistencies among radiometers.''

\subsection{Inter-channel calibration consistency}
\label{sec:interChannelConsistency}

\begin{figure}
	\includegraphics{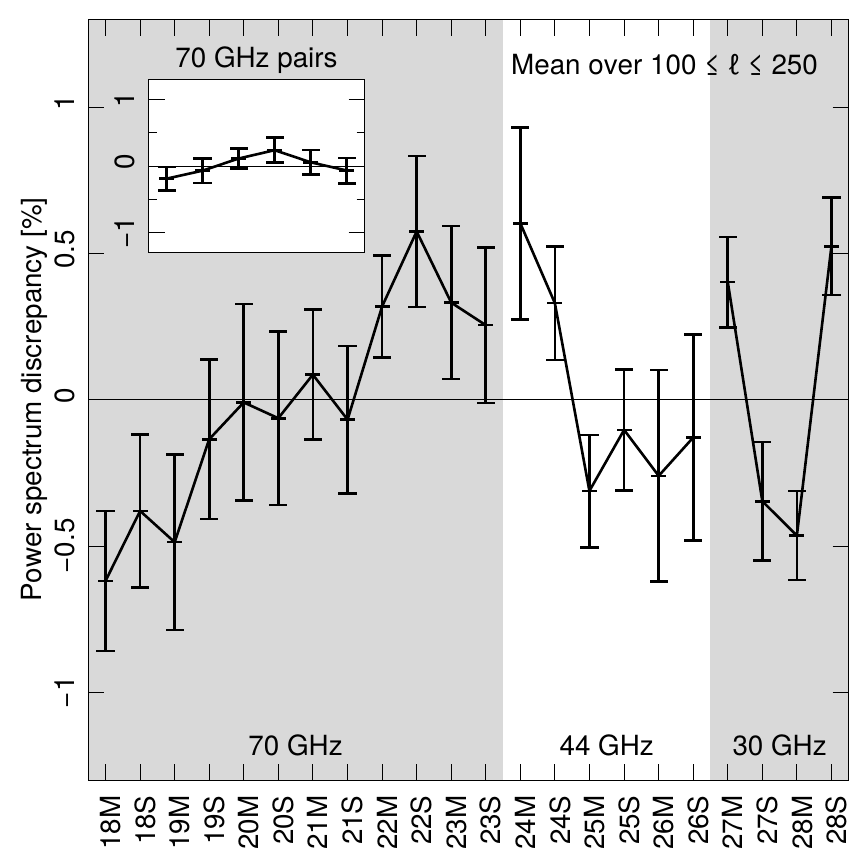}
    \caption{\label{fig:spectrumLevelRad} Discrepancy among the radiometers of
the same frequency in the height of the power spectrum $C_\ell$ near the first
peak. For a discussion of how these values were computed, see the text.
\emph{Inset}: to better understand the linear trend in the 70\,GHz radiometers,
we have computed the weighted average between pairs of radiometers whose
position in the focal plane is symmetric. The six points refer to the
combinations 18M/23M, 18S/23S, 19M/22M, 19S/22S, 20M/21M, and 20S/21S,
respectively. Note that all six points are consistent with zero within
$1\sigma$; see also Fig.~\protect\ref{fig:dipoleParametersPlot}.}
\end{figure}

In this section we provide a quantitative estimate of the relative calibration
error for the LFI frequency maps by measuring the consistency of the power
spectra computed using data from one radiometer at time. By ``relative error''
we mean any error that is different among the radiometers, in contrast to an
``absolute error,'' which induces a common shift in the power spectrum. We
have computed the power spectrum of single radiometer half-ring maps and have
estimated the variation in the region around the first peak
($100 \leq \ell \leq 250$), since this is the multipole range with the
best S/N.

The result of this analysis is shown in Fig.~\ref{fig:spectrumLevelRad}, which
plots the values of the quantity
\begin{equation}
\label{eq:interchannelConsistency}
\delta_{\rm rad} = \frac{\left<C_\ell^{\rm HR}\right>_{\rm rad}}
 {\left<C_\ell^{\rm HR}\right>_{\rm freq}} - 1,
\end{equation}
where $C_\ell^{\mathrm{HR}}$ is the cross-power spectrum computed using two
 half-ring maps, and $\left<\cdot\right>$ denotes an average over $\ell$. The
quantity $\left<C_\ell\right>_{\mathrm{freq}}$ is the same average computed
using the full frequency half-ring maps. Note that the $\delta_{\mathrm{rad}}$
slope is symmetric around zero in the 70\,GHz radiometers; this might be caused
by residual unaccounted power in the far sidelobes of the beam. The same
explanation was advanced in Sect.~\ref{sec:DipoleEstimation} to explain a
similar effect. It is interesting to note that the amplitude of the two
systematics is comparable; the trend in Fig.~\ref{fig:dipoleParametersPlot} has
a peak-to-peak variation (in temperature) of about $0.5\,\%$, while the trend
in Fig.~\ref{fig:spectrumLevelRad} has a variation (in power) of roughly
$1.0\,\%$. We combine the values of $\delta_{\mathrm{rad}}$ for those pairs of
radiometers whose beam position in the focal plane is symmetric (e.g., 18M
versus 23M, 18S versus 23S, 19M versus 22M, etc.), since in these pairs the
unaccounted power should be balanced. We have found that indeed all the six
combinations of $\delta_{\mathrm{rad}}$ are consistent with zero within
$1\,\sigma$ (see the inset of Fig.~\ref{fig:spectrumLevelRad}).

Since the cross-spectrum of two half-ring maps does not depend on the level of
uncorrelated noise, the fluctuations of $\delta_i$ around the average value
that can be seen in Fig.~\ref{fig:spectrumLevelRad} can be interpreted as
relative calibration errors. If we limit our analysis to the multipole range
$100 \leq \ell \leq 250$, we can estimate the error of the 70\,GHz map as the
error on the average height of the peaks (i.e., the value $\sigma/\sqrt{N}$,
with $\sigma$ being the standard deviation and $N$ the number of points) that
is, 0.25, 0.16, and 0.10 percent and 30, 44, and 70\,GHz, respectively.

\subsection{Inter-frequency calibration consistency}
\label{sec:interFrequencyConsistency}

In this section we carry on an analysis similar to the one presented in
Sect.~\ref{sec:interChannelConsistency}, where we compare the absolute level of
the maps at the three LFI frequencies, i.e., 30, 44, and 70\,GHz.

We make use of the full frequency maps, as well as the pair of half-ring maps
at 70\,GHz. Each half-ring map has been produced using data from one of the two
halves of each pointing period. We quantify the discrepancy between the 70\,GHz
map and another map by means of the quantity
\begin{equation}
\label{eq:interfreqRatio}
\Delta_{\ell}^{\text{70 GHz,other}} = \frac{C^{\text{HR1} \times
 \text{HR2}}_\ell}{C^{\text{HR1}\times\text{other}}_\ell} - 1,
\end{equation}
where $C^{\text{HR1} \times \text{HR2}}_\ell$ is the cross-spectrum between the
two 70\,GHz half-ring maps, and $C^{\text{HR1}\times\text{other}}_\ell$ is the
cross-spectrum between the first 70\,GHz half-ring map and the map under
analysis. In the ideal case (perfect correspondence between the spectrum of the
70\,GHz map and the other map) we expect $\Delta_\ell = 0$. As was the case for
Eq.~\eqref{eq:interchannelConsistency}, this formula has the advantage of
discarding the white noise level of the spectrum $C_{\ell}^{\text{other}}$ by
using the cross-spectrum with the 70\,GHz map, whose noise should be
uncorrelated.

\begin{figure}[tb]
	\includegraphics{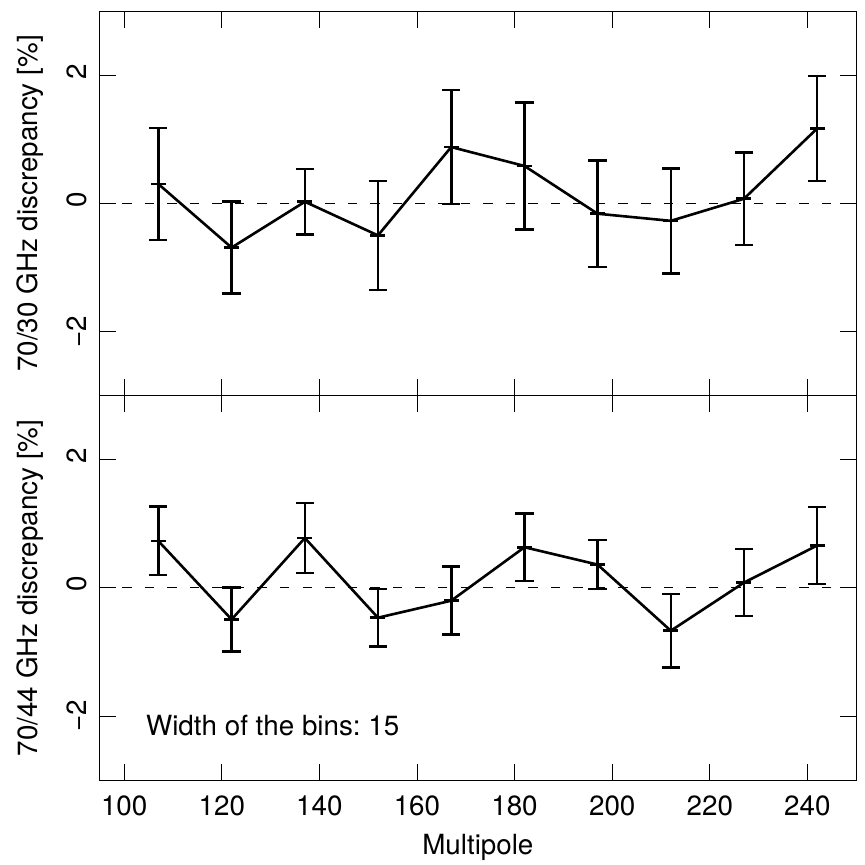}
	\caption{\label{fig:interfreqRatio} Estimate of
$\Delta_{\ell}^{\text{70 GHz,other}}$ (Eq.~\protect\ref{eq:interfreqRatio}),
which quantifies the discrepancy between the level of the 70\,GHz power
spectrum and the level of another map. \emph{Top}: comparison between the
70\,GHz map and the 30\,GHz map in the range of multipoles
$100 \leq \ell \leq 250$. The error bars show the rms of the ratio within each
bin of width 15. \emph{Bottom}: the same comparison done between the 70\,GHz
map and the 44\,GHz map. A 60\,\% mask was applied before computing the
spectra.}
\end{figure}

Over the multipole range $100 \leq \ell \leq 250$, the average
discrepancy\footnote{To reduce the impact of the Galactic signal we have
masked 60\,\% of the sky, since we found that less aggressive masks produced
significant biases in the ratios.} is $0.15\pm 0.17\,\%$ for the 44\,GHz map,
and $0.15\pm 0.26\,\%$ for the 30\,GHz map, as shown in
Fig.~\ref{fig:interfreqRatio}. Such numbers are consistent with the calibration
errors provided in Sect.~\ref{sec:interChannelConsistency}.

\subsection{Null tests}
\label{sec:nullTests}

\begin{figure*}[tb]
    \includegraphics[width=\textwidth]{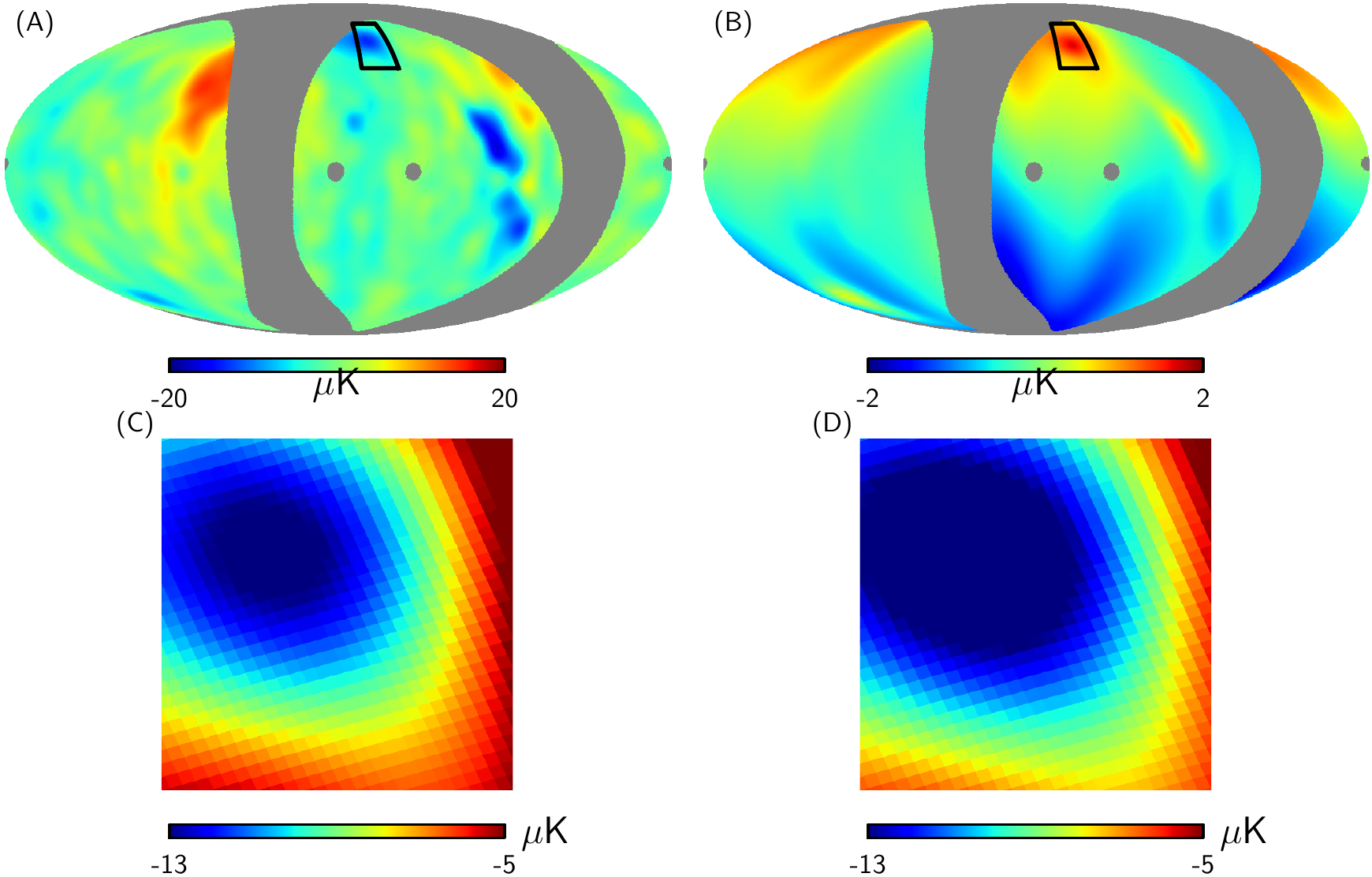}
    \caption{\label{fig:fourPiPencilMaps} Difference in the application of the full $4\pi$ beam model versus the pencil beam approximation. \textit{Panel A}: difference between survey 1 and survey 2 for a 30\,GHz radiometer (LFI-27S) with the $4\pi$ model, smoothed to $15^\circ$.  We do not show the same difference with pencil beam approximation, as it would appear indistinguishable from the $4\pi$ map. \textit{Panel B}: double difference between the $4\pi$ $1-2$ survey difference map in panel A and the pencil difference map (not shown here). This map shows what changes when one drops the pencil approximation and uses the full shape of the beam in the calibration. \textit{Panel C}: zoom on the blue spot visible at the top of the map in panel A. \textit{Panel D}: same zoom for the pencil approximation map. The comparison between panel C and D shows that the $4\pi$ calibration produces better results.}
\end{figure*}

In \OldPlanckLFICalPaper{} we provided a study of a number of null tests, with the purpose of testing the quality of the calibration. In this new data release, we have moved the bulk of the discussion to \citet{planck2014-a04}. We just show here one example, which is particularly relevant in the context of the LFI calibration validation. Figure~\ref{fig:fourPiPencilMaps} shows the variation in the quality of the maps due to the use of the full $4\pi$ convolution versus a pencil beam approximation, as discussed in Sect.~\ref{sec:beamEfficiency}. The analysis of many similar difference maps has provided us with sufficient evidence that using the full $4\pi$ beam convolution reduces the level of systematic effects in the LFI maps.

%%%%%%%%%%%%%%%%%%%%%%%%%%%%%%%%%%%%%%%%

\section{Measuring the brightness temperature of Jupiter}
\label{sec:planets}

\newcommand{\DeltaTDipoleDirac}{\Delta T_{\delta}}
\newcommand{\DeltaTDipole}{\Delta T}

\newcommand{\Pointingt}{\vec{x}_{t}}
\newcommand{\psit}{\psi_{t}}

\newcommand{\Pointing}{\vec{x}_{\mathrm{b}}}
\newcommand{\PointingE}{\vec{x}}
\newcommand{\DipoleE}{\vec{D}_{\mathrm{E}}}
\newcommand{\DipoleEv}{\vec{D}_{\mathrm{E}}}
\newcommand{\RotBeamObs}{\tens{U}}
\newcommand{\BeamRd}{B}
\newcommand{\betarad}{\beta_{\mathrm{rad}}}
\newcommand{\RadiometerM}[1]{\tens{M}_{#1}}
\newcommand{\RadiometerA}{\tens{A}}
\newcommand{\vsun}{\vec{v}_{\mathrm{Sun}}}
\newcommand{\vplanck}{\vec{v}_{\mathrm{Planck}}}
\newcommand{\PointingBx}{x_\mathrm{b}}
\newcommand{\PointingBy}{y_\mathrm{b}}
\newcommand{\PointingBz}{z_\mathrm{b}}
\newcommand{\HEALpix}{{\tt HEALPix}}
\newcommand{\Nside}{N_\mathrm{side}}

\newcommand{\PlanetPointingt}{\vec{x}_{\mathrm{p},t}}

\newcommand{\PlanetPointing}{\vec{x}_{\mathrm{p}}}
\newcommand{\PlanetPointingBeam}{\vec{y}_{\mathrm{p}}}
\newcommand{\PlanetI}{I_{\mathrm{p}}}
\newcommand{\BackgroundI}{I_{\mathrm{bg}}}
\newcommand{\PlanetIz}{I_{\mathrm{p},0}}
\newcommand{\PlanetIzAver}{\bar{I}_{\mathrm{p},0}}

\newcommand{\DeltaTantp}{\Delta T_{\mathrm{ant},p}}
\newcommand{\DeltaTantt}{\Delta T_{\mathrm{ant},t}}

\newcommand{\Tant}{T_{\mathrm{A}}}
\newcommand{\Tantp}{T_{\mathrm{A,p}}}
\newcommand{\Tantt}{T_{\mathrm{ant},t}}
\newcommand{\BeamResponset}{B_t}
\newcommand{\Background}{b}

\newcommand{\PlanetocentricPlanckLatitude}{\delta_{\mathrm{P}}}
\newcommand{\PlanetFlattening}{e_{\mathrm{p}}}

\newcommand{\DeltaPlanet}{\Delta_\mathrm{p}}
\newcommand{\DeltaPlanetRef}{\Delta_{\mathrm{p},\mathrm{ref}}}

\newcommand{\OmegaPlanet}{\Omega_{\mathrm{p}}}
\newcommand{\OmegaPlanetPolar}{\Omega_{\mathrm{p}}^{\mathrm{polar}}}
\newcommand{\OmegaPlanetPolarRef}{\Omega_{\mathrm{p}}^{\mathrm{polar},\mathrm{ref}}}

\newcommand{\OmegaBeam}{\Omega_{\mathrm{b}}}

\newcommand{\Tb}{T_{\mathrm{B}}}

\newcommand{\Fcent}{\nu_{\mathrm{cen}}}

\begin{DIFnomarkup}
\begin{table*}
\begingroup
\newdimen\tblskip \tblskip=5pt
\caption{\label{tbl:planetCrossingTimes} Visibility epochs of Jupiter.}
\nointerlineskip
\vskip -3mm
\footnotesize
\setbox\tablebox=\vbox{
   \newdimen\digitwidth 
   \setbox0=\hbox{\rm 0} 
   \digitwidth=\wd0 
   \catcode`*=\active 
   \def*{\kern\digitwidth}
   \newdimen\dotwidth % We need this to properly space the numbers below!
   \setbox0=\hbox{.} 
   \dotwidth=\wd0 
   \catcode`!=\active 
   \def!{\kern\dotwidth}
\halign{
% Template
#\hfil\tabskip 3.0em&
#\hfil\tabskip 3.0em&
#\hfil\tabskip=0pt\cr
\noalign{\doubleline}
\omit\hfil 30\,GHz\hfil&
\omit\hfil 44\,GHz$^{\rm a}$\hfil&
\omit\hfil 70\,GHz\hfil\cr
\noalign{\vskip 3pt\hrule\vskip 5pt}
31 October--2 November 2009& 24--27 October 2009& 29 October--1 November 2009\cr
30 June--3 July 2010& 31 October--2 November 2009& 1--5 July 2010\cr
14--18 December 2010& 30 June--2 July 2009& 12--16 December 2010\cr
1--4 August 2011& 8--12 July 2010& 2--10 August 2011\cr
31 August--7 September 2012& 5--8 December 2010& 5--11 September 2012\cr
21 February--1 March 2013& 15--18 December 2010& 15--24 February 2013\cr
& 1--3 August 2011& \cr
& 7--9 August 2011& \cr
& 31 August--6 September 2012& \cr
& 11--16 September 2012& \cr
& 7--12 February 2013& \cr
& 23 February--1 March 2013& \cr
\noalign{\vskip 5pt\hrule\vskip 3pt}}}
\endPlancktablewide
\tablenote {{\rm a}} The observation of Jupiter is more scattered in time for
the 44\,GHz radiometers because of their peculiar placement in the LFI focal
plane.\par
\endgroup
\end{table*}
\end{DIFnomarkup}

The analysis of the flux densities of planets for this \Planck\ data release has been considerably extended.  We only use Jupiter data for planet calibration, so we will focus the discussion on observations of this planet.  The new analysis includes all seven transits of Jupiter through each main beam of the 22 LFI radiometers. The analysis pipeline has been improved considerably by taking into account several effects not included in \OldPlanckLFICalPaper.

Planets provide a useful calibration cross-check; in particular, the measurement of the brightness temperature of Jupiter can be a good way to assess the accuracy of the calibration, as Jupiter is a remarkably bright source with a S/N per scan as high as 50 and a relatively well known spectrum.  Furthermore, at the resolution of LFI beams it can be considered a point-like source. 

\begin{figure*}
     \includegraphics[]{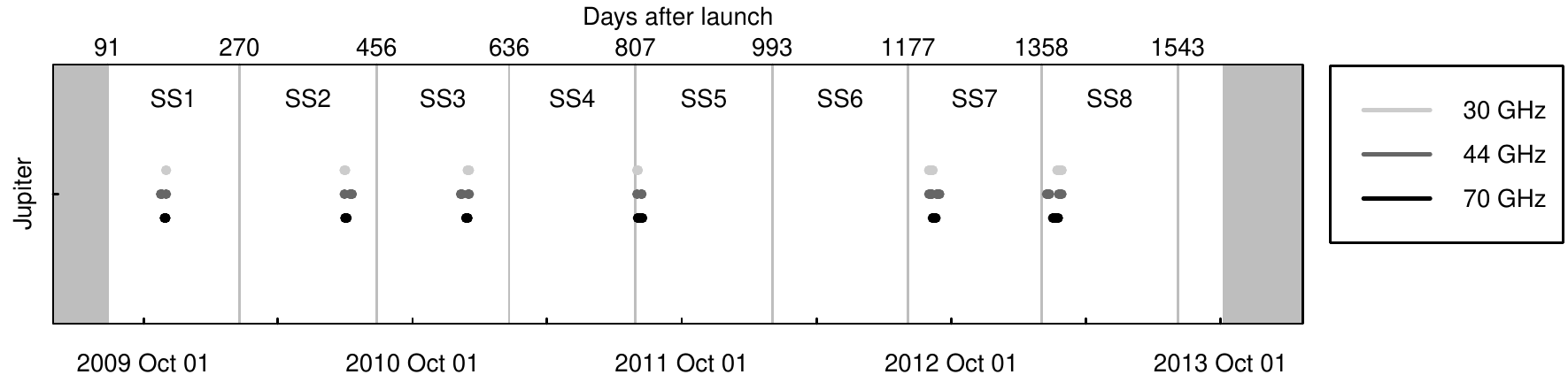}
     \caption{\label{fig:planetCrossingTimes} Visual timeline of Jupiters's crossings with LFI beams.  Here ``SS'' lables sky surveys.}
\end{figure*}

\begin{figure*}
     \includegraphics[]{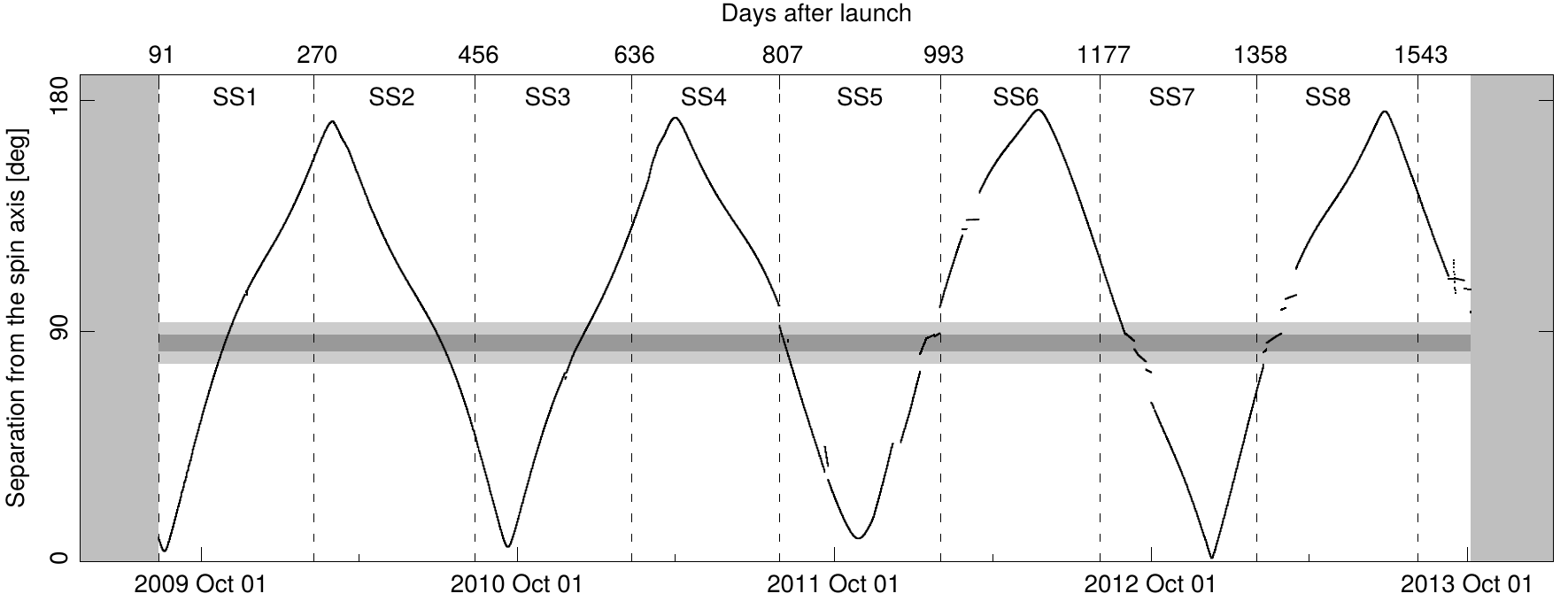}
     \caption{\label{fig:jupiterAngle} Time dependence of the angle between Jupiter's direction and the spin axis of the \Planck\ spacecraft. The darker horizontal bar indicates the angular region of the 11 LFI beam axes, and the lighter bar is enlarged by $\pm 5^\circ$.}
\end{figure*}

\subsection{Input data}

Table~\ref{tbl:planetCrossingTimes} lists the epochs when the LFI main beams crossed Jupiter, and Figs.~\ref{fig:planetCrossingTimes} and \ref{fig:jupiterAngle} give a visual timeline of these events. The first four transits occurred in nominal scan mode (spin shift $2\arcmin$, $1\degr$ per day) with a phase angle of $340\degr$, and the last three scans in deep mode (shift of the spin axis between rings of $0.5\arcmin$, $15\arcmin$ per day) with a phase angle of $250\degr$ \citep[see][]{planck2014-a01}. The analysis follows the procedure outlined in \OldPlanckLFICalPaper, but with a number of improvements:
\begin{enumerate}
\item the brightness of Jupiter has been extracted from timelines to fully exploit the time dependence in the data;
\item seven transits have been considered instead of two, which allowed us to better analyse the sources of scatter among the measurements;
\item all the data have been calibrated simultaneously;
\item different extraction methods have been exploited, in order to find the most reliable among them.
\end{enumerate}

In the following discussion, we refer to a ``timeline'' (one for each of the 22 LFI radiometers) as the list of values $(t, \PlanetPointingt, \Pointingt, \psit,\DeltaTantt),$ with $t$ the epoch of observation, $\PlanetPointingt$ the instantaneous apparent planet positions as seen from \Planck, $\Pointingt$ and $\psit$ the corresponding beam pointing directions and orientations, and $\DeltaTantt$ is the measured antenna temperature. The values of $\DeltaTantt$ provided by the LFI pipeline are calibrated and have their dipole and quadrupole signals removed. The pipeline also provides the values of $\Pointingt$ and $\psit$. We recovered $\PlanetPointingt$ from the Horizons\footnote{\url{http://ssd.jpl.nasa.gov/?horizons}} on-line service.  

Samples from each radiometer timeline have been used in this analysis only if the following conditions were met: (1) the samples have been acquired in stable conditions during a pointing period \citep{planck2014-a03}; (2) the pipeline has not flagged them as ``bad''; (3) their angular distance from the planet position at the time of the measurement is less than $5\degr$; and (4) they are not affected by any anomaly or relevant background source. We checked the last condition by visually inspecting small coadded maps of the selected samples.

\subsection{Description of the analysis pipeline}

In the following paragraphs, we describe how we improved the pipeline used to extract the brightness temperature $\Tb$ of Jupiter from the raw LFI data. Such extraction goes through a first estimation of the antenna temperature $\Tant$ and a number of corrections to take into account various systematic effects.  We present the methods used to estimate $\Tant$ in Sect.~\ref{sec:JupiterTant}, and then in Sect.~\ref{sec:JupiterTbri} we discuss the estimation of $\Tb$.  Since the computation of $\Tb$ requires an accurate estimate of the planet solid angle $\OmegaPlanet$, we discuss the computation of this factor in a dedicated part, Sect.~\ref{sec:JupiterSolidAngle}.

\subsubsection{Estimation of the antenna temperature}
\label{sec:JupiterTant}

Following \OldPlanckLFICalPaper\ and \citet{cremonese.2002.asteroid.detection}, the recovery of the instantaneous planet signal from a timeline is equivalent to the deconvolution of the planet shape from the beam pattern $\BeamResponset$ at time $t$. Since the planet can be considered a point source, the most practical way is to assume 
\begin{equation}
\label{eq:TantToTb}
\DeltaTantt= \Tantp \; \BeamResponset(\delta\PlanetPointingt) + \Background,
% \left(1-\frac\OmegaPlanet\OmegaBeam\right),
\end{equation}
where $\Tantp$ is the unknown planet antenna temperature, $\Background$ the background, and $\BeamResponset(\delta\PlanetPointingt)$ the beam response for the planet at the time of observation. Of course, $\BeamResponset$ depends on the relative position of the planet with respect to the beam, $\delta\PlanetPointingt$. If a suitable beam model is available, $\BeamResponset$ can be determined and $\Tantp$ can be recovered from least squares minimization. We use an elliptic Gaussian centred on the instantaneous pointing direction as a model for the beam, because it shows a very good match with the main beam of the {\tt GRASP} model \citep{planck2014-a05}, with peak-to-peak discrepancies of the order of a few tenths of a percent (the importance of far sidelobes is negligible for a source as strong as Jupiter).  To compute $\BeamResponset$, the pointings are rotated into the beam reference frame, since this allows for better control of the beam pattern reconstruction.\footnote{This is the opposite of \OldPlanckLFICalPaper, which used the planet reference frame.}

\subsubsection{Estimation of the brightness temperature}
\label{sec:JupiterTbri}

In \OldPlanckLFICalPaper, we computed the brightness temperature $\Tb$ from the antenna temperature $\Tant$ by means of the following formula (assuming monochromatic radiometers):
\begin{equation}
\Tb = B_\textrm{Planck}^{-1} \left(\Tant \fsl \frac\OmegaBeam\OmegaPlanet
 \left.\frac{\partial B_\textrm{Planck}}{\partial T}\right|_{T_\mathrm{CMB}}
 \right),
\end{equation}
where $B_\textrm{Planck}$ is Planck's blackbody function, $\OmegaBeam$ and $\OmegaPlanet$ the beam and planet solid angles, and $T_\mathrm{CMB} = 2.7255\,\mathrm{K}$ is the temperature of the CMB monopole.  In this 2015 \Planck\ data release, we have also introduced corrections to account for the bandpass.

The accuracy of the $\Tb$ determination is affected by confusion noise (i.e., noise caused by other structure in the maps), which we estimate from the standard deviation of samples taken between a radius of $1^\circ$ and $1.5^\circ$ (depending on the beam) and $5^\circ$ from the beam centre.  These samples are masked for strong sources or other defects. Since background maps are not subtracted, the confusion noise is larger than the pure instrumental noise. However, since the histogram is well described by a normal distribution, we have used error propagation\footnote{We can quickly derive an order of magnitude for the size of the confusion noise effect in the estimation of $\Tant$. Since the confusion noise is of the order of a few mK and the number $N$ of samples within 2 FWHM (the radius used for estimating $\Tb$) is of the order of $10^3$--$10^4$, we can expect an accuracy of the order of $1\,\textrm{mK}/\sqrt{N} \approx 0.1\,\textrm{mK}$.} to assess the accuracy of $\Tb$ against the confusion noise.

In the conversion of $\Tantp$ into $\Tb$ through Eq.~\eqref{eq:TantToTb}, the pipeline implements a number of small corrections:
\begin{enumerate}
\item detector-to-detector differences in the beam solid angle $\OmegaBeam$, accounting for $\pm6\,\%$, which is probably the most important effect;
\item changes in the solid angle of the planet, $\OmegaPlanet$, due to the change of the Jupiter--\Planck\ distance, which introduces a correction factor of up to $6.9\,\%$ percent; 
\item changes in the projected planet ellipticity, due to the planeto-centric latitude of the observer and the oblateness of the planet, to reduce observations as if they were made at Jupiter's pole;
\item blocking of background radiation by the planet, changing from about $0.7\,\%$ to $1.5\,\%$, depending on the ratio $\OmegaPlanet/\OmegaBeam$; \item a $\phi_{\rm sl}$ correction, which accounts for the fraction of radiation not included in the main beam (about $0.2\,\%$).
\end{enumerate}

\subsubsection{Determination of the solid angle of Jupiter}
\label{sec:JupiterSolidAngle}

The solid angle $\OmegaPlanet$ of Jupiter for a given planet-spacecraft distance $\DeltaPlanet$ and planeto-centric latitude $\PlanetocentricPlanckLatitude$ is given by
\begin{equation}
 \OmegaPlanet(\PlanetocentricPlanckLatitude) = \OmegaPlanetPolarRef
 \left(\frac{\DeltaPlanetRef}{\DeltaPlanet}\right)^2 \PlanetFlattening
 \sqrt{1-(1-\PlanetFlattening^2)\sin^2\PlanetocentricPlanckLatitude},
\end{equation}
where $\DeltaPlanetRef$ is a fiducial planet-spacecraft distance (for Jupiter, $\DeltaPlanetRef \approx 5.2\,\text{AU}$), $\OmegaPlanetPolarRef$ is the solid angle of the planet as seen from its pole at the fiducial distance, and $\PlanetFlattening$ $(<1)$ the ratio between the polar and equatorial radii of the planet.

Our new pipeline does not use the ellipticity, FWHM, and orientation of the elliptical beam parameters provided by \citet{planck2014-a05}, as they were derived from a marginalization over $\Tantp$ on the same Jupiter data used in this analysis. However, because of a degeneracy between $\Tantp$ and the beam parameters, the results were sensitive to details of the fitting procedure (up to about $1\,\%$), such as the radius of the area being analysed and the minimization method. Therefore, we have first determined a new set of beam parameters for each transit and computed the weighted average (using a numerical minimization), and then we used such parameters to determine $\Tantp$.

\subsection{Results}

We present here the results per radiometer and transit, and we discuss also how to combine the measurements performed using the 30, 44, and 70\,GHz radiometers to obtain three estimates of $\Tb$ at these three nominal frequencies.

\subsubsection{Brightness temperatures per radiometer and transit}

\begin{figure}
	\centering
	\includegraphics[width=\columnwidth]{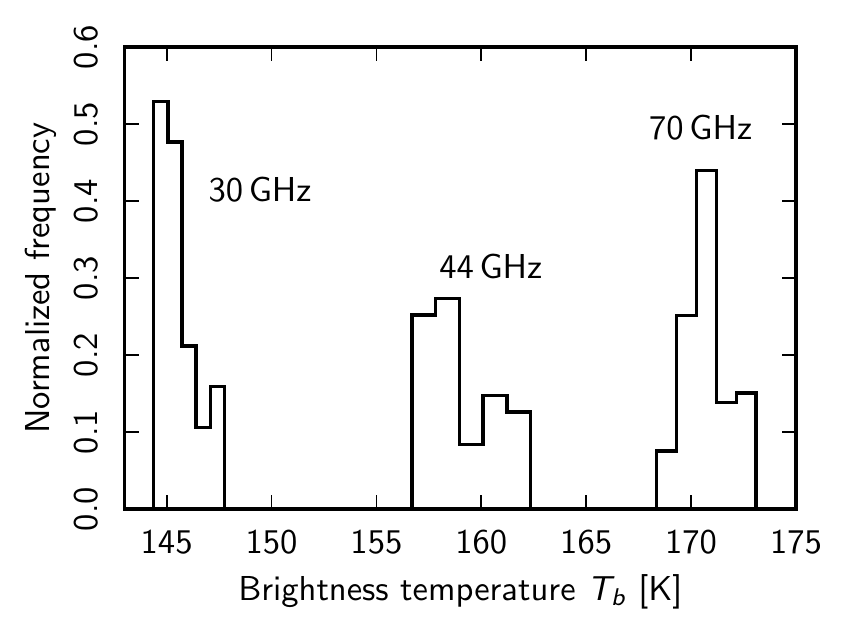}
	\caption{\label{fig:jupiterBrightnessTemperatureHist} Distribution of the values of $\Tb$ for Jupiter measured by the 22 LFI radiometers during each of the seven transits. The histogram has been produced using five bins per frequency. The lack of Gaussianity in the three distributions is evident.}
\end{figure}

In \OldPlanckLFICalPaper, based on the first two transits, we noted that confusion noise alone was not able to account for variations in $\Tb$ found for different radiometers belonging to the same frequency channel. We therefore assumed the presence of some unidentified systematics error, dominating the ultimate accuracy in the measures.

The fact that residual systematics are more important than confusion noise and background is still true in the 2015 data release. The values of $\Tb$ measured by radiometers in the same frequency channel have a spread of 0.6\,\%, 1.0\,\%, and 0.6\,\% of the average signal at 30, 44 and 70\,GHz, respectively, and are not normally distributed (see Fig~\ref{fig:jupiterBrightnessTemperatureHist}). These observed dispersions are a factor of $\sim 3$ larger than the confusion noise, and cannot be ascribed to the background, whose effect is only 1\,\% of the observed scatter. The excess dispersion must be due to a small residual systematic effect such as pointing, beam model, or mismatch of center frequency.

One possible cause of the observed dispersion in the brightness temperature is some systematic effect in the estimation of the beam parameters (see Sect.~\ref{sec:JupiterSolidAngle}).

Non-Gaussianities in the beam, as well as beam smearing were investigated by replacing the elliptical beam with band averaged beam maps derived from {\tt GRASP} calculations. The results are consistent with the elliptical beam, with residuals of at most $4\times10^{-3}\,\unit{K}$ in $\Tantp$.

We computed $\Tantp$ and $\Tb$ again using an analytical approximation based on the assumption of negligible background (which is quite a good approximation for Jupiter), and compared the results. The two methods agree at the level of $6\times10^{-7}\,\unit{K}$ for $\Tantp$, and at the level of $6\times10^{-4}\,\unit{K}$ for $\Tb$.

\subsubsection{Combination of the results and comparison with WMAP}

We now want to combine the measurements of the 22 LFI radiometers in order to have three estimates of $\Tb$ at the three LFI nominal frequencies, 30, 44, and 70\,GHz.

Such determination of $\Tb$ depends on proper knowledge of the central frequency, $\Fcent$, for each detector. This parameter is derived from the bandpasses, but these are not known exactly. It is possible to remove most of the differences among the 30\,GHz and among the 44\,GHz radiometers by slightly changing the $\Fcent$ values of the radiometers by as little as $\pm0.2\,\unit{GHz}$. However, it is still not clear how the bandshapes would have to be modified to explain such changes in $\Fcent$. For this reason, we did not include such corrections for 30\,GHz and 44\,GHz data.

\begin{figure*}[tb]
	\centering
	\includegraphics[width=0.95\textwidth]{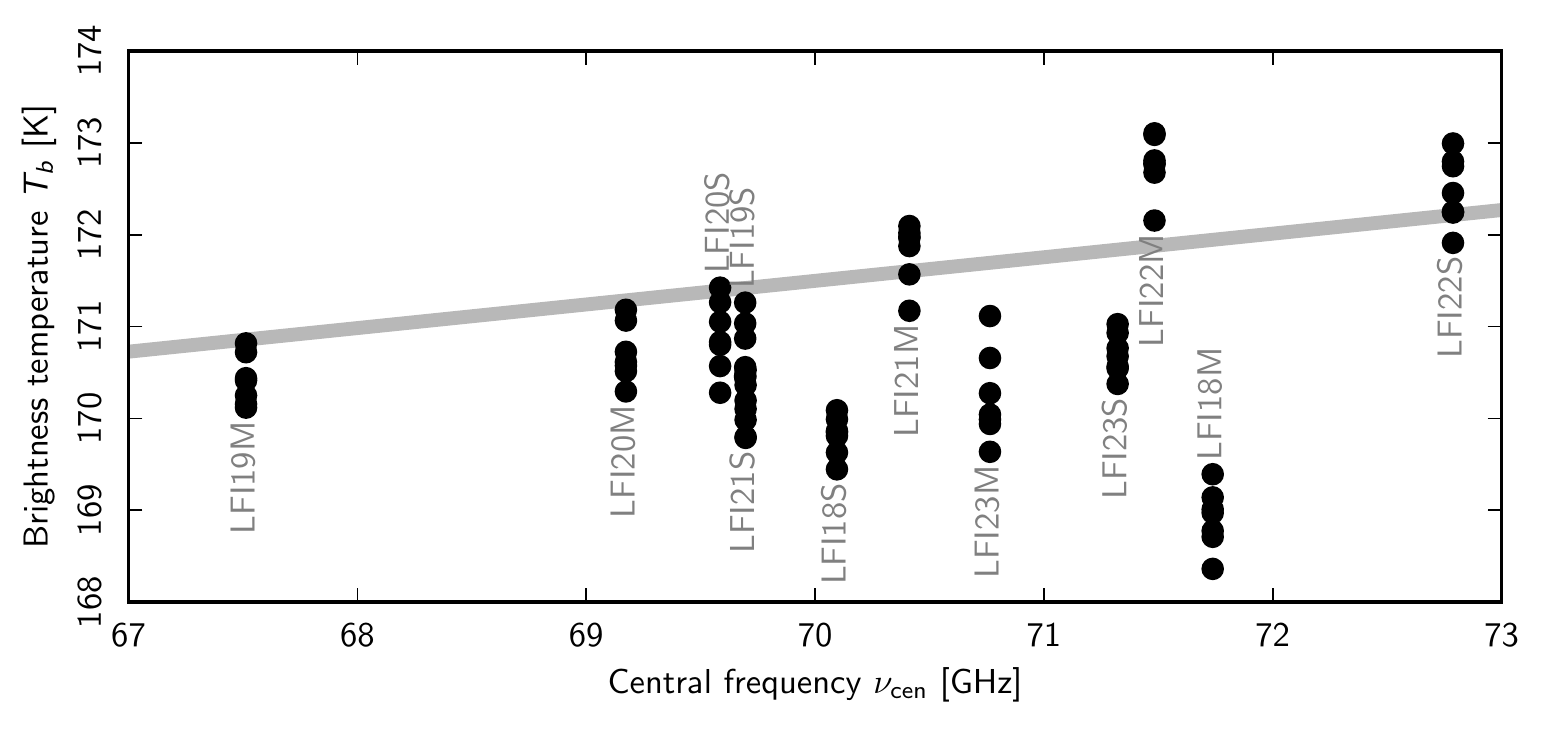}
	\caption{\label{fig:TbVsNuCen}Distribution of the values of $T_b$ for Jupiter as a function of the central frequency $\Fcent$ of each radiometer.}
\end{figure*}

At 70\,GHz the situation is complicated even more by the fact that the $\Fcent$ values of the radiometers are significantly spread over the channel bandwidth, so that each of the 70\,GHz radiometers samples a slightly different portion of the Jupiter spectrum. Indeed, comparing $\Tb$ at 70\,GHz from one transit to another after having ordered the radiometers for increasing $\Fcent$, it is possible to see quite a significant correlation (see Fig.~\ref{fig:TbVsNuCen}).  Using a linear regression of the $\Tb$ at 70\,GHz against $\Fcent$, a ``cleaned'' list of $\Tb$ values was obtained.   Their average is identical to the simple weighted average of $\Tb$ for all of the transits and of the detectors $\Tb = (171.558 \pm 0.008)\,\unit{K}$. The standard deviation of the whole set of samples reduces to just 2.7\,\%. Interestingly, the inferred slope $d\Tb/d\Fcent = (0.2570 \pm 0.0058)\,\unit{K}/\unit{GHz}$ matches very well with the one from WMAP data, $d\Tb/d\Fcent = (0.243 \pm 0.025)\,\unit{K}/\unit{GHz}$, and the correlation between transits loses most of its statistical significance. We attempted the same test for the 30\,GHz and 44\,GHz channels, but the spread in $\Fcent$ is too small to produce meaningful results. These results open the possibility of including the spectral slope directly as a free parameter of the fit in a future analysis.

\begin{table}
\begingroup
\newdimen\tblskip \tblskip=5pt
\caption{\label{tab:jupiterTemperatures} Brightness temperature of Jupiter.}
\nointerlineskip
\vskip -3mm
\footnotesize
\setbox\tablebox=\vbox{
   \newdimen\digitwidth
   \setbox0=\hbox{\rm 0}
   \digitwidth=\wd0
   \catcode`*=\active
   \def*{\kern\digitwidth}
   \newdimen\signwidth
   \setbox0=\hbox{+}
   \signwidth=\wd0
   \catcode`!=\active
   \def!{\kern\signwidth}
\halign{\tabskip 0pt\hbox to 2.0cm{#\leaderfil}\hfil\tabskip 2.0em&
    \hfil$#$\hfil\tabskip 0pt\cr
\noalign{\doubleline}
\omit $\Fcent$\hfil& \Tb\cr
\omit [GHz]\hfil& [{\rm K}]\cr
\noalign{\vskip 3pt\hrule\vskip 5pt}
\noalign{\vskip 2pt}
28.4& 145.9\pm0.9\cr
44.2& 159.8\pm1.4\cr
70.4& 171.6\pm1.0\cr
\noalign{\vskip 5pt\hrule\vskip 10pt}}}
\endPlancktable                    % ends one-column \halign
\endgroup
\end{table}

The result of this analysis produced the brightness temperatures $\Tb$ listed in Table~\ref{tab:jupiterTemperatures}. The errors are 0.6\,\%, 0.9\,\%, and 0.6\,\% at 30, 44, and 70\,GHz, respectively.

\begin{figure}
     \includegraphics{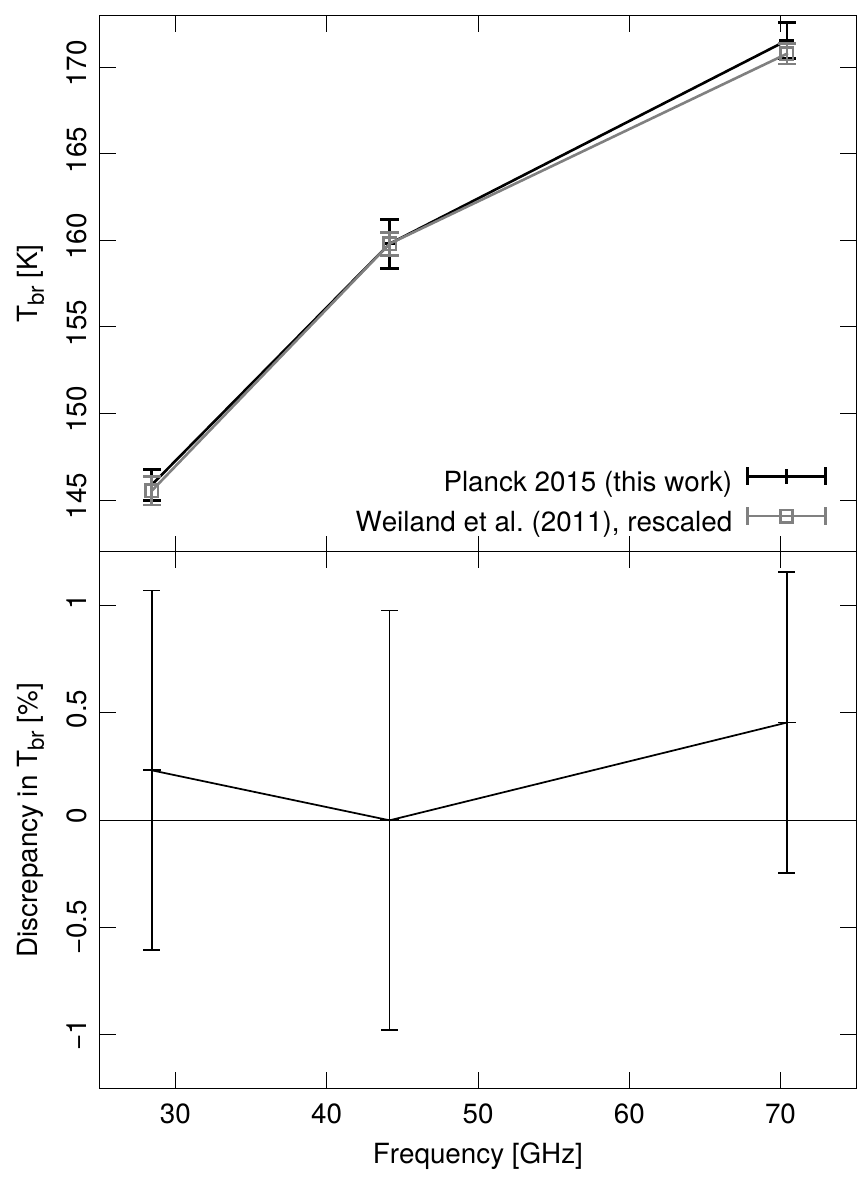}
     \caption{\label{fig:jupiterSpectrum} \emph{Top}: brightness temperature of Jupiter ($T_\mathrm{br}$) compared with the data from \protect\citet{weiland2010}, linearly rescaled in frequency to match LFI's central frequencies $\Fcent$ and corrected for difference between LFI's and WMAP's dipoles. \emph{Bottom}: deviation from unity of the ratio between LFI's estimate for $T_\mathrm{br}$ and WMAP's. The agreement is excellent among the three frequencies.}
\end{figure}

As in \OldPlanckLFICalPaper, in Fig.~\ref{fig:jupiterSpectrum} we compared Jupiter's $\Tb$ averaged in each LFI band with the spectrum provided by WMAP.  The agreement is quite good, with a difference that does not exceed $0.5\%$. In this comparison, we must note that WMAP and \Planck-LFI are calibrated on slightly different dipoles, with \Planck{} assuming an amplitude of $3364.5\,\unit{\mu K}$ \citep{planck2014-a01}, while WMAP assumed an amplitude of $3355\,\unit{\mu K}$ \citep{hinshaw2009}. In addition, WMAP central frequencies are different from those of $\Planck$. So to make the comparison, the measures provided in \citet{weiland2010} are scaled by $1.00268$ and linearly interpolated to the averaged $\Fcent$ of each LFI frequency channel.

\subsection{What does the analysis of Jupiter tells about the LFI calibration?}

The errors in our estimates of Jupiter's brightness temperature are 0.6\,\%, 0.9\,\%, and 0.6\,\%. These are consistent with (albeit slightly larger than) the numbers presented in Sects.~\ref{sec:interChannelConsistency} and \ref{sec:interFrequencyConsistency}, where our analysis on the consistency of inter-channel and inter-frequency cross-spectra in the range $100 \leq \ell \leq 250$ produced estimates between 0.1\,\% and 0.3\,\%. We believe that the larger error bars are a direct consequence of a number of facts: (1) bandpass uncertainties play a larger role in the analysis of a non-thermodynamic sky signal like Jupiter's emission in the microwave range; (2) the amount of data usable for this analysis is only a limited fraction of the overall amount used to create LFI maps; and (3) uncertainties in the main beam profile have a larger impact in the study of a point-like source such as Jupiter, with respect to the analysis presented in Sect.~\ref{sec:validationAndAccuracy}, which dealt with angular scales corresponding to $100 \leq \ell \leq 250$.

%%%%%%%%%%%%%%%%%%%%%%%%%%%%%%%%%%%%%%%%%%%%%%%%%%%%%%%%%%%%%%%%%%%%%%

\section{Conclusions}
\label{sec:conclusions}

We have described the method used to calibrate the \Planck/LFI data for the 2015 \Planck\ data release, and we have provided a quantitative analysis which shows in detail the amount of change in the calibration level due to every improvement we implemented in the pipeline since the previous data release. Compared to our 2013 release, we have improved the LFI calibration in several ways, most notably in the use of an internally consistent dipole signal as a calibrator, as well as more accurate data analysis algorithms. As a result, we have improved by a factor of 3 the accuracy of the LFI calibration, which for the 2015 release ranges from 0.20\,\% to 0.35\,\%, depending on the frequency.

An important byproduct of our analysis is a novel estimate of the solar dipole signal; using LFI data only, we have estimated the amplitude of the dipole to be $3\,365.6\pm3.0\,\unit{\mu K}$ and the direction of its axis  to be $(l, 90^\circ - b) = (264.01\pm0.05^\circ, 48.26\pm0.02^\circ)$ (Galactic coordinates). This result matches the numbers provided by \citet{hinshaw2009} within $1\sigma$. This slight difference in amplitude has the effect of a shift by $\sim 0.3\,\%$ in the overall level of the calibrated timelines. Together with the improved LFI calibration, we now find very good agreement between the level of the spectra estimated by WMAP.

\begin{acknowledgements}
The Planck Collaboration acknowledges the support of: ESA; CNES, and CNRS/INSU-IN2P3-INP (France); ASI, CNR, and INAF (Italy); NASA and DoE (USA); STFC and UKSA (UK); CSIC, MINECO, JA and RES (Spain); Tekes, AoF, and CSC (Finland); DLR and MPG (Germany); CSA (Canada); DTU Space (Denmark); SER/SSO (Switzerland); RCN (Norway); SFI (Ireland); FCT/MCTES (Portugal); ERC and PRACE (EU). A description of the Planck Collaboration and a list of its members, indicating which technical or scientific activities they have been involved in, can be found at \url{http://www.cosmos.esa.int/web/planck/planck_collaboration}.
\end{acknowledgements}

\bibliographystyle{aat}
\bibliography{Planck_bib,custom_bibliography}

\appendix

\section{Smoothing calibration curves}
\label{sec:smoothing}
In this appendix we explain in some detail the algorithm used to smooth the LFI calibration constant timelines. This task is performed by the smoothing filter discussed in Sect.~\ref{sec:Pipeline} (see also Fig.~\ref{fig:calPipelineDiagram}).

\subsection{Purpose of the smoother}

Variations in the orientation of the \Planck\ spacecraft during the mission caused its instruments to observe the calibration signal (the CMB solar dipole) with varying amplitude. This in turns induced variations in the accuracy of the reconstruction of the calibration constant $K = G^{-1}$ (Eq.~\ref{eq:radiometerEquation}), as shown in Fig.~\ref{fig:gainCurve}. Such variations are mainly due to statistical errors and are in general unrelated to the true stability of the LFI detectors. Therefore, before applying such calibration constants to the raw data measured by the LFI radiometers, we have applied a low-pass filter that smooths most of the high-frequency fluctuations.

In \OldPlanckLFICalPaper{} we employed a simple smoothing filter, which used wavelets to clean the stream of calibration constants of high-frequency fluctuations (note that no smoothing was applied to the 30\,GHz radiometers, since the calibration algorithm used for them made this step unnecessary).

In the 2015 \Planck\ data release, we have improved our smoothing filter in order to take into account sudden jumps in the calibration constant that are caused by a genuine change in the state of the radiometer; we call these ``real jumps.'' Such real jumps are not due to statistical effects and it is therefore incorrect to include such jumps when smoothing the data. The actual smoothing algorithm used in the 2015 \Planck-LFI data release works as follows:
\begin{enumerate}
\item determine if there are sudden variations in the calibration constants that might be of non-statistical origin and make a list of them;
\item split the stream of calibration constants into sub-streams, using as boundaries the jumps found in the previous step;
\item apply a low-pass filter to each sub-stream defined in the previous step.
\end{enumerate}
In the following paragraphs we will provide more details about the implementation of these steps.

\subsection{Detecting jumps in the calibration constants}

We discuss here how our data analysis code is able to determine the presence of sudden jumps in the calibration constant that are due to some real change in the state of the radiometer. This task is not trivial, since we must look for sudden variations in a stream of numbers (the $K$ values) that is dominated by statistical noise; moreover, the rms of such data changes with time, since it is correlated with the amplitude of the dipole signal $\Delta T_\mathrm{dip}$ (see Fig.~\ref{fig:gainCurve}). Within a region of high rms, it is therefore possible to mistake a sudden change in the value of $K$ due to statistical fluctuations with an intrinsic change in the radiometer's calibration. We therefore have developed a figure of merit which allows us to disentangle these two families of jumps. 

\begin{figure}
    \centering
    \includegraphics[width=\columnwidth]{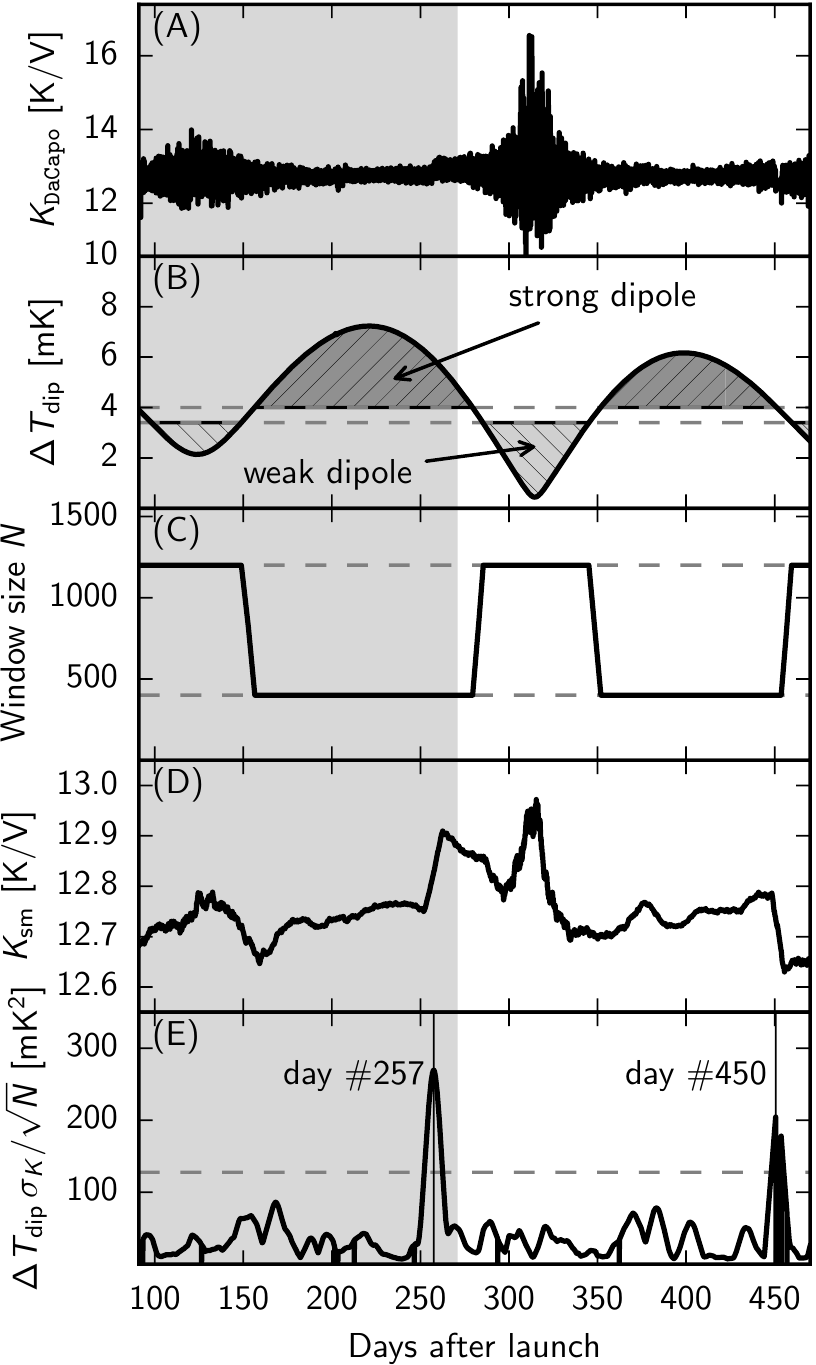}
    \caption{\label{fig:smootherFindJumps} Example of application of the algorithm for the detection of jumps in the stream of $K=G^{-1}$ values (Eq.~\protect\ref{eq:radiometerEquation}) for LFI-27M during \Planck's first year. \textit{Panel A:} the set of calibration constants $K$ computed by \texttt{Da Capo} for LFI27M. The $x$-axis here is cropped to the first two sky surveys (one year of data). \textit{Panel B:} the amplitude of the dipole signal, $\delta T_{\rm dip}$, as seen within each ring by the \Planck\ spacecraft. The two horizontal lines mark the thresholds for regions where the signal is considered either ``strong'' or ``weak.'' \textit{Panel C:} the size $N$ of the window used to compute the moving average of $K$ (there are roughly $N=30$ values of $K$ per day). In regions where the dipole signal is weak or strong, the window width is 1200 or 400 samples, respectively; outside such regions, we use a linear interpolation between these two values. \textit{Panel D:} the result of applying a moving average with the variable window size (Panel C) to the series of data shown in Panel A. This is \emph{not} the smoothed series used for calibration; only the rms of the moving average is used (see next panel). \textit{Panel E:} the figure of merit used to detect jumps is the product of the dipole amplitude and the rms of the moving average. The threshold used to detect jumps in this particular example (LFI-27M) is equal to the 99th percentile of such values (grey dashed line). In this case, two jumps have been found (days 257 and 450).}
\end{figure}

In order to implement our figure of merit, we have defined a procedure to quantify the level of statistical fluctuations in the data. Such a procedure is similar to a smoothing filter and we describe it with the aid of Fig.~\ref{fig:smootherFindJumps}, which shows its application to some real data (the values of $K$ for radiometer LFI-27M, calculated during the first two sky surveys). As already described in Sect.~\ref{sec:Pipeline}, the variation in the statistical noise in $K$ (panel A) is related to the strength of the dipole signal, i.e., the amplitude of $\Delta T_\mathrm{dip}$ (panel B). Therefore, in order to properly weight the importance of variations in the value of $K$, we apply a moving average to the stream of $K$ values, where the amplitude of the window (panel C) depends on the value of $\Delta T_\mathrm{dip}$. The result of the moving average ($K_\mathrm{sm}$) is shown in panel D of Fig.~\ref{fig:smootherFindJumps}. For each window, the code computes the rms, $\sigma_{K}$, of the $N$ values. This quantity depends both on the statistical noise and on the presence of real jumps in the value of $K$. Therefore, we use the expression
\begin{equation}
\label{eq:smoothingFigureOfMerit}
\Delta T_\mathrm{dip} \times \frac{\sigma_K}{N}
\end{equation}
as a figure of merit for the determination of the presence of real jumps. This quantity is proportional to the rms of the moving average, but it is weighted by the amplitude of the calibration signal; the stronger the latter, the more likely that a real jump is present in the $N$ samples.

The threshold used with Eq.~\eqref{eq:smoothingFigureOfMerit} is defined in terms of percentiles, specifically, we consider all the data that are greater than the $n$-th percentile to mark the presence of a real jump. The value for $n$ depends on the radiometer: for 30\,GHz radiometers it is 99; for 44\,GHz it is 99.9; and for 70\,GHz it is 99.5. Such values have been determined by considering the quality of the null\footnote{The null tests we used in this process are typically the difference between survey maps; in the case of an ideal, perfectly calibrated instrument, such differences should yield a map where all the pixels are zero.} test (see Sect.~\ref{sec:smoothingValidation}).

\subsection{The smoothing algorithm}

Once the positions of the jumps have been determined, the code applies a smoothing algorithm to each subset of values of the original $K$ (produced by \texttt{Da Capo}) between two consecutive jumps. The algorithm applies a low-pass filter in the Fourier domain that retains only 5\,\% of the lowest frequencies. After this step, in order to further reduce the noise, we apply a moving average to the result, where each sample is weighted by the value of $\Delta T_\mathrm{dip}$.

\subsection{Validation of the algorithm}
\label{sec:smoothingValidation}

\begin{figure*}
    \centering
    \includegraphics[width=\textwidth]{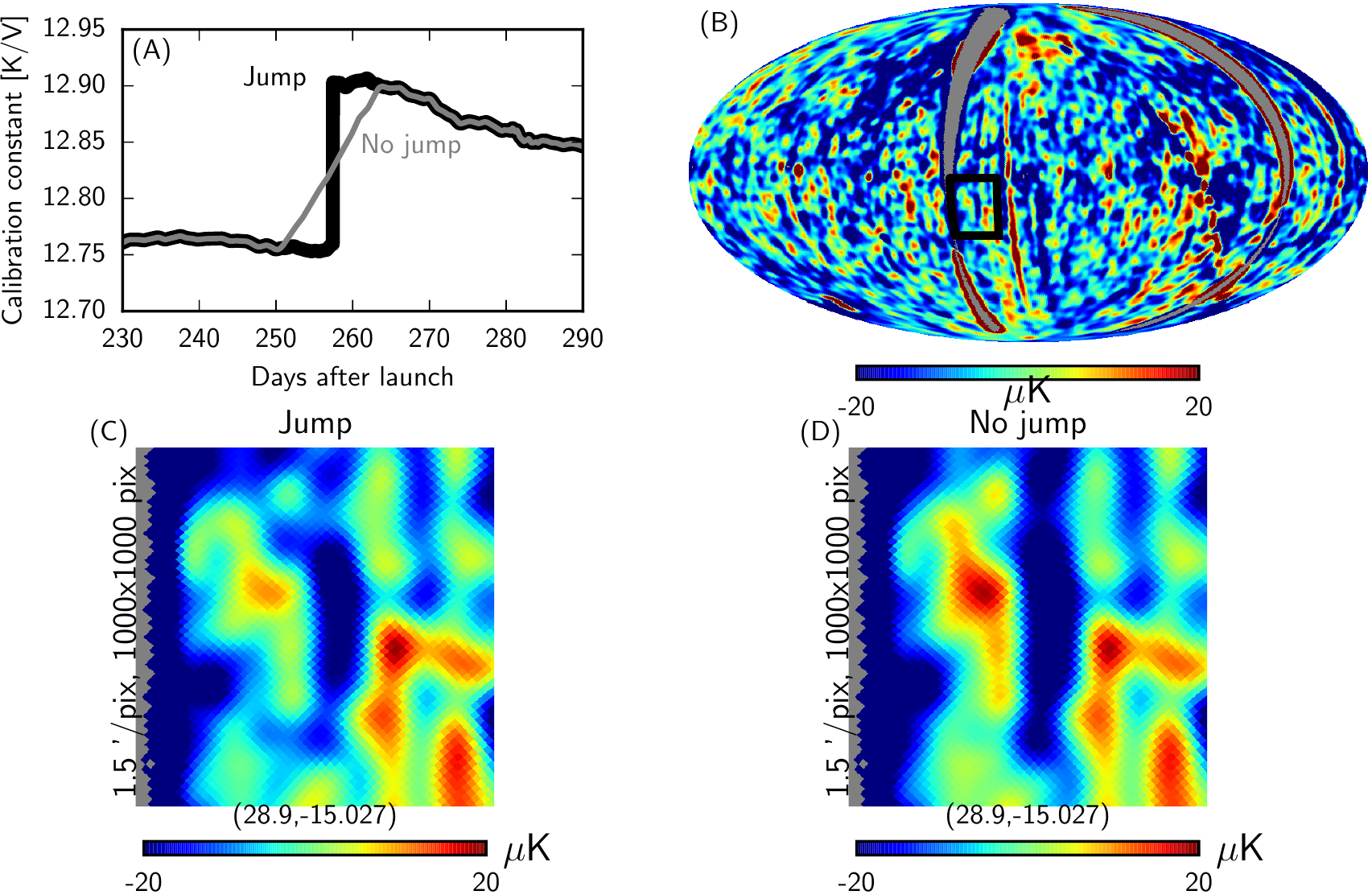}
    \caption{\label{fig:smootherMaps} Comparison between the behaviour of the smoothing code in the nominal case and in the case where the jump near day 257 is considered to be a statistical fluctuation. \textit{Panel A}: value of the calibration constant in the two cases. Note that fast variations have been removed from the data in order to make the plot clearer. If the jump is not considered to be real by the algorithm, the smoothing stage introduces an increasing slope in the values (thin grey line). \textit{Panel B}: map of the difference between the maps from Surveys~1 and 2 (the jump happened during the first survey), in ecliptic coordinates. Residual systematic effects produce stripes that are either aligned with the direction of the scan (i.e., perpendicular to the ecliptic plane) or with the Galactic plane. The black square highlights a region in the map that has been observed during the jump near day 257. \textit{Panel C}: zoom into the region highlighted in panel B (the difference between maps from Surveys~1 and 2), when the jump near day 257 has been considered real (see Panel A, thick black line). \textit{Panel D}: the same as the previous panel, but data have been calibrated assuming no real jump near day 257 (Panel A, thin grey line). Features are sharper in the latter case, and therefore we can conclude that the former calibration produces better results.}
\end{figure*}

Since an incorrect identification of a real jump is likely to produce stripes in maps, we have run a number of null tests in order to optimize the free parameters of the smoothing algorithm (e.g., the threshold used to detect jumps and the widths of the moving average windows).  We have calibrated the data using a number of combinations of parameters and have produced single-survey sky maps (i.e., maps obtained using six months of data). We have then differenced them, under the hypothesis that a perfect calibration would produce a map where the value of every pixel is consistent with zero. Fig.~\ref{fig:smootherMaps} shows an example of this analysis; this shows that the jump found by the code near day 257 is likely to be a real jump, because not considering it as such leads to a stronger stripe in survey-difference maps.

\raggedright
\end{document}